\documentstyle[12pt,A4]{article}

\input epsf
\epsfverbosetrue
\begin{document}
\begin{titlepage}
\begin{tabbing}
\hspace{11cm} \= HIP -- 1998 -- 19 / TH \\
\> LTH -- 422\\
\> \today
\end{tabbing}

\begin{centering}
\vfill
{\Large\bf Four-quark flux distribution and binding \\in lattice SU(2)}
\vspace{1cm}

P. Pennanen\footnotemark[1],
A. M. Green\footnotemark[2]$^,$\footnotemark[3]

{\em Helsinki Institute of Physics \\ $\,^2$Department of Physics\\
P.O. Box 9, FIN-00014 University of Helsinki, Finland }

and C. Michael\footnotemark[4]

{\em Theoretical Physics Division, Dept. of Math. Sciences,
University of Liverpool, Liverpool, UK.}

\setcounter{footnote}{1}
\footnotetext{E-mail: {\tt petrus@hip.fi}}
\setcounter{footnote}{3}
\footnotetext{{\tt anthony.green@helsinki.fi}}
\setcounter{footnote}{4}
 \footnotetext{{\tt cmi@liv.ac.uk}}
\renewcommand{\thefootnote}{\arabic{footnote}}
\vspace{1.5cm}

\begin{abstract}  The full spatial distribution of the 
color fields of two and four static quarks is 
measured in lattice SU(2) field theory at separations 
up to 1 fm at $\beta=2.4$.  The four-quark case is equivalent to a 
$q\bar{q}q\bar{q}$ system in SU(2) and is relevant to meson-meson 
interactions. By subtracting two-body flux tubes from the four-quark 
distribution we isolate the flux contribution connected with the four-body 
binding energy. This contribution is further studied using a model for the 
binding energies.
Lattice sum rules for two and four quarks are used to verify the results.

\noindent

\end{abstract}

\end{centering}
\end{titlepage}

\section{Introduction}

Monte Carlo simulations of lattice gauge theory are among the most powerful 
tools for investigating non-perturbative phenomena of QCD such as confinement. 
The potential $V(R)$ between two static quarks at separation $R$ in 
quenched QCD is a simple manifestation of confinement and 
has been studied intensively. At large $R$ the potential rises linearly 
as predicted by the hadronic string model. One can also measure the spatial
distribution of the color fields around such static quarks in order to
get a detailed picture of the confining flux tube.  In Refs.
\cite{gre:96,pen:97b}, which contain references to earlier work,  this was
done for the  ground state and two excited
states of the two-quark potential. Transverse and longitudinal profiles of
chromoelectric and -magnetic fields  were compared with vibrating string
and dual QCD models for the flux  tube, with the latter model
reproducing quite well the shape of the energy profile measured on a
lattice. Instead of SU(3), the gauge group used was SU(2), which  is
more manageable with present-day computer resources and is expected to
have very similar features of confinement. This is reflected in the fact
that the flux tube models considered do not distinguish between SU(2)
and SU(3) and in the small $N_c$ dependence observed in the spectrum of 
pure gauge theories \cite{tep:97}.

A more complicated situation is encountered with multi-quark systems, which 
are abundant in nature and whose understanding from first principles, i.e.
from QCD, is at present very limited. This is mainly due to the failure of
perturbation theory in this intermediate energy domain and the heavy 
computer requirements for Monte Carlo simulations. The simplest 
multi-quark system, this meaning more than a single meson or baryon, 
consists of four quarks and occurs e.g. in meson-meson 
scattering and bound states. Understanding the four-quark interaction would 
be the first step 
in deriving nuclear physics from QCD. Previously, static four quark systems 
have been extensively simulated in SU(2) lattice gauge theory and a 
phenomenological potential model containing a many-body interaction term $f$ 
has been developed to explain the observed binding energies 
\cite[references therein]{glpm:96,pen:96b}. Here binding energies, which
have values up to $\approx 100$ MeV, mean $E_4-[V_2(a)+V_2(b)]$, where $E_4$ 
is the 
total energy of four quarks and $V_2(a)+V_2(b)$ the energy of the lowest lying
pairings '$a$' and '$b$' of the quarks. This so-called $f$-model with 
four independent parameters, and including 
the effect of excited gluonic states, has been found in Refs. 
\cite{gre:97,gre:98} to reproduce 100 measured energies of the six types of 
four-body geometries we have simulated.

In order to gain more insight into the binding of multi-quark systems we now
look at the microscopic properties of the color fields around
four static quarks. We are not aware of any serious theoretical model for the 
fields in this case. For this first study we treat a geometry where the quarks 
sit at the corners of a square. This geometry was chosen mainly because a 
simple version of the $f$-model using only two-body ground state potentials 
reproduces the observed binding energies. 

This paper is organised as follows: The method we use to measure 
the fields is first presented in Sect. \ref{smethod} along with the 
details of our simulation and data analysis techniques. The resulting 
potentials and binding energies are discussed in Sect. \ref{se}.
These are input for the two- and four-body lattice sum rules presented in 
Sect. \ref{ssum} which relate the energies to sums over flux distributions
and help us to see when the measurement of the latter is accurate. 
Using the results from the sum rule check as a guide, flux distributions 
before and after subtracting two-body flux tubes from the four 
quark distribution are shown in Sect. \ref{sdist}.  
In Sect. \ref{sf} we analyse the fields using the simple $f$-model, and 
Sect. \ref{sconc} contains our conclusions. 

\section{Measurement method \label{smethod}}

\subsection{Color fields \label{scolmeas}}

The method used to study the color fields on a lattice is to measure
the correlation of a plaquette $\Box\equiv{1 \over N} {\rm Tr}
U_{\Box}$ 
with the Wilson loop $W(R,T)$ that represents the static
quark and antiquark  at separation $R$.  When the plaquette is located
at $t=T/2$ in the   $\mu,\ \nu$ plane, the following expression
isolates, in the limit  $T \to \infty$, the contribution of the color
field at position ${\bf r}$:
 \begin{equation}  \label{fmnT}
 f_R^{\mu \nu}({\bf r})=\left[{\langle
W(R,T)
  \Box^{\mu \nu}_{\bf r}\rangle}
-\langle W(R,T)\rangle \langle \Box^{\mu \nu}\rangle
\over {\langle W(R,T)\rangle} \right].
\end{equation}
Here $\langle \Box \rangle$ is taken in the gauge vacuum.
Like all the other expectation values, it is averaged over all lattice sites.

In the naive continuum limit these contributions are related to the mean
squared fluctuation of the Minkowski color fields by
 \begin{equation}
\label{fmn}
 f_R^{ij}({\bf r})\rightarrow  -\frac{a^4}{\beta}B^2_k({\bf r}) \quad
{\rm with} \ i,\ j,\ k\   {\rm cyclic \ \ \ and} \quad
f_R^{i4}({\bf r})\rightarrow \frac{a^4}{\beta}E^2_i({\bf r}).
\end{equation}

The squares of the longitudinal
and transverse electric and magnetic fields are identified as as
 \begin{equation} \label{ae2}
 {\cal E}_x = f^{41},\; {\cal E}_{y} = f^{42},\;  {\cal E}_{z} = f^{43},\; 
 {\cal B}_x = f^{23},\;  {\cal B}_{y} = f^{31},\;  {\cal B}_{z} = f^{12},
 \end{equation}
where the indices $1,2,3,4$ correspond to the directions $x,y,z,t$. 

These can then be combined naively to give the action density
 \begin{equation} \label{ae} 
S({\bf r})=\sum_i ({\cal E}_i + {\cal B}_i)
 \end{equation}
 \noindent and the energy density
\begin{equation}
 \label{ae3}
 E({\bf r})=
\sum_i ({\cal E}_i - {\cal B}_i )
 \end{equation}
of the gluon field.

In the special case of two quarks lying on the same lattice axis, chosen
here as the $x$-axis, we can identify the squares of the longitudinal
and transverse electric and magnetic fields as
 \begin{equation} \label{ae4}
 {\cal E}_L =  {\cal E}_x,\; {\cal E}_T =  {\cal E}_{y,z}\; {\rm and} \;
{\cal B}_L = {\cal B}_x,\;  {\cal B}_T = {\cal B}_{y,z}.
 \end{equation}

\begin{figure}[h]
\hspace{0cm}\epsfxsize=200pt\epsfbox{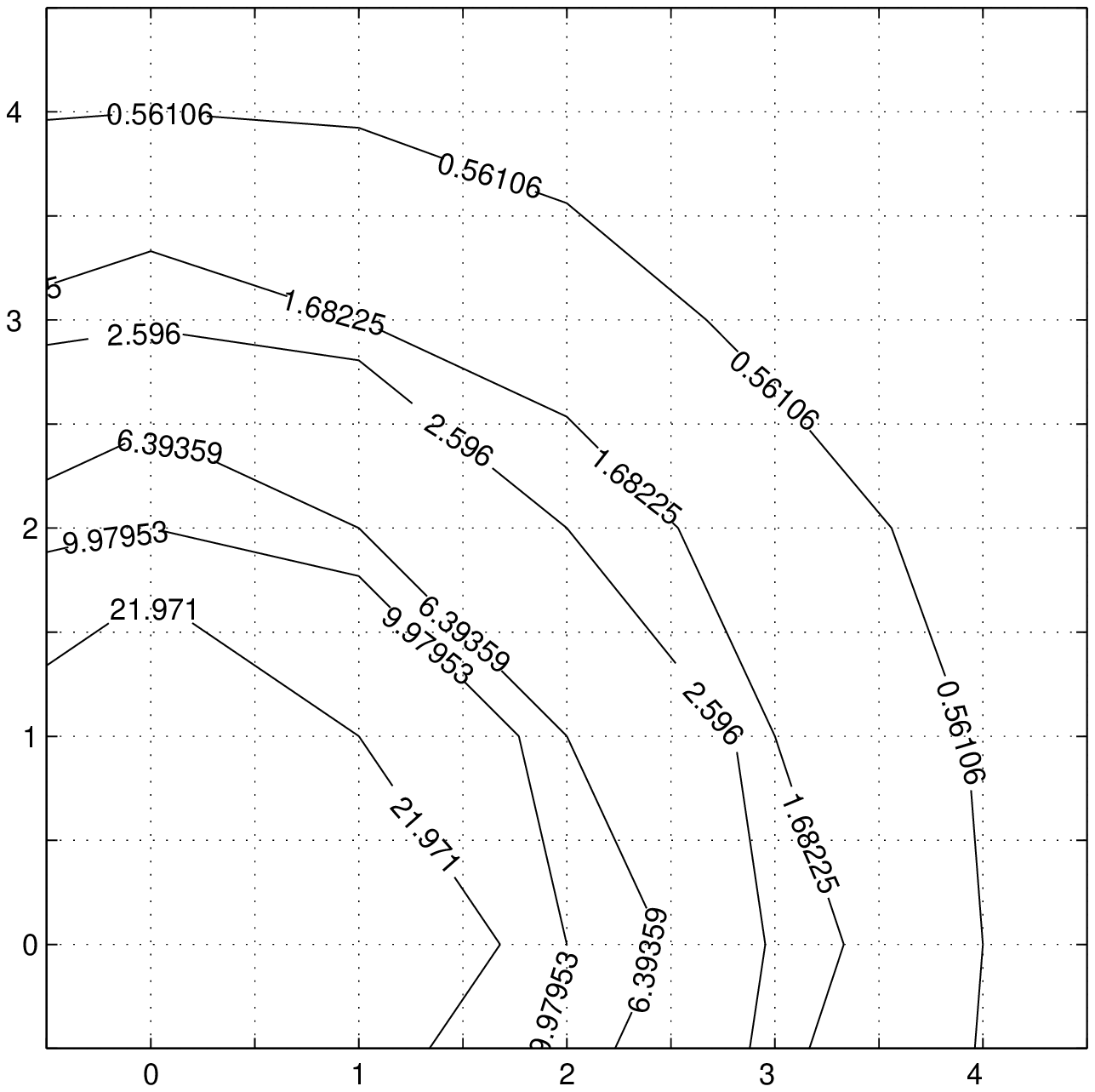}\epsfxsize=200pt\epsfbox{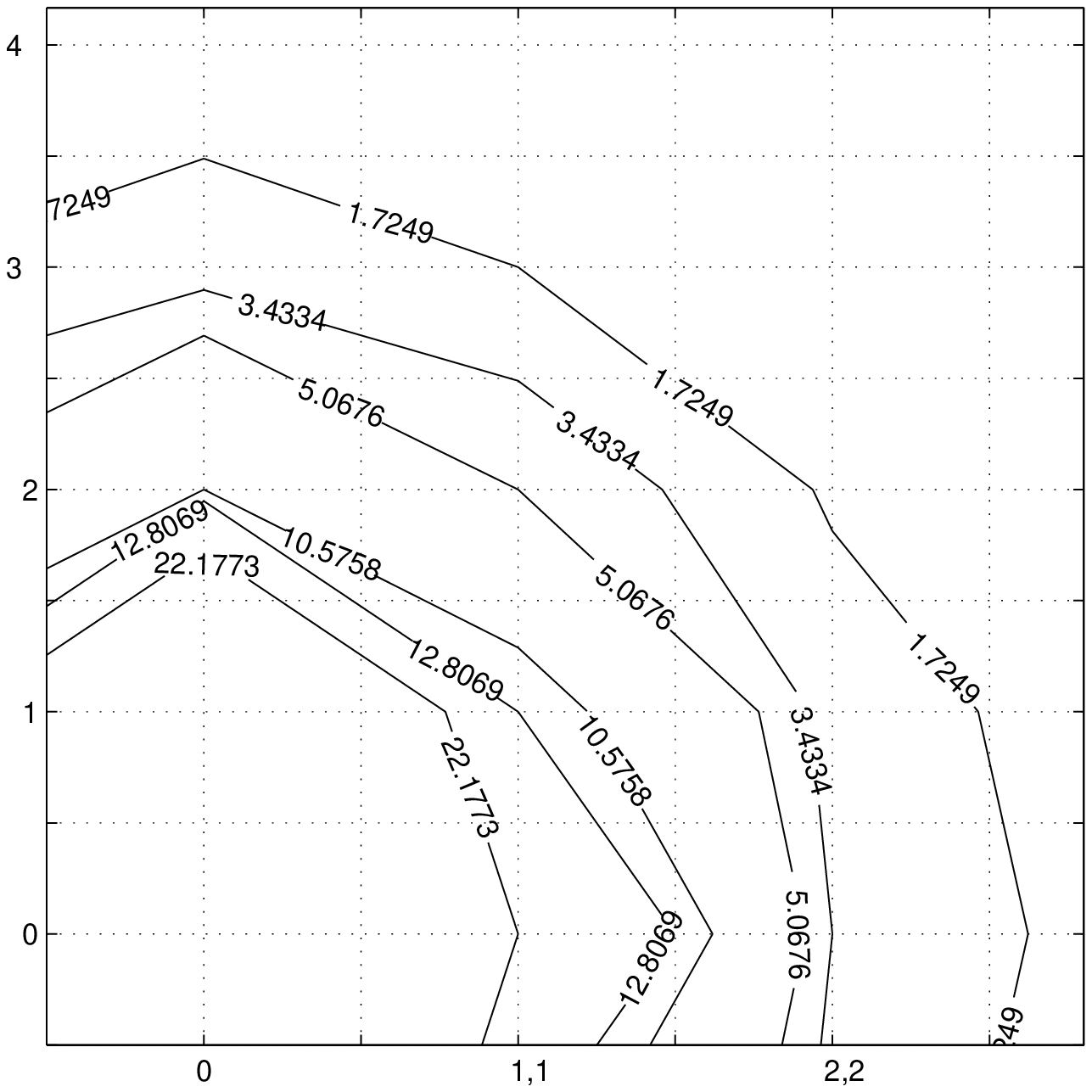} 
 \caption{Lack of rotational invariance as illustrated by the action density at $T=3$ 
in the transverse plane at the quark for a) a two-body on-axis flux tube with 
separation $R=2$ and b) a two-body diagonal tube with separation $R=2,2$. }
 \label{frotvar}
\end{figure}

Because the lattice breaks rotational symmetry, the fields were 
measured everywhere in space instead of only on the transverse lattice axis 
as in 
previous simulations. In Sect.~\ref{sf} the flux tubes for quarks at the 
opposite corners of a square are also needed. However, assuming rotational 
invariance and interpolating 
on-axis results to off-axis (diagonal) points would introduce some error into
the subtraction of
two-body distributions from the four-body ones and render the results less
reliable. The measured lack of rotational invariance of a $R=2$ on-axis flux 
tube is 
illustrated in Fig.~\ref{frotvar} a) for the action density at $T=3$ 
in the transverse plane through a color source (i.e. at the quark). The 
contour lines are drawn using
interpolation in units of GeV/fm$^3$. These values in physical units are
obtained 
by scaling the dimensionless lattice values by $\beta/a^4$, which equals 
$\approx 2418$ GeV/fm$^3$ in this case. 
In Fig.~\ref{frotvar} a) the rotational invariance is seen 
to be good except at the shortest distances; e.g. the value of the action 
density at point $(1,1)$ is achieved on-axis at a distance some 15\% 
longer, while the $(2,2)$ value is achieved at about the same
distance. 
 
A similar plot for the corresponding off-axis tube is shown in 
Fig.~\ref{frotvar} b). Because of the diagonal orientation of the tube on 
the lattice, the lattice spacings in the figure are different
in the horizontal and vertical directions; on the horizontal axis they are
$\sqrt{2}$ times the normal lattice spacing on the vertical axis. This is
because the direction perpendicular to the line connecting the diagonal
quarks, and in the same plane as the quarks, is also diagonal
 the lattice. For this off-axis
situation the lack of rotational invariance seems to persist to longer 
distances. For 
example, the value marked by the outermost contour line
at a distance of $\sqrt{10}$ is obtained at a distance about 10\% longer on the
horizontal axis. This suggests that significant error
can be introduced if an on-axis flux tube is interpolated to an off-axis 
situation, or an off-axis tube is measured only on the transverse lattice 
axis.

The parts of the fields symmetrical with respect to the quarks were 
averaged in the measurement. For the two-body on-axis case this meant
16-fold averaging; each transverse plane has eight-fold symmetry, and 
the transverse planes with equal distance from the center of the quarks
are the same. In the case of an off-axis flux tube the symmetry is only
eight-fold due to the different lattice spacings in the two directions.
For four quarks at the corners of the square we again have 16-fold symmetry;
the quark plane is divided into eight symmetrical parts, and the parts above
and below this plane are the same.

\begin{figure}[h]
\hspace{0cm}\epsfxsize=150pt\epsfbox{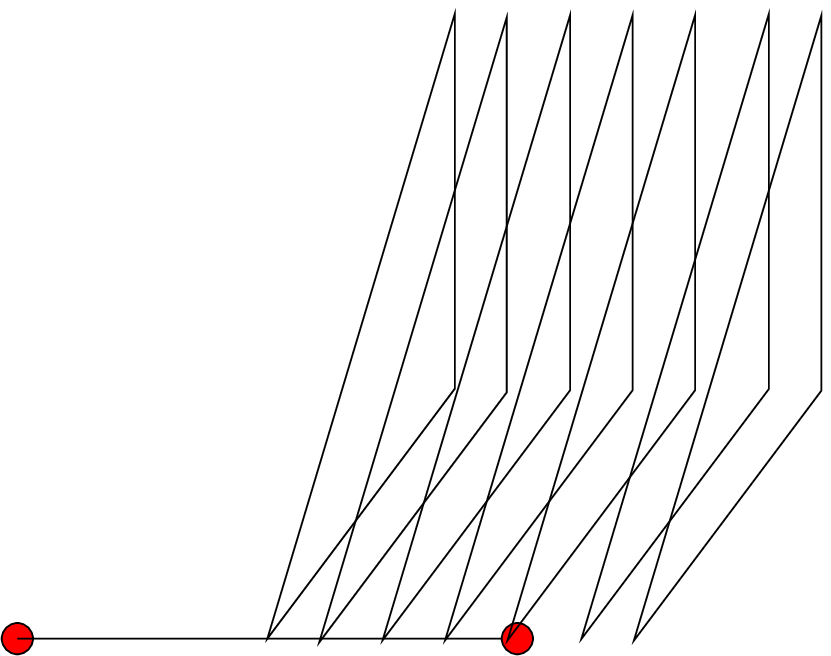}\epsfxsize=150pt\epsfbox{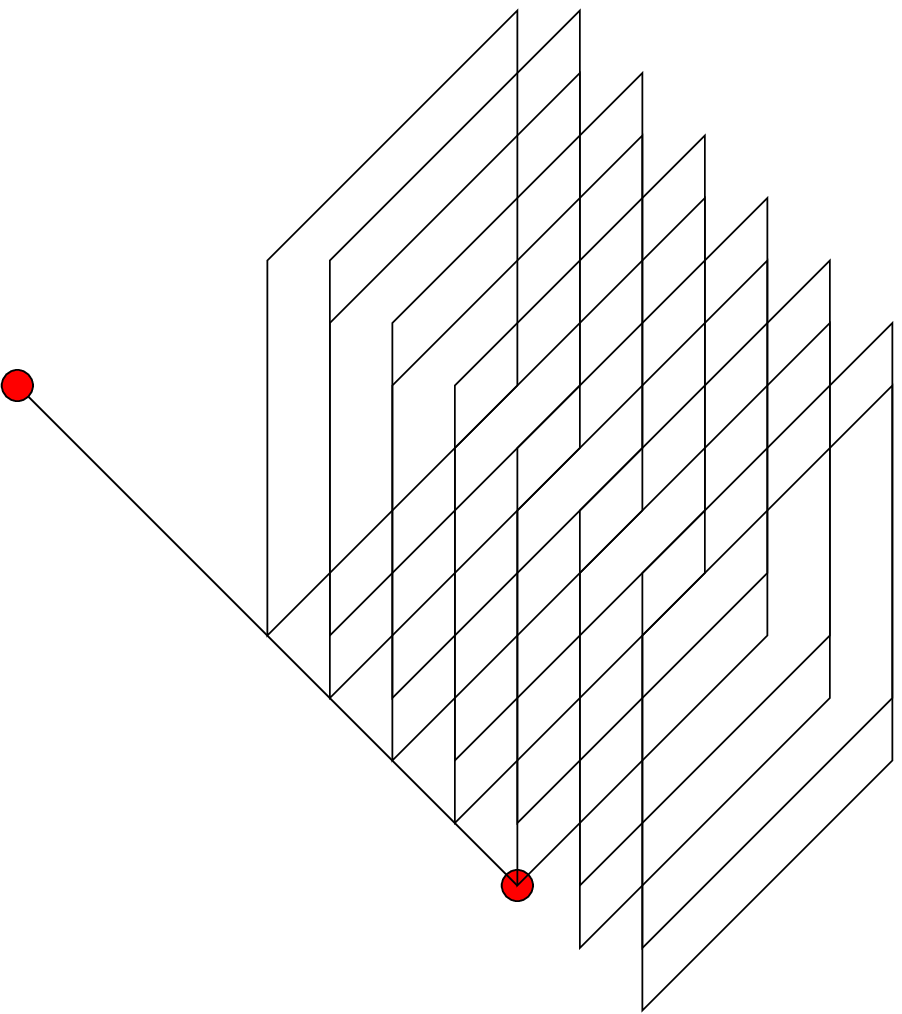}\epsfxsize=150pt\epsfbox{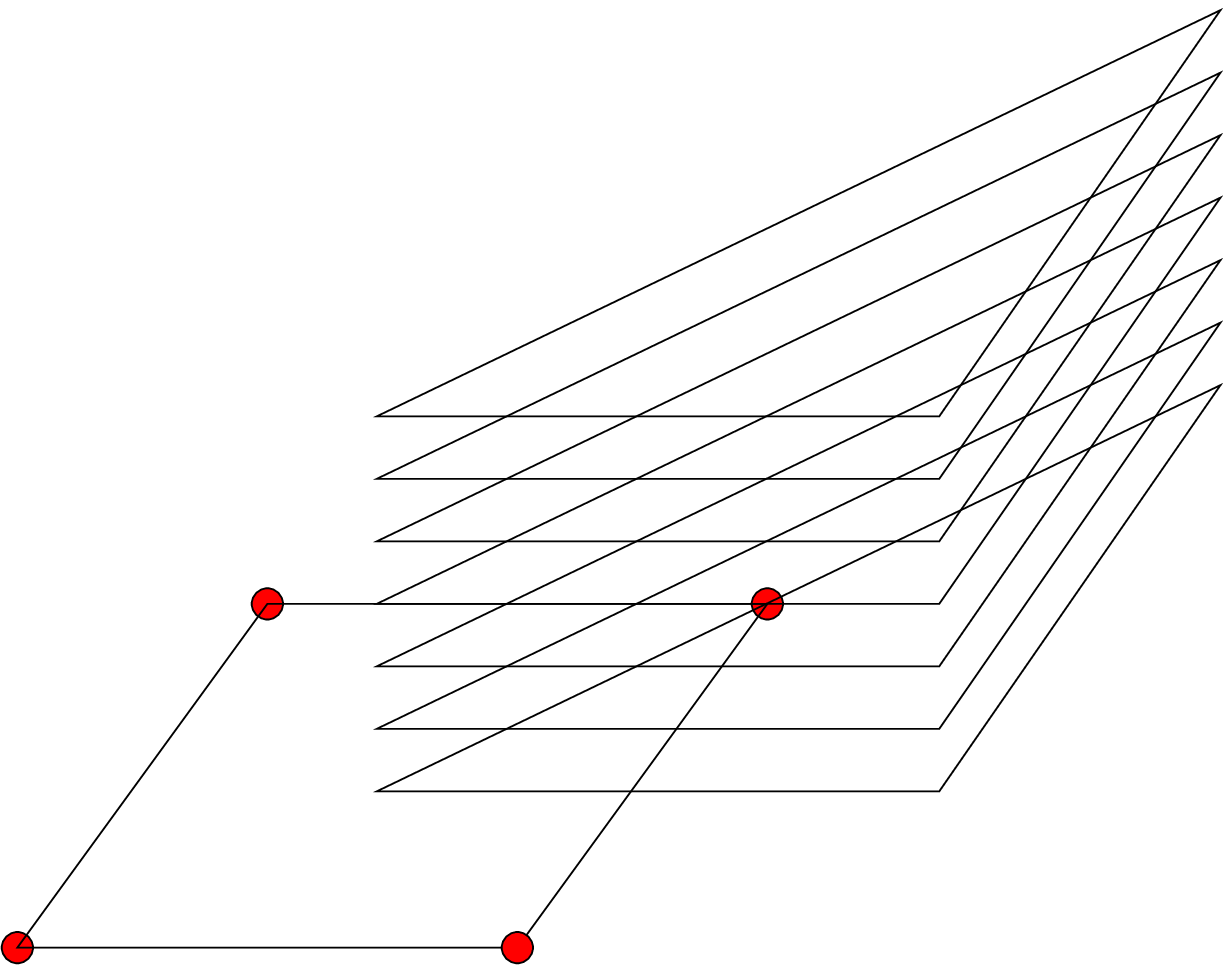} 
 \caption{Area used for flux distribution measurement in the a) on-axis, b) diagonal and c) four-quark cases. The quarks are placed at $R=8$. }
 \label{fmeas}
\end{figure}

The quark distances we measured were $R=2,4,6,8$. For all these values,
the energy and flux distribution measurement was performed for \\
a) two quarks on a lattice axis separated by $R$ lattice units, \\
b) two quarks on an axis diagonal with respect to the lattice axis and separated by $\sqrt{2}R$ 
units and \\
c) four quarks at the corners of a square with side length $R$. \\
Fig.~\ref{fmeas} shows the measured areas in these three cases.
In the on-axis case the 
microscopically measured volume consisted of 7 transverse planes at zero to
six lattice units away from 
the center point in between the quarks, each covering a $6\times 6$ area
with the region above the diagonal line removed because of symmetry.
For the diagonal case 12 (diagonal) transverse planes
of size $4\times 6$, starting $6\sqrt{2}$ lattice units away from the center
point were measured. In the four-quark case we had 7 planes parallel
to the quark plane and zero to six units outside it, each covering
a $9\times 9$ area with the region above the diagonal line again removed. 
For the smaller $R=2$ system only 5, 8 planes were included in the on-axis
and diagonal cases respectively, while in the four-quark case each plane
covered only a $7\times 7$ area. 

In addition to extracting the detailed structure of the color fields in 
space, there is also interest, when discussing sum rules, in the 
integrated values of these fields. Therefore we added the contributions from
all measured points to get these integrated values. This is referred
to as ``sum 1'' in Sect.~\ref{ssum} of this paper. In the simulations we also
calculated the correlation of the total sum of all plaquettes on the lattice
and the Wilson loop, and this will be referred to as ``sum 2'' below. The 
latter
sum, therefore, includes a much larger volume than sum 1 and so should be a
more realistic estimate. However, its error is expected to be larger due
to the larger number of points.

\subsection{Lattice operators for quarks \label{sloper}}

To explore the color fields around static quarks we need to find efficient
lattice operators to represent the creation and destruction of the quarks.
Here ``efficient'' means that the operators have a large overlap with the
state we want to study and small overlap with other states.
We use the same approach as previously when such 
operators were constructed for the measurement of the energies of two- and 
four-body systems. Each spatial link on the lattice is fuzzed, i.e.
replaced by a normalised sum of $c$ times the link itself plus the 
surrounding four spatial U-bends or ``staples''. Previous experience 
shows that $c=4$ is suitable. This is performed iteratively
a number of times (the {\em fuzzing level}) until the operator is efficient.

By performing the measurements on lattices with different levels of fuzzing 
we obtain a variational basis, which is important for the minimization of
the excited state contamination to the ground state signal. As we do not need 
information on the 
first excited two-body state with the same symmetry as the ground state, we 
are not worried by the fact that this second state, which we obtain 
after diagonalising our basis, also contains sizeable effects from the higher 
excited 
states as it effectively shields the ground state from too much contamination.
The first excited state ($A_{1g}'$) with ground-state symmetry has been 
studied in Ref. \cite{pen:97b} using a three-state basis. 

In the two quark case we initially used fuzzing levels 2 
and 13. This choice gives good estimates of all 2-quark energies, even for the 
$R=8$ off-axis quark pair of length $8\sqrt{2}$ -- the largest quark separation
for the geometries considered. However, for the 2-quark flux tube profiles
a problem emerged for this longest diagonal tube. At the midpoint of the tube 
the profile exhibited a valley -- a feature not seen in any of the 
shorter tubes. This apparently arises, since the operator representing the 
diagonal flux tube 
is constructed from two L-shaped paths and our highest fuzzing level 13 was 
apparently not
able to adequately reach the center of the L-shaped paths with side length 8 
at $T=3,4$ -- for higher $T$'s no useful signal was obtained. 
Changing the 
higher fuzzing level from 13 to 40 in a test run somewhat alleviated the 
problem, but our 
estimate of 
excited state contamination calculated from the energies and presented below 
in Table~\ref{th} of Sect. \ref{se} did not show a significant decrease with
this change in the fuzzing level. This further highlights how the 
inadequacy of the variational basis was only visible in the flux distribution
and not in the energies, i.e. variational principles can give good
energies but poor wavefunctions. 

Unfortunately this change of fuzzing levels was still not enough to get a 
realistic signal in all cases for $R=8$. This would have required using 
a variational basis where the paths to be fuzzed were not simply L-shaped
but closer to the diagonal in shape. Therefore, in the following, we did not
use the $R=8$. For $R=6$ the transverse shape of the action in 
the diagonal flux tube was qualitatively correct, but even so it was still 
some $30\%$ lower in the middle than expected from the on-axis result. 

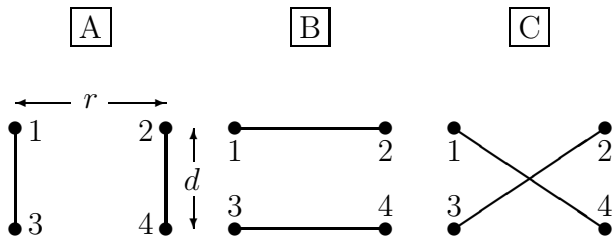
\begin{figure}[htb]
{
\newcommand{\thlen}{\setlength{\unitlength}{0.75pt}}
\newcommand{\quarksTR}{\multiput(0,0)(40,60){2}{\circle*{6}}
        \multiput(60,-20)(40,60){2}{\circle*{6}} }
\newcommand{\quarksNP}{\multiput(0,0)(0,40){2}{\circle*{6}}
        \multiput(60,-20)(40,20){2}{\circle*{6}} }
\newcommand{\quarksT}{\multiput(0,0)(40,60){2}{\circle*{6}}
        \multiput(60,20)(40,-20){2}{\circle*{6}} }
\newcommand{\quarksR}{\multiput(0,0)(0,40){2}{\circle*{5}}
        \multiput(60,0)(0,40){2}{\circle*{5}} }
\newcommand{\cubelabels}[3]{\put(28,-12){\makebox(0,0)[tr]{#1}}
        \put(85,-12){\makebox(0,0)[tl]{#2}}
        \put(105,20){\makebox(0,0)[cl]{#3}} }
\thlen
\newsavebox{\axes}
\savebox{\axes}{ {\thinlines
        \put(0,0){\vector(0,1){65}}
        \put(0,0){\vector(2,1){70}}
        \put(0,0){\vector(3,-1){85}} }}
\newsavebox{\grid}
\savebox{\grid}{ {\thinlines
        \multiput(0,0)(40,20){2}{\line(0,1){40}}
        \multiput(60,-20)(40,20){2}{\line(0,1){40}}
        \multiput(0,0)(60,-20){2}{\line(2,1){40}}
        \multiput(0,40)(60,-20){2}{\line(2,1){40}}
        \multiput(0,0)(0,40){2}{\line(3,-1){60}}
        \multiput(40,20)(0,40){2}{\line(3,-1){60}} }}
\setlength{\unitlength}{0.95pt}
\begin{center}
\setlength{\tabcolsep}{13pt}
\renewcommand{\arraystretch}{3.5}
\begin{tabular}{cccc}
 & \fbox{A} & \fbox{B} & \fbox{C} \\
\raisebox{15pt}{} &
\begin{picture}(60,60)
\quarksR
\put(30,48){\makebox(0,0)[b]{$r$}}
\put(22,50){\vector(-1,0){22}}
\put(37,50){\vector(1,0){23}}
\put(67,20){\makebox(0,0)[l]{$d$}}
\put(71,27){\vector(0,1){13}}
\put(71,13){\vector(0,-1){13}}
\put(8,-2){\makebox(0,0)[b]{3}}
\put(52,-2){\makebox(0,0)[b]{4}}
\put(8,35){\makebox(0,0)[b]{1}}
\put(52,35){\makebox(0,0)[b]{2}}
\thicklines
\multiput(0,0)(60,0){2}{\line(0,1){40}}
\end{picture} &
\begin{picture}(60,60)
\quarksR
\put(0,5){\makebox(0,0)[b]{3}}
\put(60,5){\makebox(0,0)[b]{4}}
\put(0,27){\makebox(0,0)[b]{1}}
\put(60,27){\makebox(0,0)[b]{2}}
\thicklines
\multiput(0,0)(0,40){2}{\line(1,0){60}}
\end{picture} &
\begin{picture}(60,60)
\quarksR
\put(0,5){\makebox(0,0)[b]{3}}
\put(60,5){\makebox(0,0)[b]{4}}
\put(0,27){\makebox(0,0)[b]{1}}
\put(60,27){\makebox(0,0)[b]{2}}
\thicklines
\put(0,0){\line(3,2){60}}
\put(0,40){\line(3,-2){60}}
\end{picture} \\
\end{tabular}
\end{center}
}
\caption{Three ways to pair four quarks in the case of two colors. 
\label{fpair}}
\end{figure}

For the case of four quarks the variational basis 
is obtained from the different ways to pair the four quarks, shown 
in Fig.~\ref{fpair}, all at the same fuzzing level. For $R=2,4,8$ ($6$) we 
used fuzzing level 13 (40). With three basis states in hand
we might have obtained better information on the first excited state, whose
wavefunction is essentially $(|A\rangle-|B\rangle)/\sqrt{2}$. However, for
the ground state [basically $(|A\rangle+|B\rangle)/\sqrt{2}$] the two and 
three basis state results are identical as was found earlier for the 
energies~\cite{glpm:96}. Thus we used only two basis 
states $A,B$ in most of the runs. 

\subsection{Variance reduction}

As there are many observables, each involving delicate cancellations, getting 
a good signal requires a large amount of computer time. One way to
achieve
this more easily is the so-called multihit or link integration method 
\cite{par:83}, where the statistical fluctuation of links in the Wilson loops 
is reduced by replacing the links by their thermal average. For calculating 
the expectation value of the link $U_{n\vec{\mu}}$ we only need to consider 
the part of the action involving this link -- for the usual Wilson action,
which uses just the plaquette operator, this is the sum $W$ of the 
six U-bends (staples) surrounding the link. In the case of SU(2) it
can be shown that
\begin{equation}
<U_{n\vec{\mu}}> = \frac{\int dU_{n\vec{\mu}} U_{n\vec{\mu}} e^{-\frac{1}{N}\beta
{\rm Tr}(UW^{\dagger})}}{\int dU_{n\vec{\mu}} e^{-\frac{1}{N}\beta
{\rm Tr}(UW^{\dagger})}}  = \frac{1}{d}\frac{I_{2}(\beta d)}{I_{1}(\beta d)}W,\; \;\; d = {\rm det} W,
\end{equation}
where the $I_n$'s are modified Bessel functions \cite{bro:81}. Their values
were integrated numerically and stored as an array of 5000 points, the values
given by analytical integration differing in the 7th or 8th decimal place. 
Using denser arrays did not change the Wilson loop correlations to an accuracy
of 6 decimal places. The expectation value of a link is 
a real number times an SU(2) matrix, and the real numbers have to be stored
for calculating correlations. The multi-hit algorithm cannot be used 
concurrently for 
links which are sides of the same plaquette, as the surrounding staples
are kept fixed. 

In our case we used multihit only on the time-like links of the Wilson loops. 
It is also possible to multihit all links of the Wilson loop and also the 
plaquette, but then the algorithm needs to be modified with several 
exceptions to avoid the problem just mentioned; this is discussed in 
Ref. \cite{hay:96}. The variance reduction we observe is presented in Table
\ref{tmhit}
for fuzzing levels 0,16,40 -- the results for levels 2,13 being similar. These 
test runs used a $16^3\times 32$ lattice. 
As expected, the observed error reduction 
increases with time separation, as more links are then multi-hit. The reduction
is also larger when the observables involve delicate cancellations; 
the error on the flux distributions calculated using Eq. \ref{fmnT} is 
reduced
more than the error on the Wilson loop. In fact, as seen in the first four 
rows of the table, for the latter the effect
is negligible. 
A rough estimate given in 
Ref.~\cite{mic:85} of the error reduction for an unfuzzed Wilson loop by a 
factor of 
$(0.8)^n$ with $n$ links multihit, giving 0.26 for $T=3$ and 0.17 for $T=4$, 
is seen to be larger than what we observe for potentials obtained by 
diagonalizing a basis consisting of fuzzed loops.

An interesting question is the effect on the errors of multi-hit versus
the choice of the variational basis. This is compared
in Table \ref{tmhitsum} for the flux observables presented in the last eight 
rows of 
Table~\ref{tmhit}. In Table~\ref{tmhitsum} ``average error'' refers to the 
average of 
the errors on the flux distribution measurements in Table~\ref{tmhit}
(and the corresponding one for fuzz levels 2,13). The error reduction is 
calculated both as the ratio of the average error with or without multihit
and as the average of the error reductions of the field measurements in 
Table~\ref{tmhit}. The bottom row shows for the two choices of fuzzing levels 
the ratios of the time consumptions and the average errors.
The choice of the variational basis can be seen
to reduce errors by a magnitude comparable to the multi-hit algorithm. 
Switching off the use of multi-hit for the fuzzing level 2,13 case 
leads to a reduction in computing time by a factor of 0.90. This means that
not using multi-hit on the time-links of the Wilson loop takes, in this case,
50\% more computing time to achieve the same accuracy. This is a significant 
saving, but not as large as we first hoped would be achieved.

\begin{table}[htb]
\begin{center}
\begin{tabular}{l|cccc}
               & \multicolumn{3}{c}{error} \\
observable     & without & with & reduction   \\ \hline
Potential, $R=4$, $T=3$ & 0.19 \% & 0.19 \% & 0.99 \\
Potential, $R=4$, $T=4$ & 0.22 \% & 0.22 \% & 1.00 \\
Potential, $R=8$, $T=3$ & 0.34 \% & 0.33 \% & 0.98 \\ 
Potential, $R=8$, $T=4$ & 0.56 \% & 0.52 \% & 0.93 \\ \hline
Action, $R=4$, $T=3$    & 1.19 \% & 0.73 \% & 0.61  \\
Energy, $R=4$, $T=3$    & 2.21 \% & 1.59 \% & 0.72  \\ 
Action, $R=4$, $T=4$    & 1.63 \% & 1.02 \% & 0.63  \\
Energy, $R=4$, $T=4$    & 1.94 \% & 1.30 \% & 0.67  \\ 
Action, $R=8$, $T=3$    & 3.70 \% & 2.91 \% & 0.79  \\
Energy, $R=8$, $T=3$    & 6.68 \% & 6.30 \% & 0.94  \\ 
Action, $R=8$, $T=4$    & 6.38 \% & 4.89 \% & 0.77  \\
Energy, $R=8$, $T=4$    & 8.02 \% & 6.72 \% & 0.84  \\ 
\end{tabular}
\caption{Error reduction with multihit for potentials and flux in the center
of $R=4,8$ flux tubes with fuzz levels 0,16,40. Errors are scaled to 1000 
measurements. \label{tmhit}}
\end{center}
\end{table}

\begin{table}[htb]
\begin{center}
\begin{tabular}{l|c|cc|cc}
           &        & \multicolumn{2}{c}{average error} \\
Basis      & time   & without & with    & ratio & average of reductions  \\ \hline
0 16 40    & 7803 s & 3.97 \% & 3.18 \% & 0.80 & 0.75 \\
2  13      & 4108 s & 4.81 \% & 3.70 \% & 0.77 & 0.78 \\ \cline{1-4}
ratio  & 0.52   & 0.83    & 0.86      \\ 
\end{tabular}
\caption{Error reduction with multihit and error reduction for a different
variational basis, all for the same number of measurements (see text). \label{tmhitsum}}
\end{center}
\end{table}

When the fields at or next to the color sources are measured, the plaquette
touches the Wilson loop. In this case we cannot have any common link 
multihit in the Wilson loop and not multihit in the plaquette,
as then we would use two different values of the same link in the same
observable. Therefore, for correct measurements of the quark self-energies,
which involve these links, 
we need to store versions of the Wilson loops with the appropriate links
not multihitted. Neglecting this complication has resulted in erroneous
measurements at the color sources in previous works \cite{bal:94}.

Previously, a group in Wuppertal has measured four-quark flux distributions 
in SU(2) (unpublished and private communication). They employed a higher 
$\beta$ value and used larger 
lattices. However, their work is less suited
for understanding the binding as the multihit algorithm was not switched
off when a plaquette touched the Wilson loop, leading to unreliable
self-energy measurement as discussed above. In addition, their diagonal flux 
tube was not measured directly, but instead the
results for the on-axis tube, measured only on the transverse axis and not
in full space, were interpolated to an off-axis situation. Also, no variational
basis was employed for determining the two-body ground state. In 
view of the 
problem with our variational basis for the diagonal paths mentioned above it 
is not clear if the interpolation from the on-axis case does indeed produce 
worse results for the diagonal tube for large $R$'s. 

\subsection{Details of the simulation and analysis}

The correlations in Eq.~\ref{fmnT} were measured on a $20^3\times 32$ lattice 
with maximal time
separation of the Wilson loops set to six lattice spacings. We averaged over 
all positions and orientations of the loops to improve statistics. 
The measurements were separated by one heatbath and three over-relaxation 
sweeps. Each measurement generated 4 MB of data and consumed 90 minutes of 
CPU time on a Cray C94 vector supercomputer. Sixteen or eight measurements 
were averaged 
into one block for $R=2,4$ and $R=6,8$ respectively, and 28 of these blocks 
were used for the final analysis. 
There the errors were estimated by using 100 bootstrap samples. 

\section{Energies and excited state contamination \label{se}}

The observed two-body potentials and four-quark binding energies, presented
below in Tables~\ref{tsumcheck}--\ref{tsumcheck2}, agree with previous 
results~\cite[references therein]{glpm:96}.

Since we use plaquettes in the middle of fuzzed Wilson loops in the time 
direction,
we would like to know the excited state contamination at $t=T/2$. To estimate
this contamination we use the method introduced for two-body potentials in
Ref. \cite{mic:96b}. From the 
Wilson loop ratios at each $R$-value, we define the effective two-body 
potential $V(T)= -\ln [W(T)/W(T-1)]$, since its rate of approach to a plateau 
as $T \to
\infty$ enables us to estimate the excited state contamination to the
ground state. A measure of this contamination is defined as 
\begin{equation}
h(t)={ c_1 \over c_0} e^{-(V_1-V_0)t}\ , 
\end{equation}
which should be $\ll 1$. Here $V_0$ is the ground state potential and $V_1$ the
potential of the first excited state with the same symmetry, and the $c_i$'s
come from expanding a link operator that represents the creation or
annihilation of two quarks at separation $R$ as 
$|R\rangle=c_0|V_0\rangle+c_1|V_1\rangle+\dots$ in terms of transfer matrix
eigenstates. In practice $h$ is calculated from
 \begin{equation} |h(t=T/2)| \approx \frac{\lambda}{\lambda-1}
\sqrt{V(T-1)-V(T)} =
\lambda\frac{V(T-1)-V(T\rightarrow\infty)}{\sqrt{V(T-1)-V(T)}}.
 \end{equation}
 Here the $T\rightarrow\infty$ extrapolated potential is defined as
 $$
V(T\rightarrow\infty)\equiv V(T)-\lambda\frac{V(T-1)-V(T)}{1-\lambda},\
 \lambda\equiv e^{-(V_1-V_0)}.
 $$
Table \ref{th} shows the excited state contamination for the ground
state of the two-body potential. The contamination
at $t=1$ is calculated both from $T=1,2$ and $T=3,4$, the difference
in the values reflecting the error in our estimates. 

The contamination in the
flux at $T$ is measured by $h(T/2)$, which should be small (e.g. $< 0.1$).
This suggests problems in the $R=8$ case, as already discussed in
Sect.~\ref{sloper}. In this case for $T=4$ the contamination is smaller, but 
the signal is then too noisy. In general, the consistency of results at 
larger $T$'s
suggests that the effect from excited states is negligible. 

\begin{table}[htb]
 \begin{center}
\begin{tabular}{l|l|c|c|c|c}
          & $R$ & $t=1$ (a) & $t=1$ (b) & $t=1.5$     & $t=2$ \\ \hline
\multicolumn{5}{l}{Fuzz levels 2 and 13} \\
two-body  & 2   & 0.024   & 0.039   & 0.013 & 0.007  \\
          & 4   & 0.065   & 0.071   & 0.030 & 0.019  \\
          & 8   & 0.138   & 0.093   & 0.048 & 0.034 \\ 
          & 2,2 & 0.035   & 0.048   & 0.015 & 0.009  \\
          & 4,4 & 0.074   & 0.058   & 0.024 & 0.015  \\
          & 8,8 & 0.273   & 0.078   & 0.051 & 0.039 \\
\multicolumn{5}{l}{Fuzz levels 2 and 40} \\
two-body  & 4   & 0.079   & 0.105   & 0.038 & 0.028  \\
          & 6   & 0.085   & 0.138   & 0.049 & 0.030  \\
          & 8   & 0.159   & 0.155   & 0.061 & 0.051 \\ 
          & 4,4 & 0.094   & 0.098   & 0.037 & 0.025  \\
          & 6,6 & 0.113   & 0.107   & 0.043 & 0.025  \\
          & 8,8 & 0.227   & 0.104   & 0.046 & 0.040 \\ 
\end{tabular}
 \caption{Excited state contamination as measured by $h$. $t=1$ (a) refers to values calculated using $T=1,2$ energies, $t=1$ (b) to values using $T=3,4$ energies. \label{th}}
\end{center}
\end{table}

\section{Sum rules for four static quarks \label{ssum}}

When a plaquette is used to probe  the color flux with the Wilson gauge 
action, exact
identities can be  derived for the integrals over all space of the flux
distributions.  These sum rules~\cite{mic:87,mic:96} relate
spatial sums  of the color fields measured using Eq.~\ref{fmnT} to the 
energies of the system via generalised $\beta$-functions, which show
how the bare  couplings of the theory vary with the generalised lattice
spacings $a_\mu$ in four directions. One can think of the sum rules
as providing the  appropriate anomalous dimension for the color flux
sums.  This normalises the color flux and provides a guide for
comparing  color flux distributions measured at different $a$-values.
The full  set of sum rules~\cite{mic:96} allow these generalised
$\beta$-functions  to be determined at just one $\beta$-value~\cite[and
references therein]{pen:97b,mic:96b}.  

A starting point for the sum rules is the identity
\begin{equation}
  -{d E \over d \beta}=<1\,|\sum \Box |\,1> - <0\,|\sum \Box |\,0>
 = \sum \Box_{1-0},
  \label{esr1}
\end{equation}
derived in Ref. \cite{mic:87}, which holds for ground-state energies $E$ 
obtained from the correlation of Wilson loops in the limit of large 
time separation. In Eq. \ref{esr1} the symbol $\Box$ is the plaquette action 
${1 \over N}{\rm Tr}U_{\Box}$ which is summed over all plaquettes in a 
time slice, and the subscript $1-0$ refers to the difference of this sum in 
a state containing the observable system (1) and in the vacuum (0). 
For potentials between static sources the energy $E$ includes an unphysical 
lattice self-energy contribution which diverges in the 
continuum limit. 

These relations can be trivially extended to the case of four static 
quarks.
For a general configuration of four quarks the dimensionless energy 
$E(X,Y,Z,\beta)$ is a function of the coupling $\beta$ multiplying the 
plaquette action and distances in lattice units $X,Y,Z$ with
physical lengths being $x = Xa$, $y = Ya$, $z = Za$, where $a$ is the 
lattice spacing. To remove the $\beta$-derivative from Eq. \ref{esr1} we need 
to use the independence of a physical energy $E_p/a$ of $a$ as 
$a\rightarrow 0$ when $x,y,z$ are kept constant.
That is, combining 
 \begin{eqnarray}
0 & = &  \left. {dE_p[x,y,z,\beta(a)]/a \over da}\right|_{x,y,z}\label{esrinv} \\
  & = &  -{E_p \over a^2} 
 -\left.{X \over a^2}{\partial E_p \over  \partial X}\right|_{Y,Z}
 -\left.{Y \over a^2}{\partial E_p \over  \partial Y}\right|_{X,Z}
 -\left.{Z \over a^2}{\partial E_p \over  \partial Z}\right|_{X,Y}
 +{ 1 \over a} {d\beta \over da}\,\left.
{\partial  E \over \partial \beta}\right|_{X,Y,Z} \nonumber
\end{eqnarray}
with Eq.~\ref{esr1} we get
\begin{equation}
E(x,y,z) + {\partial E \over  \partial \ln X} 
+ {\partial E \over  \partial \ln Y} + {\partial E \over  \partial \ln Z}
+E_0 = -{d\beta \over d \ln a} \, \sum \Box_{1-0}, \label{esum}
\end{equation}
where, unlike the physical energy, $E_0$ is a contribution from the unphysical 
self-energy that depends on $\beta$ and is not independent of $a$ in
the continuum limit.  

In the general case there are lattice spacings $a_i$ for all four directions 
$i=1,\ldots,4$, 
and couplings $\beta_{ij},\; i>j$ for all 6 orientations of a 
plaquette. As in Sect. \ref{smethod}, plaquettes with orientation 
41,42,43,23,31,12 are labelled with 
${\cal E}_x,{\cal E}_y,{\cal E}_z,{\cal B}_{x},{\cal B}_{y},{\cal B}_{z}$ 
respectively. When $a_i=a$ for all $i$, derivatives of the couplings with 
respect to the lattice spacings fall into two classes
\begin{equation}
{\partial \beta_{ij} \over \partial \ln a_k}= 
S \ \hbox{if}\ k=i \ \hbox{or}\ j
\ \ \hbox{and} \ \
{\partial \beta_{ij} \over \partial \ln a_k}= 
U \ \hbox{if}\ k\ne i \ \hbox{or}\ j.
\end{equation}
Using these equations and the invariance of 
$\frac{1}{a_0}E_p[X,Y,Z,\beta_{ij}(a_k)]$ with respect to $a_0$, $a_x$, $a_y$, 
$a_z$ -- in analogy to Eq.~ \ref{esrinv} --  we get
\begin{eqnarray}
E + E_0^0 & = & -\sum  S ({\cal E}_x+{\cal E}_y+{\cal E}_z) + 
              U ({\cal B}_{z}+{\cal B}_{y}+{\cal B}_{x}) \label{sume} \\
X {\partial E \over \partial X} + E_0^X & = & -\sum  S {\cal E}_x +
U {\cal E}_y + U {\cal E}_z + S {\cal B}_{z} + S {\cal B}_{y} + U {\cal B}_{x} \label{sumx}\\
Y {\partial E \over \partial Y} + E_0^Y & = & -\sum  U {\cal E}_x +
S {\cal E}_y + U {\cal E}_z + S {\cal B}_{z} + U {\cal B}_{y} + S {\cal B}_{x} \label{sumy} \\
Z {\partial E \over \partial Z} +E_0^Z & = & -\sum  U {\cal E}_x +
U {\cal E}_y + S {\cal E}_z + U {\cal B}_{z} + S {\cal B}_{y} + S {\cal B}_{x}. \label{sumz}
\end{eqnarray}
As for the $E_0$ in Eq.~\ref{esum}, the $E_0^i$'s on the LHS of
these equations are self-energy 
contributions independent of $X,Y,Z$.  Due to the isotropic nature of the 
self-energies we expect $E_0^X=E_0^Y=E_0^Y$. The negative sign on the RHS 
arises from our sign convention for the plaquette.
In the case of a planar geometry, like the square we are now measuring, there
is no extent in the direction perpendicular to the plane. If we choose 
this direction to be $z$, then Eq.~\ref{sumz} only has a self-energy term 
on the LHS. 

In Ref. \cite{pen:97b} the generalized $\beta$-functions 
$b\equiv\partial \beta/\partial \ln a=2(S+U)$ and $f \equiv
(U-S)/(2\beta)$ were determined from two-body potentials and flux 
distributions using sum rules. From the best estimates of $b=0.312(15)$ and
$f=0.65(1)$ at $\beta=2.4$ we get $S=-1.638(25),\; U=1.482(25)$ for Eqs. 
\ref{sume}--\ref{sumz}.
Therefore, using the results for self-energies and -actions from Ref. 
\cite{pen:97b}, we get $E_0^0=0.14(5)$ -- a number of interest when discussing
Tables~\ref{tsumcheck},\ref{tsumcheckb}. With these values in hand we can 
use the 
above sum rules as a check on our flux distribution measurement. 

Tables~\ref{tsumcheck},\ref{tsumcheckb} shows the observed energies and 
corresponding energy 
sums for two and four quarks, respectively.
Here ``sum 1'' means a sum over our microscopic measurements of 
the flux 
distribution, whereas ``sum 2'' denotes 
the correlation between the sum of all the plaquettes on the lattice 
and the Wilson loop(s). When our estimate of $E_0^0$ ($2E_0^0$) is added to the
two (four) -body energies, 
the energy sums
can be seen to agree with the observed energies especially for $T=2$. 

The term $E_0^0$ can be removed by considering 
differences of flux-distributions, since then the self-energies cancel. Here we
consider two such differences to be used later for a model of the binding 
energies;
a) diagonal flux tubes subtracted from one-half times the flux tubes along 
the sides -- $[F(AB)-F(C)]$ in Eq.~\ref{efsum} below -- and b) one-half times 
the flux tubes along the sides of the square 
subtracted from the four-body distribution -- $FB(4)$ in Eq.~\ref{efsum}. 
These are shown in Table~\ref{tsumcheck2}.
In the first rows the differences of diagonal and on-axis potentials 
$V(R)-V(R,R)$ are compared to the difference of corresponding energy sums 1
and 2 in Table~\ref{tsumcheck}.
The last rows contain four-body binding energies with a similar comparison.
The agreement of these sums and 
the corresponding energy differences suggests correctness
of our flux distribution measurement and subtractions, and proper 
cancellation of the self-energy distributions.
One might expect the agreement of sum 2 with the energies to be slightly better
than that of sum 1, since the area of our microscopic
measurement always leaves a small tail-end of the signal unmeasured. In 
practice larger errors on sum 2 overcome this benefit in many cases.
All the
errors in these tables are from a bootstrap analysis. 

We use Table~\ref{tsumcheck2} as a guide in the following for choosing 
the best $T$ value at which to look at the flux distribution measurement.
The $R=4,6,8$ four-body binding energies and corresponding flux sums 
agree much better at $T=2$ than at higher $T$'s, where the large errors make
the signal often consistent with zero. Therefore we use $T=2$
for these $R$'s and $T=3$ for $R=2$.

\begin{table}[htb]
 \begin{center}
\begin{tabular}{lll|ccc}
          & $R$ & observable &  $T=2$    & $T=3$     & $T=4$ \\ \hline
two-body  & 2   & potential& 0.56347(41) & 0.56246(47) & 0.56217(51) \\
          &     & sum 1    & 0.4307(36)  & 0.4668(44)  & 0.4504(54) \\ 
          &     & sum 2    & 0.443(17)   & 0.490(24)   & 0.487(32) \\ 
          & 2,2 & potential& 0.67123(79) & 0.66954(97) & 0.6689(11)  \\
          &     & sum 1    & 0.5216(77)  & 0.574(10)   & 0.548(12) \\ 
          &     & sum 2    & 0.540(28)   & 0.610(42)   & 0.601(57) \\ 
          & 4   & potential& 0.78314(55) & 0.77806(71) & 0.77594(85) \\
          &     & sum 1    & 0.6057(92)  & 0.685(14)   & 0.657(17) \\ 
          &     & sum 2    & 0.640(26)   & 0.740(41)   & 0.692(49) \\ 
          & 4,4 & potential& 0.9267(10)  & 0.9178(13)  & 0.9144(15) \\
          &     & sum 1    & 0.687(25)   & 0.822(37)   & 0.771(50) \\ 
          &     & sum 2    & 0.759(57)   & 0.913(76)   & 0.914(15) \\ 
          & 6   & potential& 0.9454(16)  & 0.9368(18)  & 0.9336(20) \\
          &     & sum 1    & 0.730(30)   & 0.832(52)   & 0.828(77) \\ 
          &     & sum 2    & 0.780(81)   & 0.90(12)    & 0.86(17) \\ 
          & 6,6 & potential& 1.1567(32)  & 1.1346(32)  & 1.1268(39) \\
          &     & sum 1    & 0.961(74)   & 1.08(12)    & 1.07(16) \\ 
          &     & sum 2    & 1.01(17)    & 1.06(26)    & 0.86(36) \\ 
          & 8   & potential& 1.1293(15)  & 1.1077(22)  & 1.0968(34) \\ 
          &     & sum 1    & 0.874(50)   & 1.008(65)   & 0.92(13) \\ 
          &     & sum 2    & 0.88(10)    & 1.16(21)    & 1.19(30) \\ 
          & 8,8 & potential& 1.5829(34)  & 1.4893(54)  & 1.4343(87) \\ 
          &     & sum 1    & 1.15(11)    & 1.29(23)    & 0.85(49) \\ 
          &     & sum 2    & 1.30(22)    & 1.31(50)    & --0.15(95) \\ \hline
\end{tabular}
 \caption{Measured energies and energy sums for two quarks (see text). 
\label{tsumcheck}}
\end{center}
\end{table}

\begin{table}[htb]
 \begin{center}
\begin{tabular}{lll|ccc}
          & $R$ & observable &  $T=2$    & $T=3$     & $T=4$ \\ \hline
four-body & 2   & energy   & 1.06879(76) & 1.06602(85) & 1.06537(91) \\
          &     & sum 1    & 0.815(10)   & 0.882(14)   & 0.858(18) \\ 
          &     & sum 2    & 0.835(32)   & 0.927(46)   & 0.925(64) \\ 
          & 4   & energy   & 1.5111(11)  & 1.5030(14)  & 1.4996(16) \\
          &     & sum 1    & 1.161(32)   & 1.375(45)   & 1.323(54) \\ 
          &     & sum 2    & 1.208(56)   & 1.423(71)   & 1.390(92) \\ 
          & 6   & energy   & 1.8613(39)  & 1.8387(43)  & 1.8338(65)  \\
          &     & Sum 1    & 1.38(10)    & 1.72(18)    & 1.62(41) \\ 
          &     & sum 2    & 1.37(19)    & 1.46(31)    & 1.00(60) \\ 
          & 8   & energy   & 2.2421(29)  & 2.1953(56)  & 2.177(18)  \\
          &     & sum 1    & 1.71(14)    & 1.77(46)    & 3.4(1.2) \\ 
          &     & sum 2    & 1.61(22)    & 1.66(75)    & 2.7(1.6) \\ \hline
four-body & 2   & energy   & 1.2706(11) & 1.2650(13)  & 1.2638(12) \\
1st excited&    & sum 1    & 0.974(12)  & 1.074(21)   & 1.037(30) \\ 
state     &     & sum 2    & 0.991(40)  & 1.094(65)   & 1.078(87) \\ 
          & 4   & energy   & 1.6630(10) & 1.6503(16) & 1.6445(24) \\
          &     & sum 1    & 1.305(31)  & 1.439(59)  & 1.23(11) \\ 
          &     & sum 2    & 1.396(50)  & 1.62(10)   & 1.41(15) \\ 
          & 6   & energy   & 1.9542(32) & 1.9255(40) & 1.91389(45) \\
          &     & sum 1    & 1.495(96)  & 1.78(20)   & 2.02(44) \\ 
          &     & sum 2    & 1.63(17)   & 2.08(38)   & 2.60(72) \\ 
\end{tabular}
 \caption{Measured energies and energy sums for four quarks. \label{tsumcheckb}}
\end{center}
\end{table}

\begin{table}[htb]
 \begin{center}
\begin{tabular}{lll|ccc}
          & $R$    & observable &  $T=2$    & $T=3$     & $T=4$ \\ \hline
two-body  & 2--2,2 & potential& --0.10775(38) & --0.10708(50) & --0.10672(59)\\
          &        & sum 1   & --0.0910(44) & --0.1072(64) & --0.0974(85) \\ 
          &        & sum 2   & --0.097(12) & --0.120(19) & --0.114(27) \\ 
          & 4--4,4 & potential& --0.14359(48) & --0.13978(65) & --0.13846(76)\\
          &        & sum 1   & --0.082(18) & --0.136(26) & --0.114(38) \\ 
          &        & sum 2   & --0.118(33) & --0.173(40) & --0.173(51) \\ 
          & 6--6,6 & potential& --0.2113(17) & -- 0.1978(17)& --0.1932(26)\\ 
          &        & sum 1   & --0.231(50) & --0.250(91) & --0.24(11) \\
          &        & sum 2   & --0.229(92) & --0.17(15) & --0.0(2) \\ 
          & 8--8,8 & potential& --0.4536(21) & --0.3817(36) & --0.3375(71) \\ 
          &        & sum 1   & --0.271(77) & --0.28(20) & 0.07(42) \\ 
          &        & sum 2   & --0.42(15) & --0.16(36) & --1.3(8) \\ \hline
four-body & 2   & energy& --0.05816(6) &  --0.05889(10) & --0.05897(13) \\
binding   &     & sum 1 & --0.0467(41) & --0.0511(74) & --0.0424(99)  \\ 
          &     & sum 2 & --0.0504(29) & --0.0543(67) & --0.0492(82) \\ 
          & 4   & energy& --0.05521(9) & --0.05309(27) & --0.05229(44) \\
          &     & sum 1 & --0.051(18) & 0.004(33) & 0.009(42)  \\ 
          &     & sum 2 & --0.073(12) & --0.056(40) & 0.00(6) \\ 
          & 6   & energy& --0.02957(93) & --0.0348(13) & --0.0335(37) \\
          &     & sum 1 & --0.080(58) & 0.06(12) & --0.04(34) \\ 
          &     & sum 2 & --0.19(7) & --0.33(16) & --0.73(53) \\ 
          & 8   & energy& --0.01662(95) & --0.0201(33) & --0.016(16) \\
          &     & sum 1 & --0.034(67) & --0.25(41) & 1.5(1.0)  \\ 
          &     & sum 2 & --0.14(11) & --0.65(53) & 0.3(1.6) \\ \hline
four-body & 2   & energy& 0.14363(30) & 0.14007(41) & 0.13946(58)  \\
1st excited&    & sum 1 & 0.1125(61)  & 0.141(14) & 0.136(22)  \\ 
state     &     & sum 2 &  0.1057(98) & 0.113(26) & 0.104(37)  \\ 
          & 4   & energy& 0.09670(23) & 0.09416(43) & 0.0926(12)  \\
          &     & sum 1 & 0.094(18) & 0.069(38) & --0.089(94)  \\ 
          &     & sum 2 & 0.115(23) & 0.139(54) & 0.03(10) \\ 
          & 6   & energy& 0.06329(53) & 0.0519(15) & 0.0466(36)   \\
          &     & sum 1 & 0.034(53) & 0.12(14) & 0.37(42)  \\ 
          &     & sum 2 & 0.074(64) & 0.29(26) & 0.88(59) \\ 
\end{tabular}
 \caption{Differences of measured energies and energy sums (see text)
\label{tsumcheck2}}
\end{center}
\end{table}

The sum rules in Eqs.~\ref{sumy}, \ref{sumz} can also be used as checks of the 
measurements if the system has no extent in the $y,z$ directions, 
respectively. In the case of two on-axis quarks we average the transverse 
directions $y,z$ so that we get a radial and an azimuthal component. Therefore
we have to add Eqs.~\ref{sumy},\ref{sumz} to get a zero sum rule. 
For the two-body diagonal and four-body cases we directly use Eq.~\ref{sumz}.
Results are shown in Table~\ref{tsumcheck3}, from where we can see
that $E_0^Y+E_0^Z=0.31(2)$ using the $R=2,4$ on-axis values and 
$E_0^Z=0.15(1)$ using $R=2,4$ diagonal and four-body values. These 
estimates agree with the expectation $E_0^X=E_0^Y=E_0^Z$. However, we are 
not aware of a reason for $E_0^0$ to be consistent with these as seems to be 
observed. In Table~\ref{tsumcheck4} we have subtracted sums of different 
observables 
to cancel these constant contributions. The ``two-body'' part of the table
shows on-axis two-body tubes subtracted from each other, and the ``four-body''
part has off-axis tubes subtracted from the four-body distribution because
the sum rule for the off-axis and four-quark cases is the same. 
This results in cancellations by an order of magnitude leaving sums that are,
in most cases, consistent with zero.

\begin{table}[htb]
 \begin{center}
\begin{tabular}{lll|ccc}
          & $R$ & observable &  $T=2$   & $T=3$     & $T=4$ \\ \hline
two-body  & 2   & sum 1    & 0.2947(55) & 0.3037(73)& 0.2817(88)  \\ 
          &     & sum 2    & 0.282(15)  & 0.282(22) & 0.252(29) \\ 
          & 4   & sum 1    & 0.3239(84) & 0.339(12) & 0.298(19) \\ 
          &     & sum 2    & 0.331(17)  & 0.340(26) & 0.246(34) \\ 
          & 6   & sum 1    & 0.271(25)  & 0.319(42) & 0.394(79) \\ 
          &     & sum 2    & 0.295(79)  & 0.49(11)  & 0.85(18) \\ 
          & 8   & sum 1    & 0.468(46)  & 0.416(65) & 0.34(12) \\ 
          &     & sum 2    & 0.530(98)  & 0.64(21)  & 0.66(34) \\ \hline
          & 2,2 & sum 1    & 0.1490(63) & 0.1479(86)& 0.121(11) \\ 
          &     & sum 2    & 0.138(15)  & 0.129(23) & 0.096(30) \\ 
          & 4,4 & sum 1    & 0.1845(96) & 0.216(16) & 0.177(26) \\ 
          &     & sum 2    & 0.181(21)  & 0.185(27) & 0.128(41) \\ 
          & 6,6 & sum 1    & 0.139(36)  & 0.198(76) & 0.35(11) \\ 
          &     & sum 2    & 0.126(75)  & 0.27(13)  & 0.59(23) \\ 
          & 8,8 & sum 1    & 0.266(79)  & 0.41(13)  & 0.34(53)  \\ 
          &     & sum 2    & 0.36(13)   & 0.70(25)  & --0.05(97) \\ \hline
four-body & 2   & sum 1    & 0.2640(64) & 0.2832(85)& 0.2685(98) \\ 
          &     & sum 2    & 0.255(14)  & 0.270(21) & 0.256(26) \\ 
          & 4   & sum 1    & 0.285(14)  & 0.330(19) & 0.263(32) \\ 
          &     & sum 2    & 0.292(22)  & 0.325(38) & 0.197(60) \\ 
          & 6   & sum 1    & 0.165(60)  & 0.33(11)  & 0.58(25) \\ 
          &     & sum 2    & 0.12(11)   & 0.38(19)  & 1.05(38) \\ 
          & 8   & sum 1    & 0.370(95)  & 0.65(25)  & 0.98(74) \\ 
          &     & sum 2    & 0.43(16)   & 0.87(55)  & 2.2(1.5)  \\ \hline
\end{tabular}
 \caption{Zero sum rule (see text). \label{tsumcheck3}}
\end{center}
\end{table}

\begin{table}[htb]
 \begin{center}
\begin{tabular}{lll|ccc}
          & $R$ & observable &  $T=2$   & $T=3$     & $T=4$ \\ \hline
two-body  & 4-2 & sum 1    & 0.029(86) & 0.036(13) & 0.016(17) \\
          &     & sum 2    & 0.049(23) & 0.059(36) & --0.007(41) \\
          & 6-2 & sum 1    & --0.024(27)& 0.015(43) & 0.113(79)  \\ 
          &     & sum 2    & 0.013(79) & 0.021(11) & 0.60(18) \\
          & 8-2 & sum 1    & 0.173(46)  & 0.112(66) & 0.06(12) \\ 
          &     & sum 2    & 0.249(97) & 0.36(21) & 0.41(35) \\
          & 6-4 & sum 1    & --0.053(26) & --0.020(35) & 0.096(77) \\ 
          &     & sum 2    & --0.036(84) & 0.15(11)  & 0.61(18) \\
          & 8-4 & sum 1    & 0.144(46)  & 0.077(68) & 0.04(12) \\ 
          &     & sum 2    & 0.20(10) & 0.30(21) & 0.31(35) \\
          & 8-6 & sum 1    & 0.197(49)  & 0.097(78) & --0.06(14) \\ 
          &     & sum 2    & 0.24(12) & 0.15(22) & --0.19(36) \\ \hline
four-body & 2-2,2   & sum 1& --0.0341(73) & --0.013(11) & 0.026(14) \\ 
          &     & sum 2    & --0.020(17)  & 0.013(26) & 0.065(37)   \\ 
          & 4-4,4   & sum 1& --0.084(13)  & --0.101(29) & --0.090(49)   \\ 
          &     & sum 2    & --0.070(31) & --0.045(50) & --0.059(73)   \\ 
          & 6-6,6   & sum 1& --0.114(39) & --0.07(12) & --0.12(21)    \\ 
          &     & sum 2    & --0.138(85) & --0.16(23) & --0.14(47)   \\ 
          & 8-8,8   & sum 1& --0.16(13)  & --0.18(29) & 0.3(1.3) \\ 
          &     & sum 2    & --0.30(20)  & --0.52(54)  & 2.3(2.3)   \\ \hline
\end{tabular}
 \caption{Zero sum rule after subtraction (see text). \label{tsumcheck4}}
\end{center}
\end{table}

The limit $T\rightarrow \infty$ will always isolate the ground state 
contribution, but large $T$ values give large errors. However, the variational
approach we use allows an accurate signal to be obtained from small 
$T$ values as the excited state contribution to the ground state signal
is to a large extent removed. The remaining excited state contamination can 
be measured 
with $h$ as discussed in Sect.~\ref{se}. In the case of distributions 
corresponding to the binding energy of four quarks the sum rule checks show 
that we have the best signal at $T=2$ in most cases.

\section{Color field distributions \label{sdist}}

The ground-state {\em energies} of four quarks in a square geometry are the 
same
when two (A,B) or three (A,B,C) basis states are used \cite{glpm:96}. 
Since we did not know if this was true also for the ground state of the {\em 
color
fields}, we initially started simulating with all three basis states. 
Another reason for this was an attempt to get a signal for the first excited 
state
of four quarks. It was then found that, as for the ground state energy, 
the color field ground state 
was the same for the two and three basis states. 
Therefore we carried out most simulations with only two basis states. 

We have visualized the spatial distribution of the color fields for two 
and four
quarks using successive transparent isosurfaces,
whose color gives the relative error. These color figures in GIF and EPS
formats are available via WWW at
{\tt http://www.physics.helsinki.fi/\~{}ppennane/pics/}.
The color field combinations corresponding to the action, energy and 
the energy sum of Eq.~\ref{sume} are shown. The distribution
around four quarks -- to be discussed later -- is shown before and after 
subtracting the fluxtubes along the sides of the square. 

As discussed in Sect.~\ref{ssum}, the $S{\cal E}+U{\cal B}$ combination 
of the color fields in Eq.~\ref{sume} corresponds to the distribution 
of the measured energy of the system. An observable easier to measure 
(involving one less delicate cancellation) is the action ${\cal E}+{\cal B}$. 
Below
we will present both the energy (Eq.~\ref{sume}) and the action distributions
by choosing various slices cutting through the different
spatial distributions. 

\subsection{Two quarks \label{s2q}}

\begin{figure}[hbtp]
\hspace{0cm}\epsfxsize=200pt\epsfbox{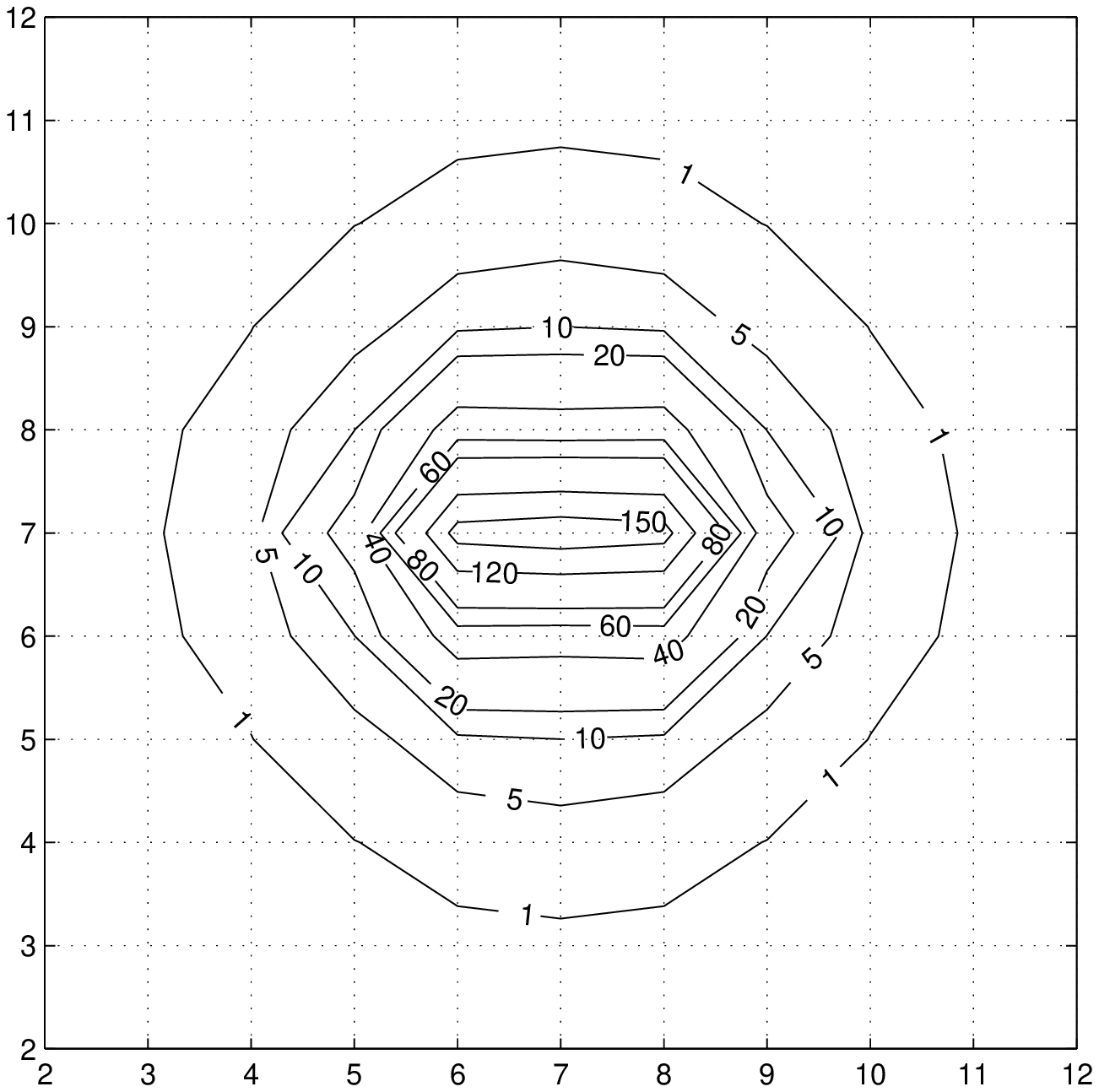}\epsfxsize=200pt\epsfbox{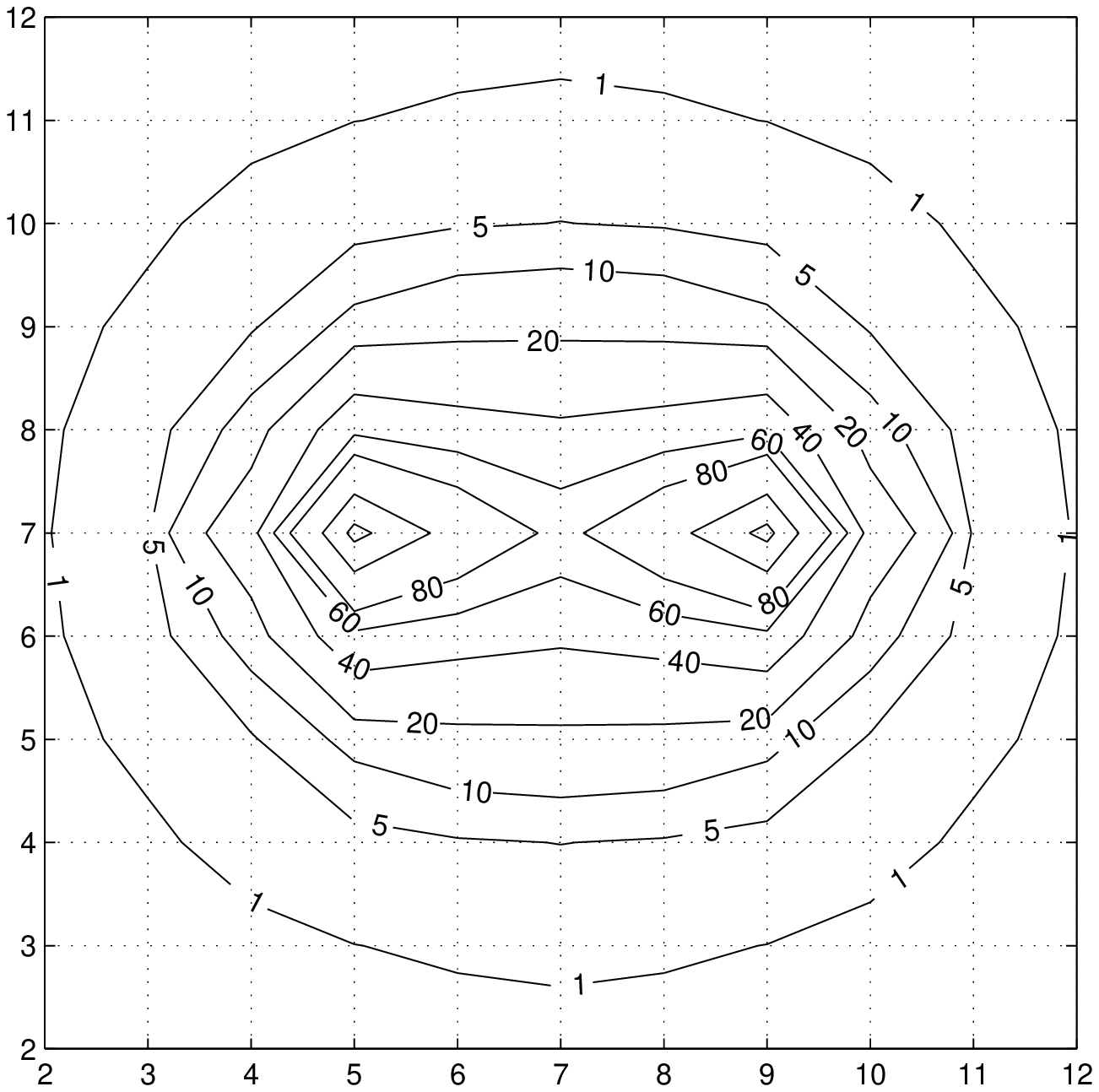} 
\hspace{0cm}\epsfxsize=200pt\epsfbox{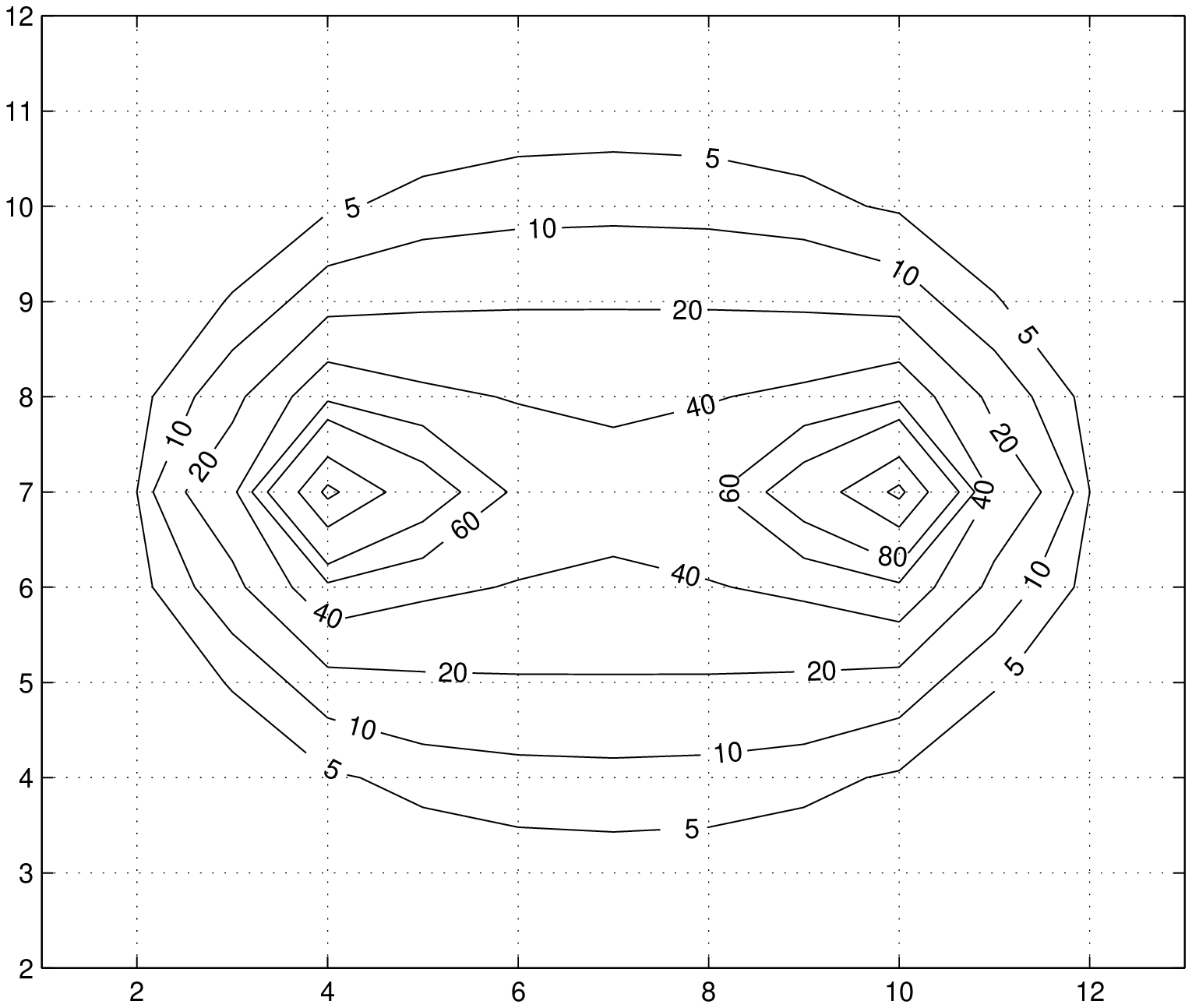}\epsfxsize=200pt\epsfbox{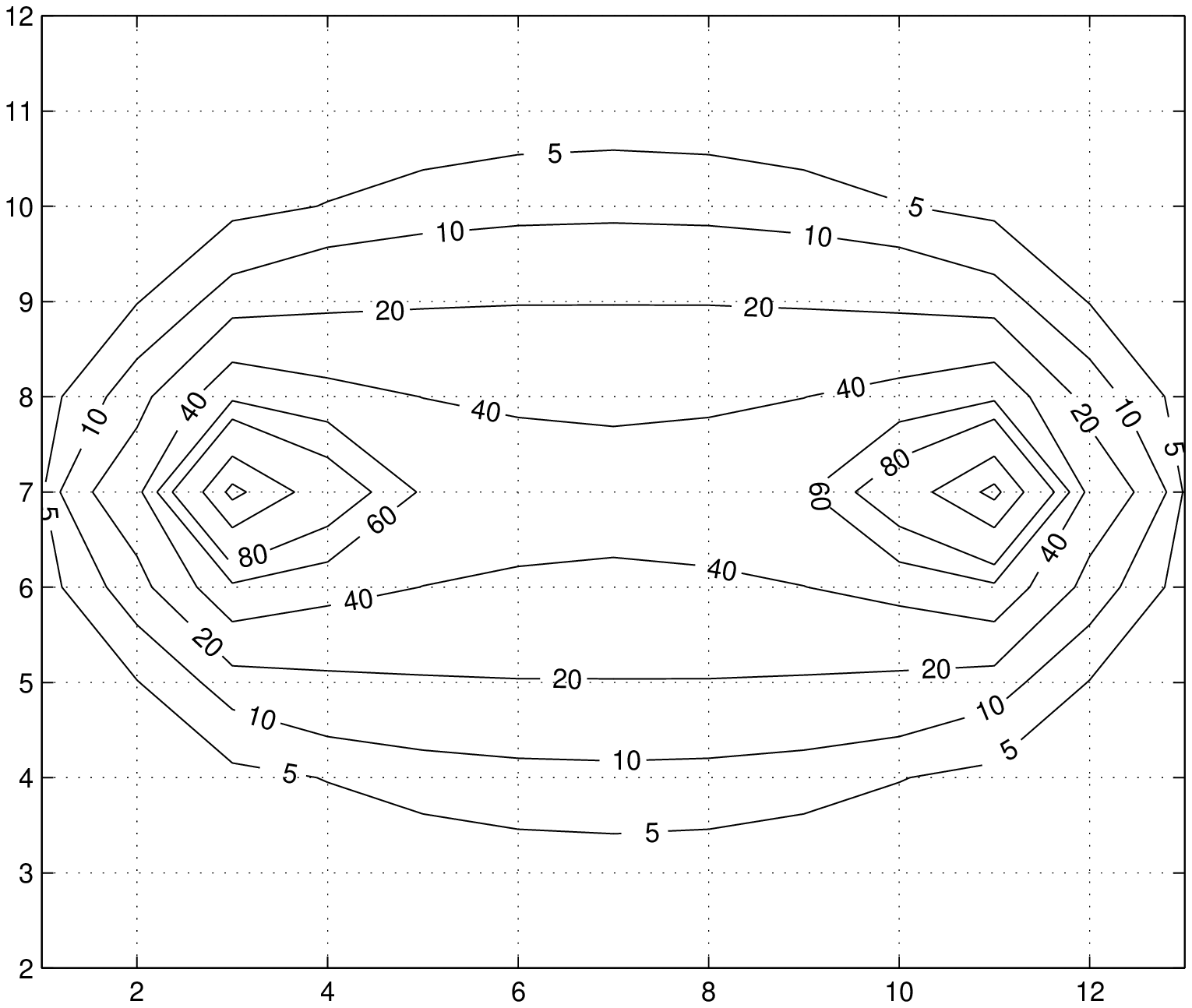} 
\caption{Two-quark action density at $T=3$ 
for a) $R=2$, b) $R=4$, c) $R=6$ and d) $R=8$. The units of isosurfaces are
GeV/fm$^3$. \label{ftwo}}
\end{figure}

\begin{figure}[hbtp]
\hspace{0cm}\epsfxsize=200pt\epsfbox{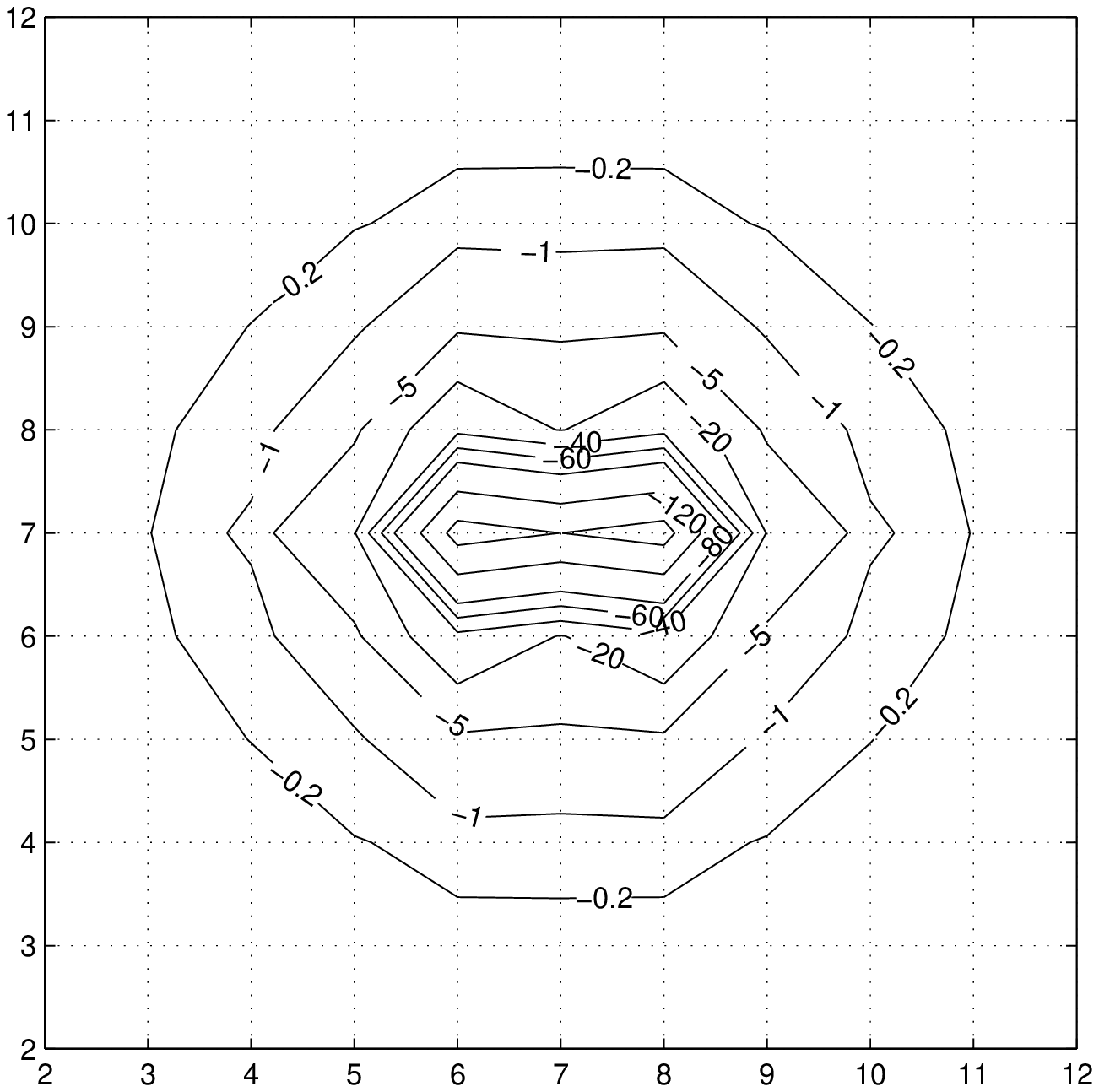}\epsfxsize=200pt\epsfbox{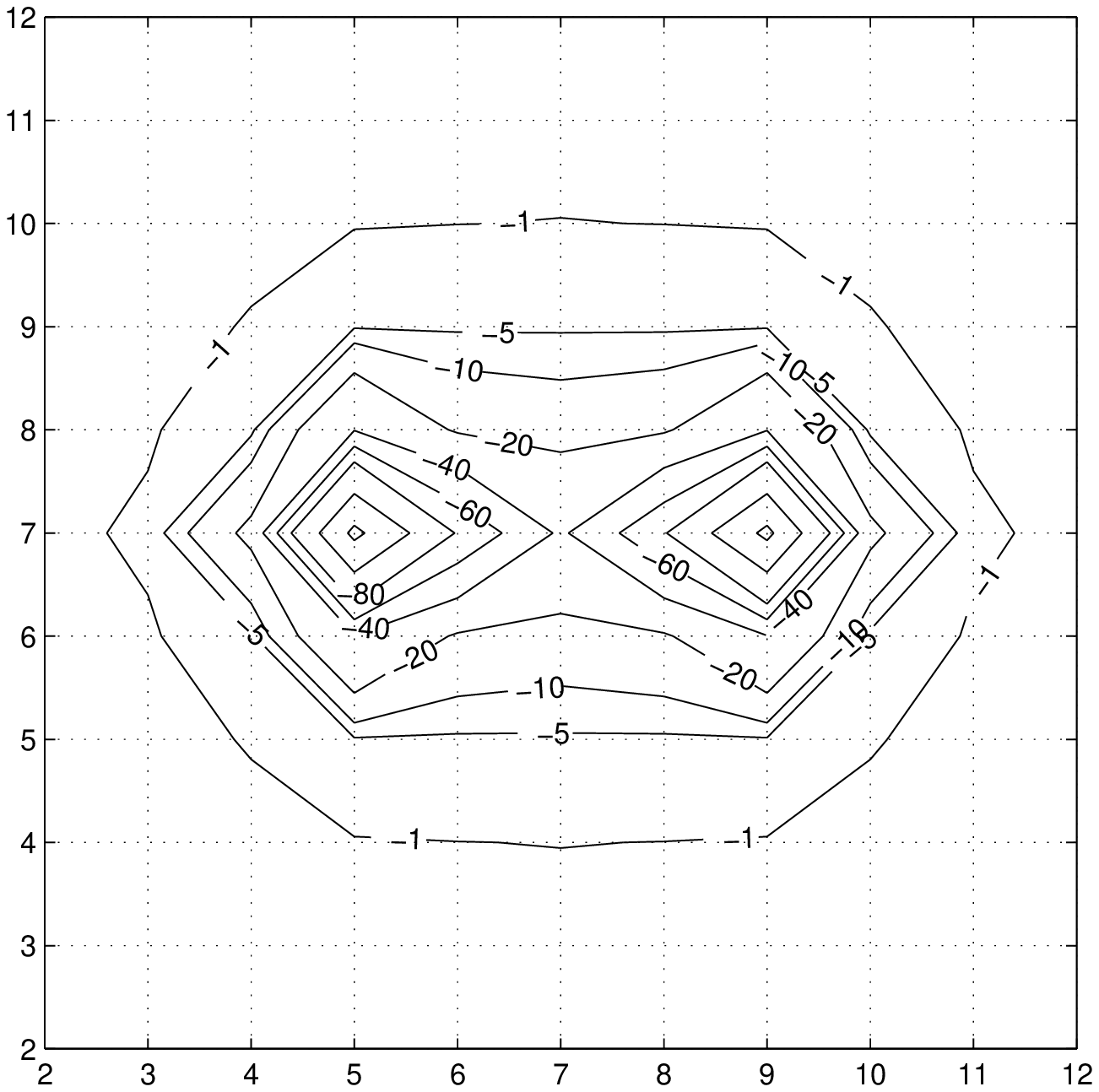} 
\hspace{0cm}\epsfxsize=200pt\epsfbox{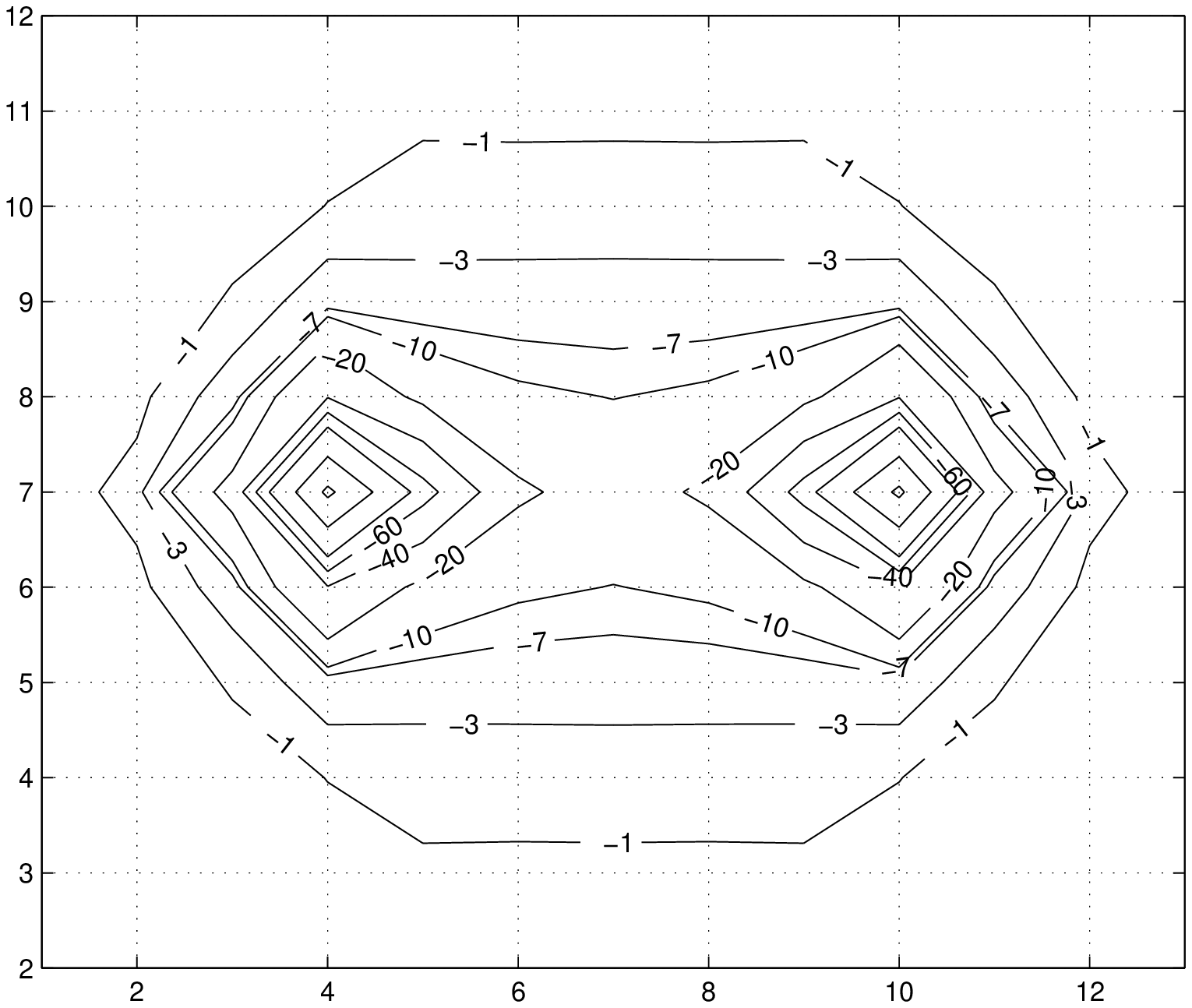}\epsfxsize=200pt\epsfbox{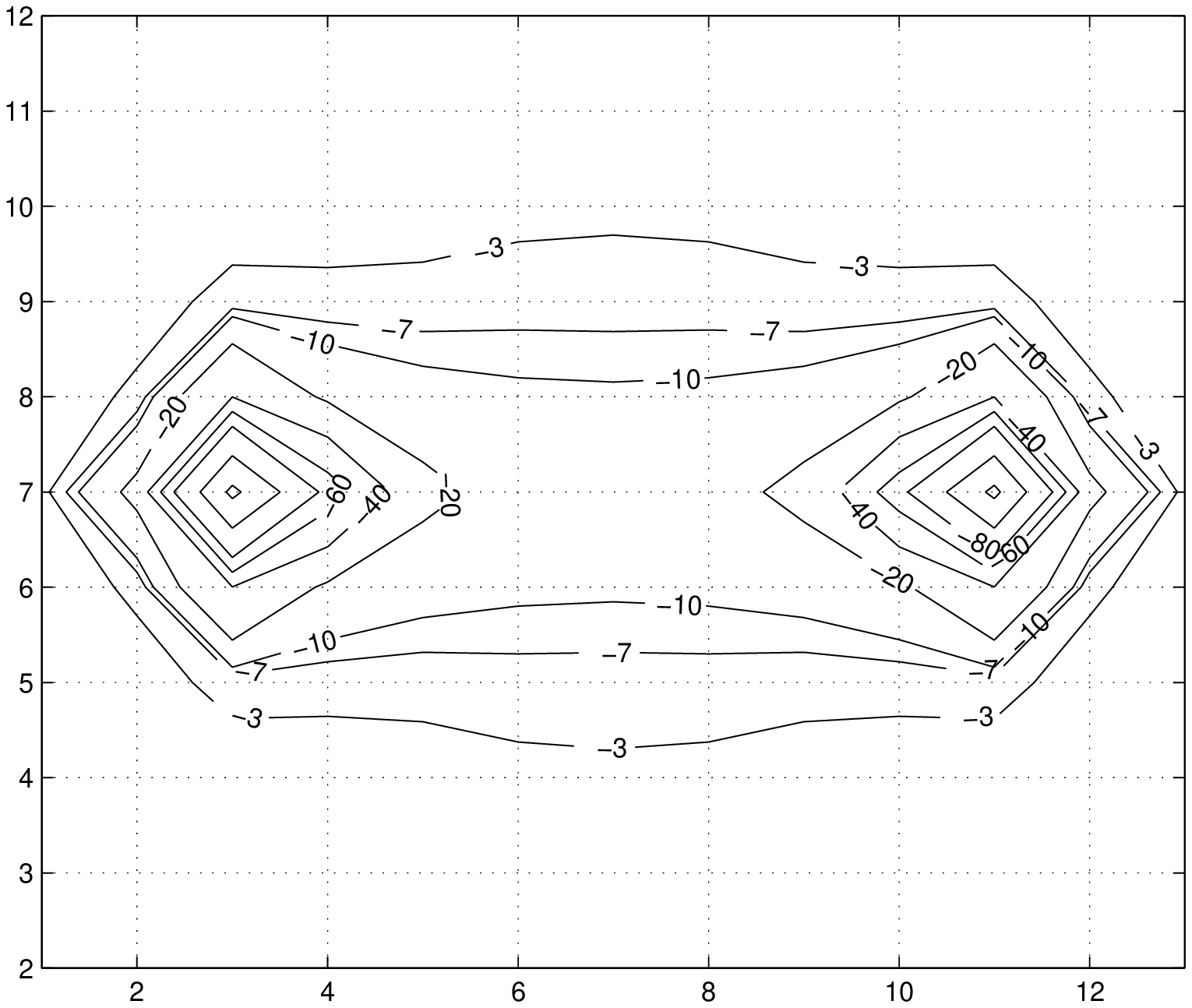} 
\caption{As in Fig.~\protect\ref{ftwo} but for the energy density 
(as in Eq.~\ref{sume})} 
\label{ftwo9}
\end{figure}

For comparison with the four-quark distributions below, the action 
distribution around two quarks on a lattice axis is presented in 
Fig.~\ref{ftwo} and the energy (as in Eq.~\ref{sume}) distribution in
Fig.~\ref{ftwo9}.

In these figures several points should be noted:

1) Both in the action and energy a flux tube structure clearly emerges
as $R$ increases.

2) The action density is that given in Eq.~\ref{ae} and is positive. However,
for historical reasons, it is the "negative" of the energy density
that is plotted throughout this paper i.e. 
$\sum(S {\cal E} + U {\cal B})$ in the notations of Eq.~\ref{ae3} and 
\ref{sume}.

3) At any given point, the magnitude of the energy field is a factor of 
about four less than that for the action.

4) The attractive potential between two quarks is responsible for the 
contours about a given quark  being deformed. It is seen that these 
contours are more spread out in the direction of the second quark.

It should be added that all of these features are well known and can be
found in Refs.~\cite{pen:97b,hay:96,bal:94}. The reason for repeating them 
here is to enable a comparison to be made with the four quark case to be 
discussed next.

\subsection{Four quarks -- before subtraction}

\begin{figure}[hbtp]
\hspace{0cm}\epsfxsize=200pt\epsfbox{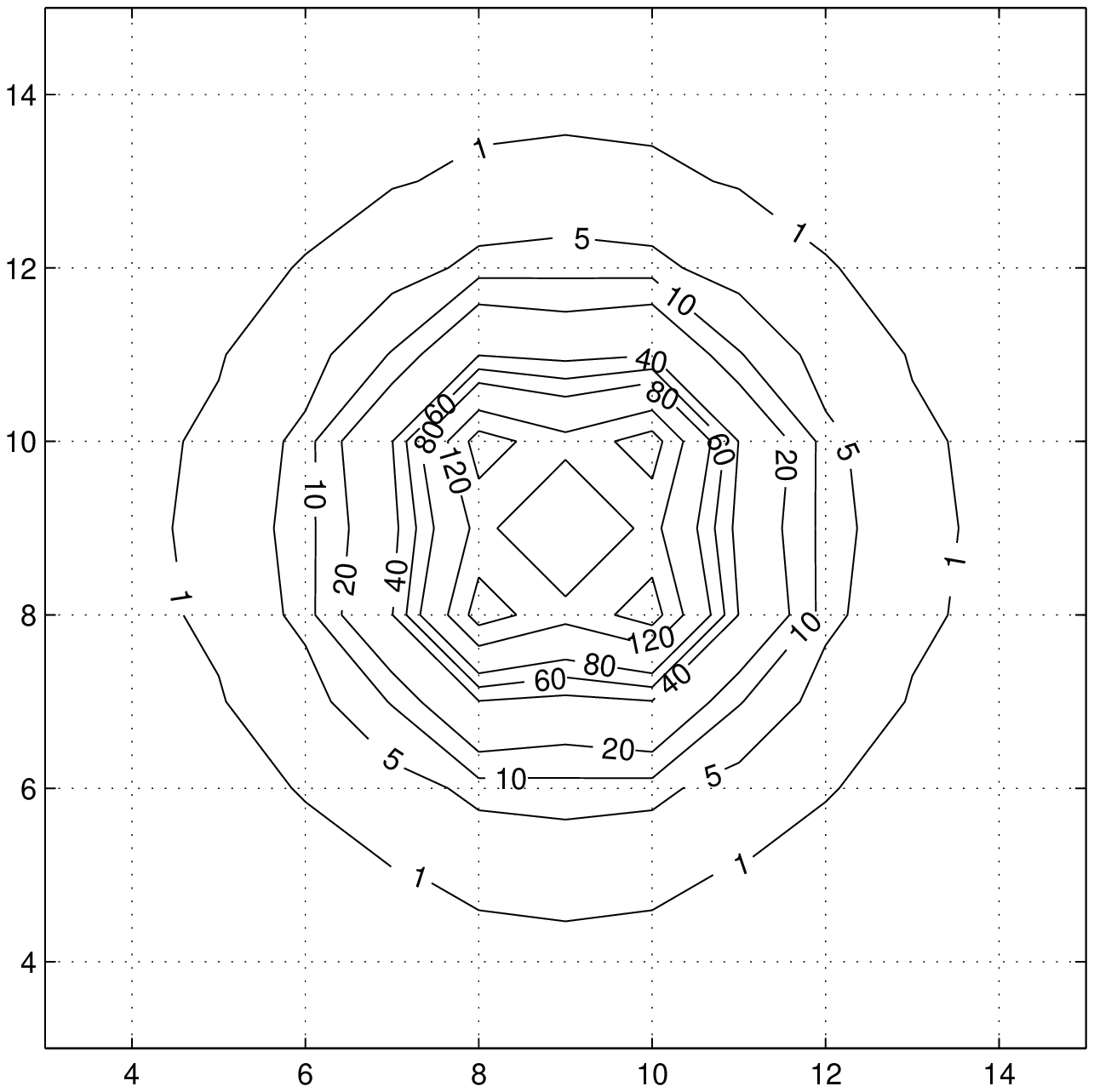}\epsfxsize=200pt\epsfbox{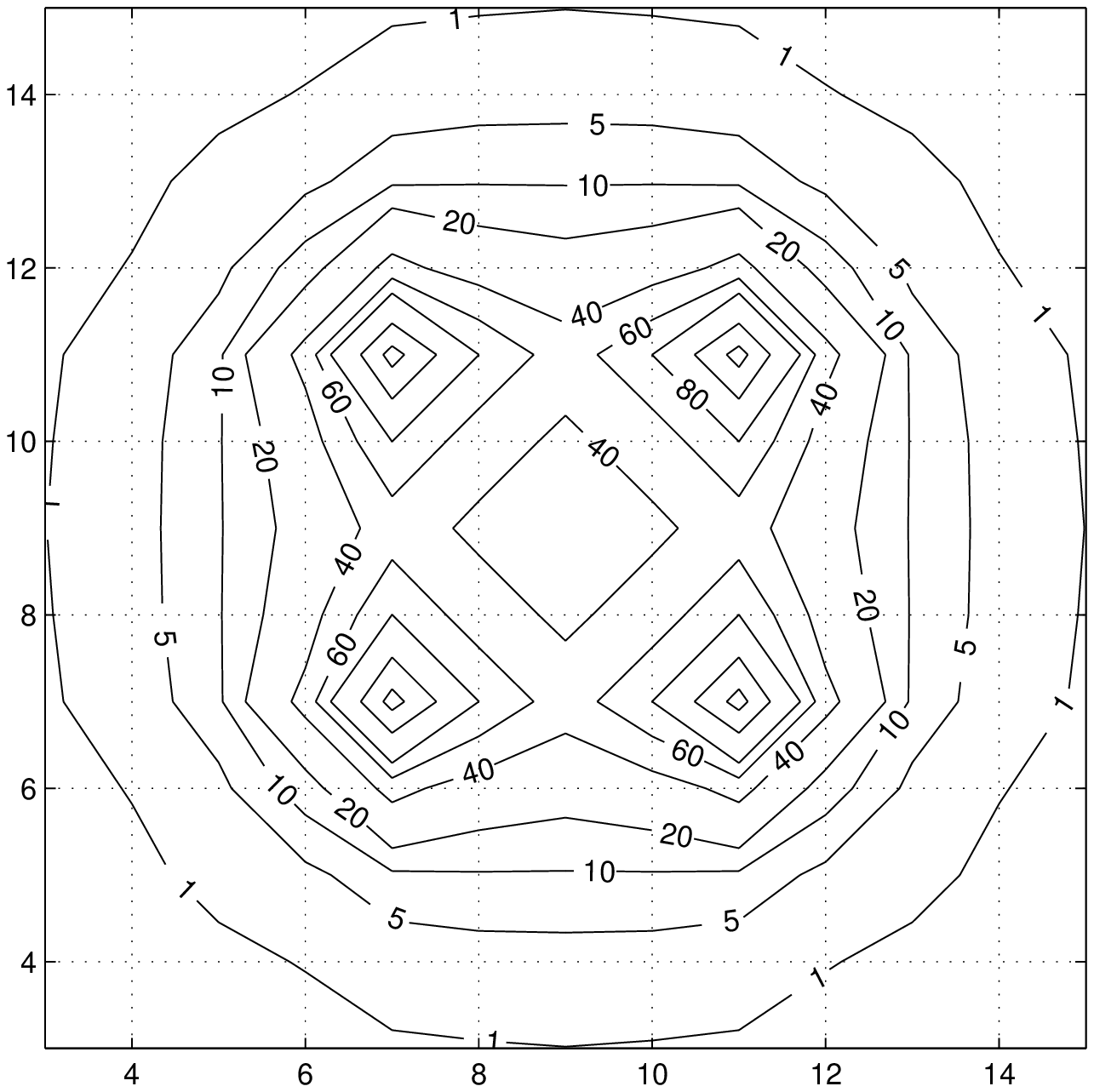} 
\hspace{0cm}\epsfxsize=200pt\epsfbox{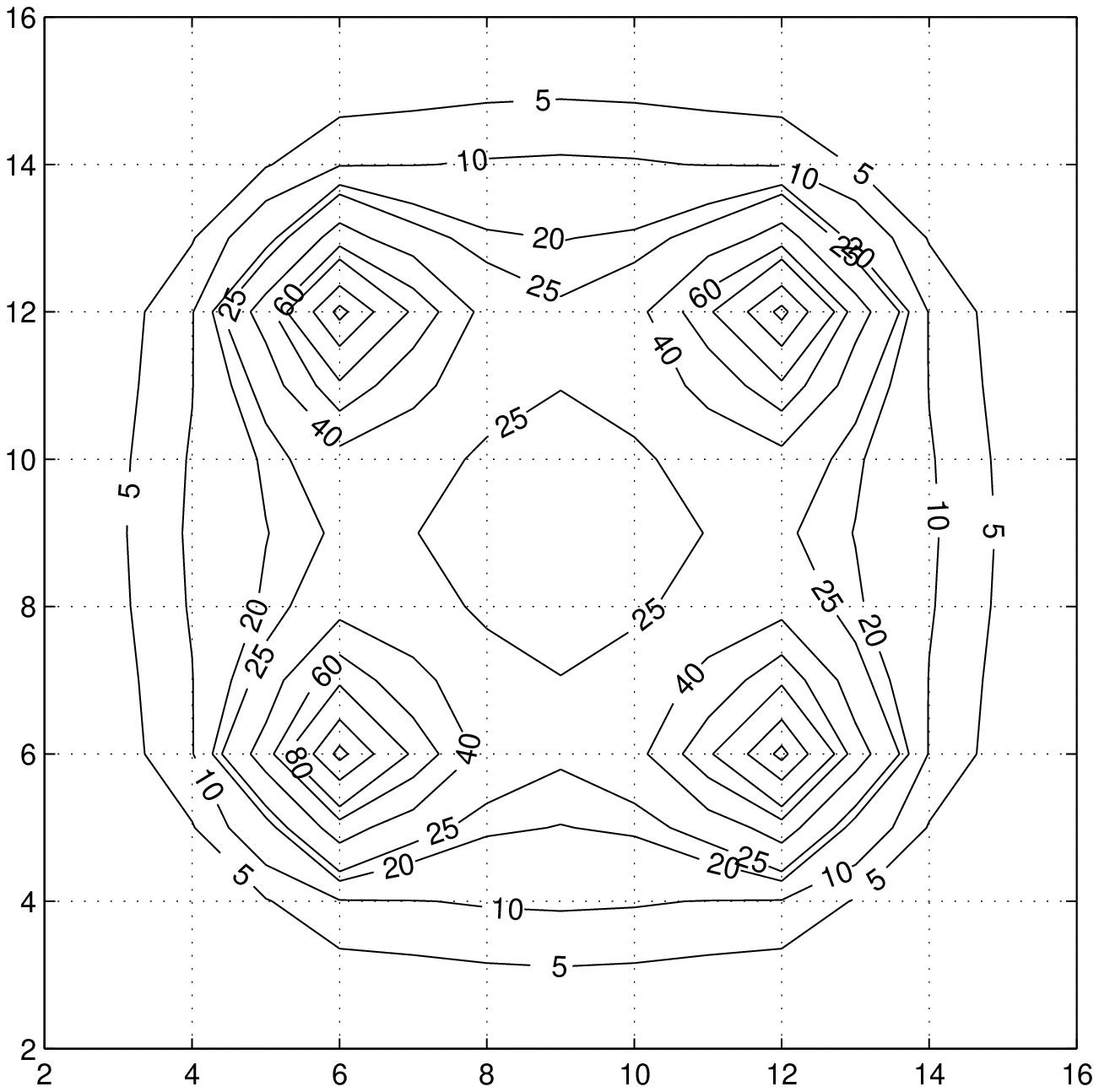}\epsfxsize=200pt\epsfbox{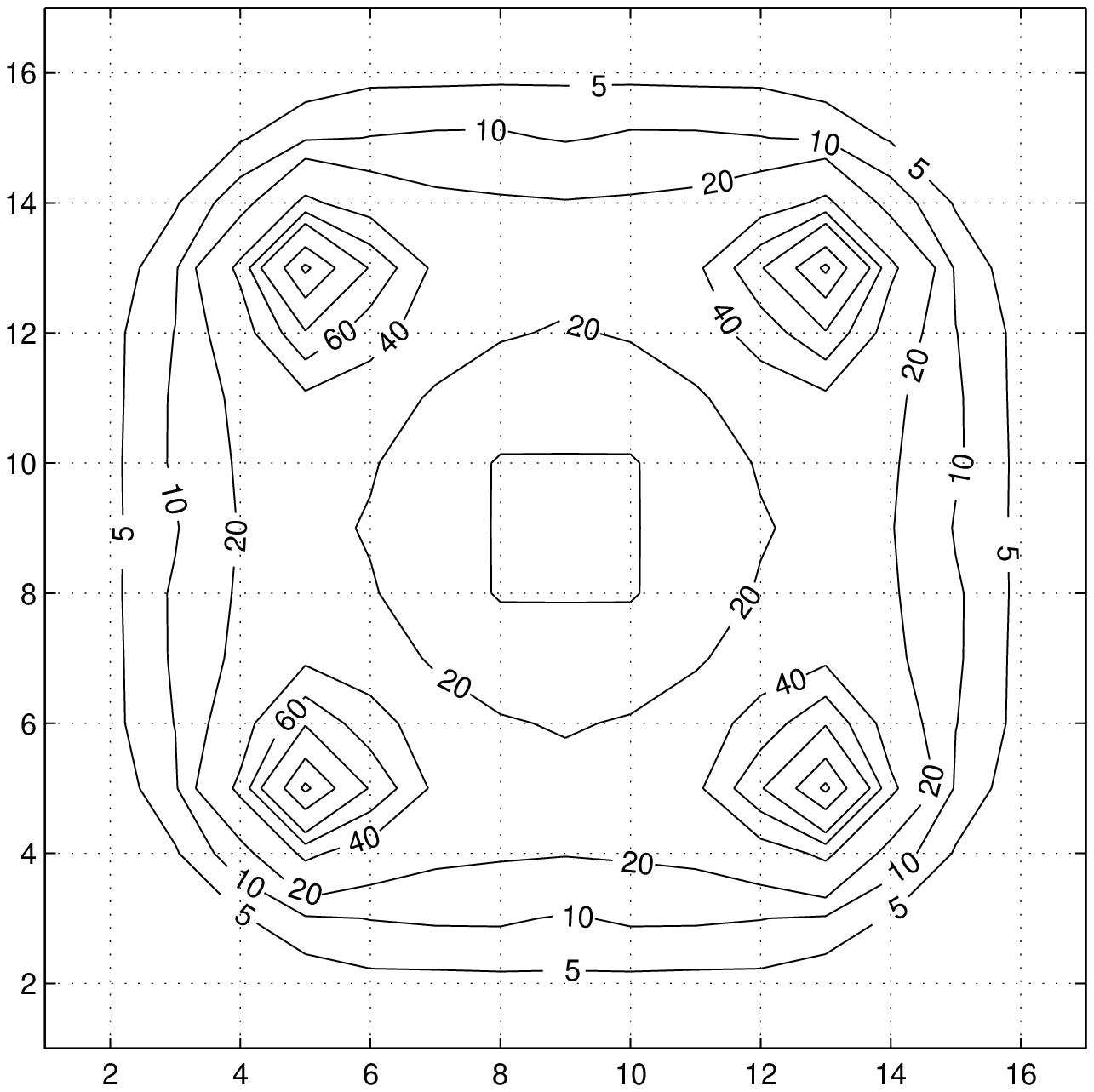} 
\caption{Four-quark action density at $T=3$ in the plane where the quarks lie
for a) $R=2$, b) $R=4$, c) $R=6$ and d) $R=8$.} \label{fqb}
\end{figure}

\begin{figure}[htbp]
\hspace{0cm}\epsfxsize=200pt\epsfbox{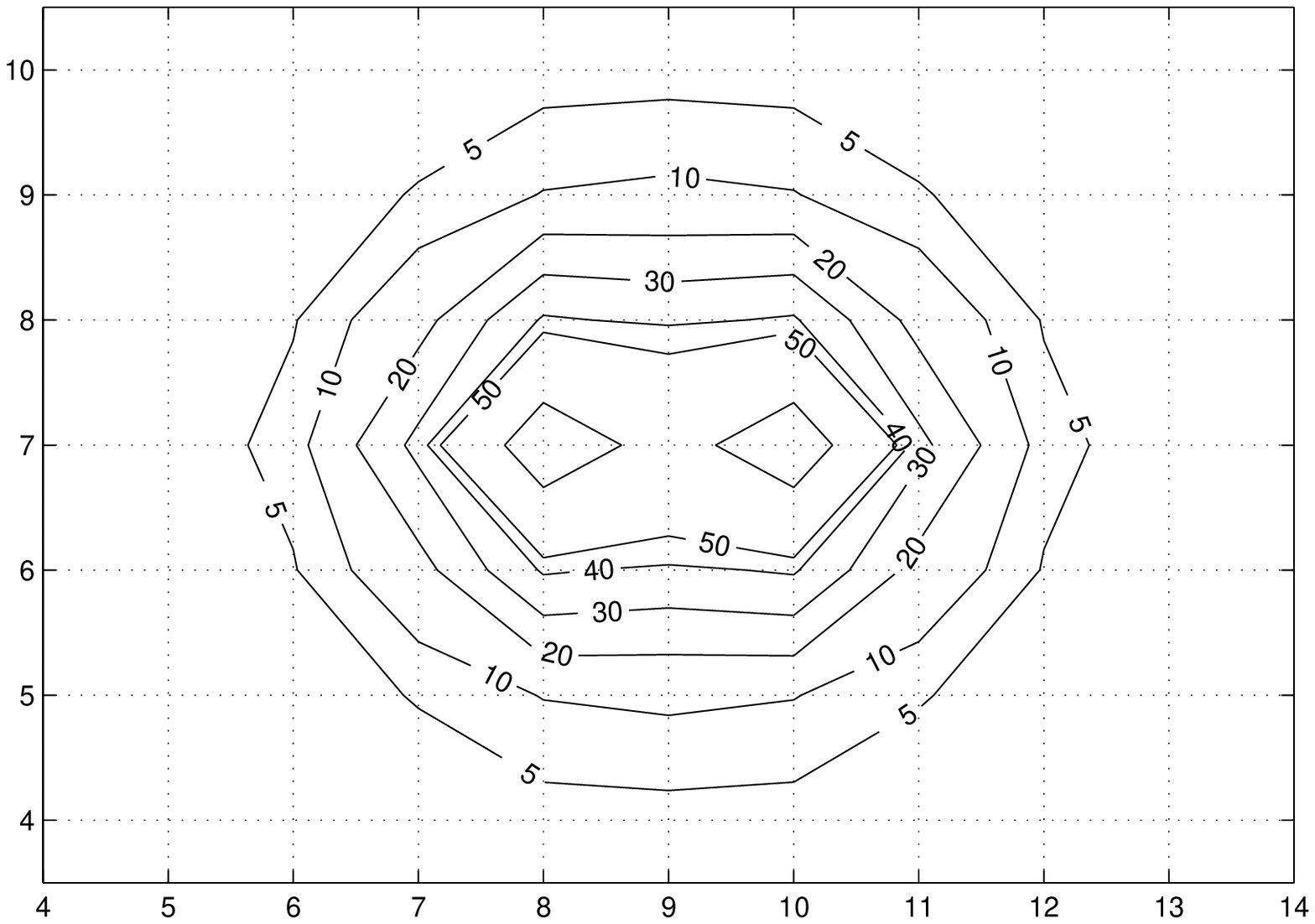}\epsfxsize=200pt\epsfbox{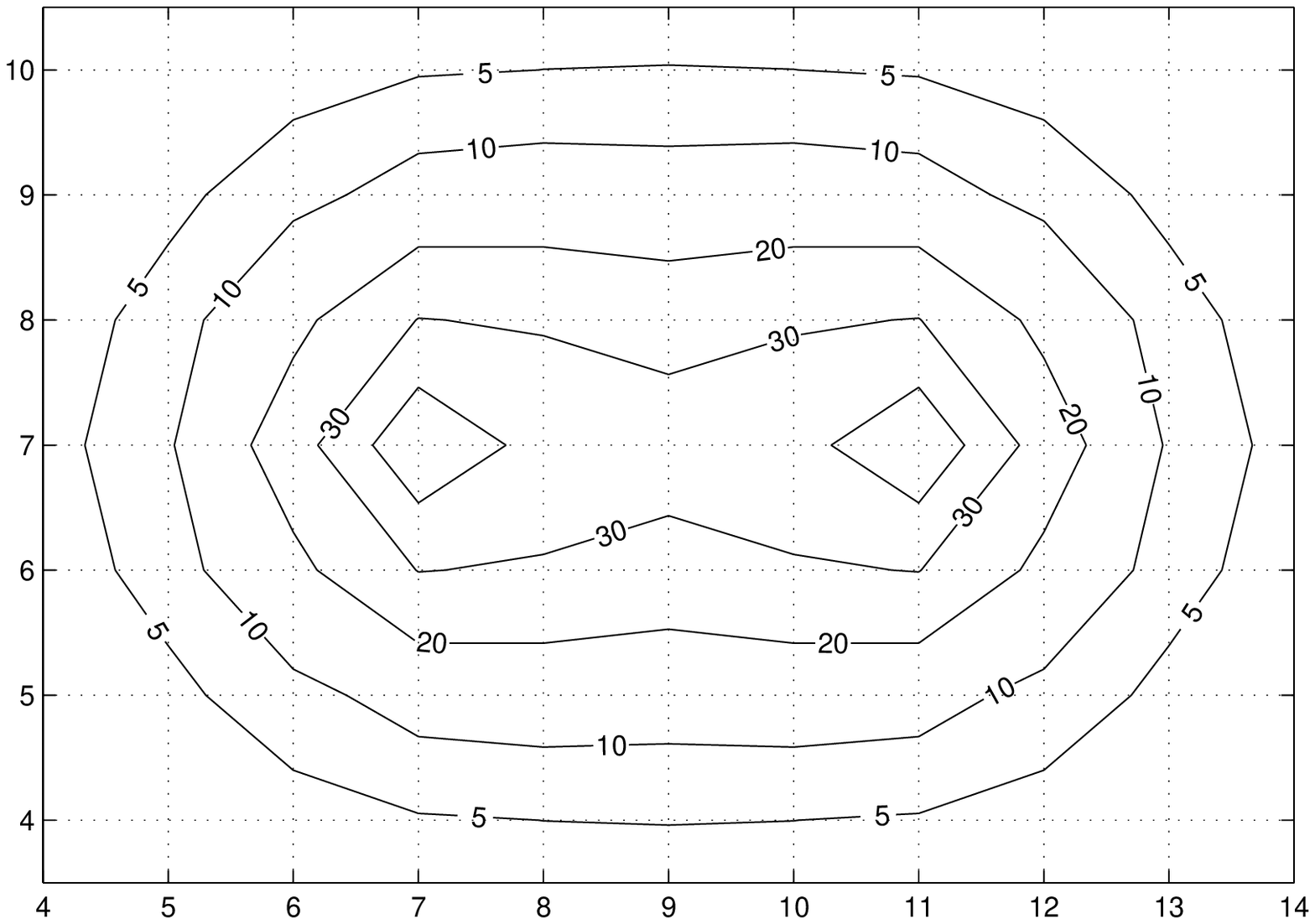} 
\hspace{0cm}\epsfxsize=200pt\epsfbox{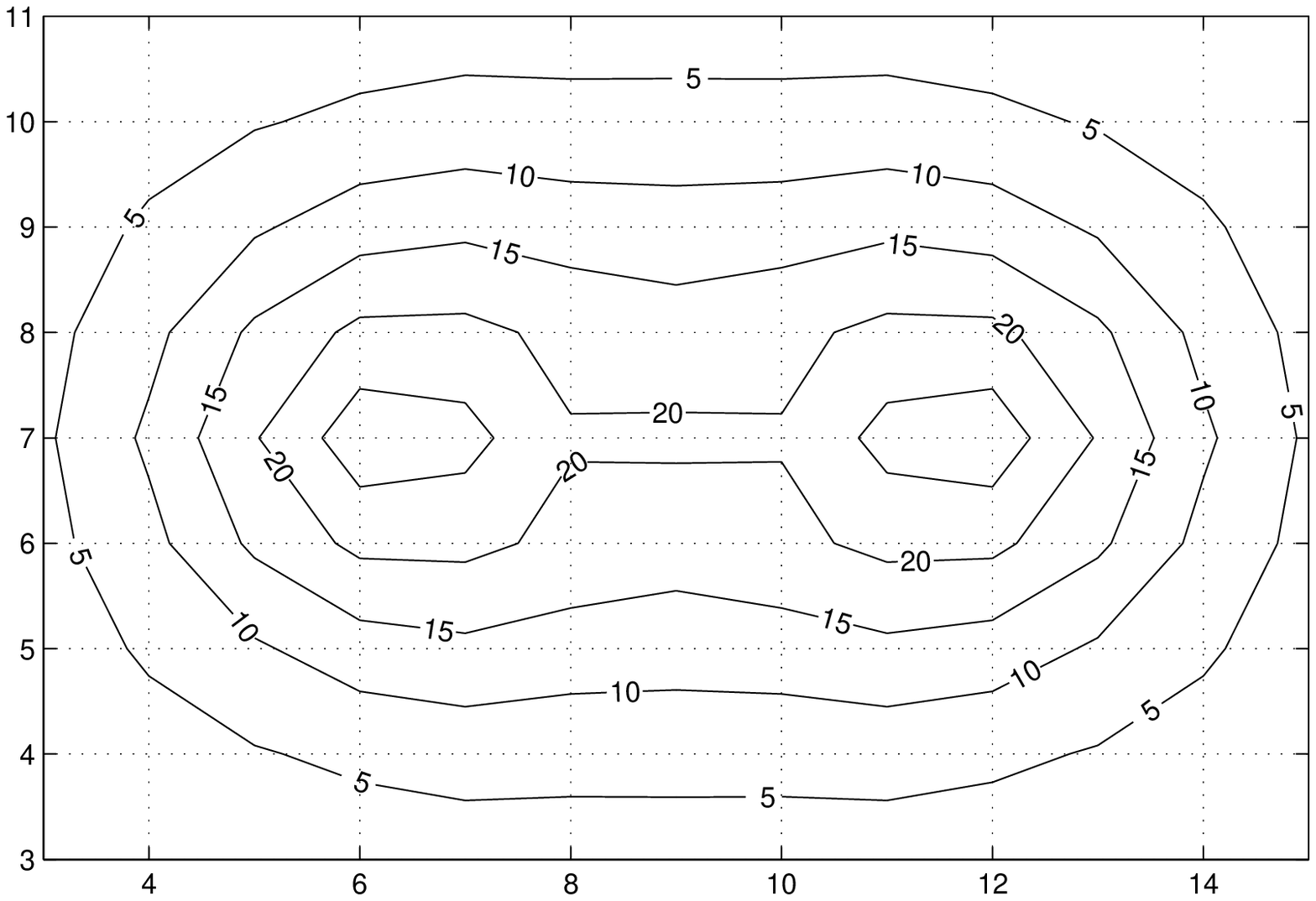}\epsfxsize=200pt\epsfbox{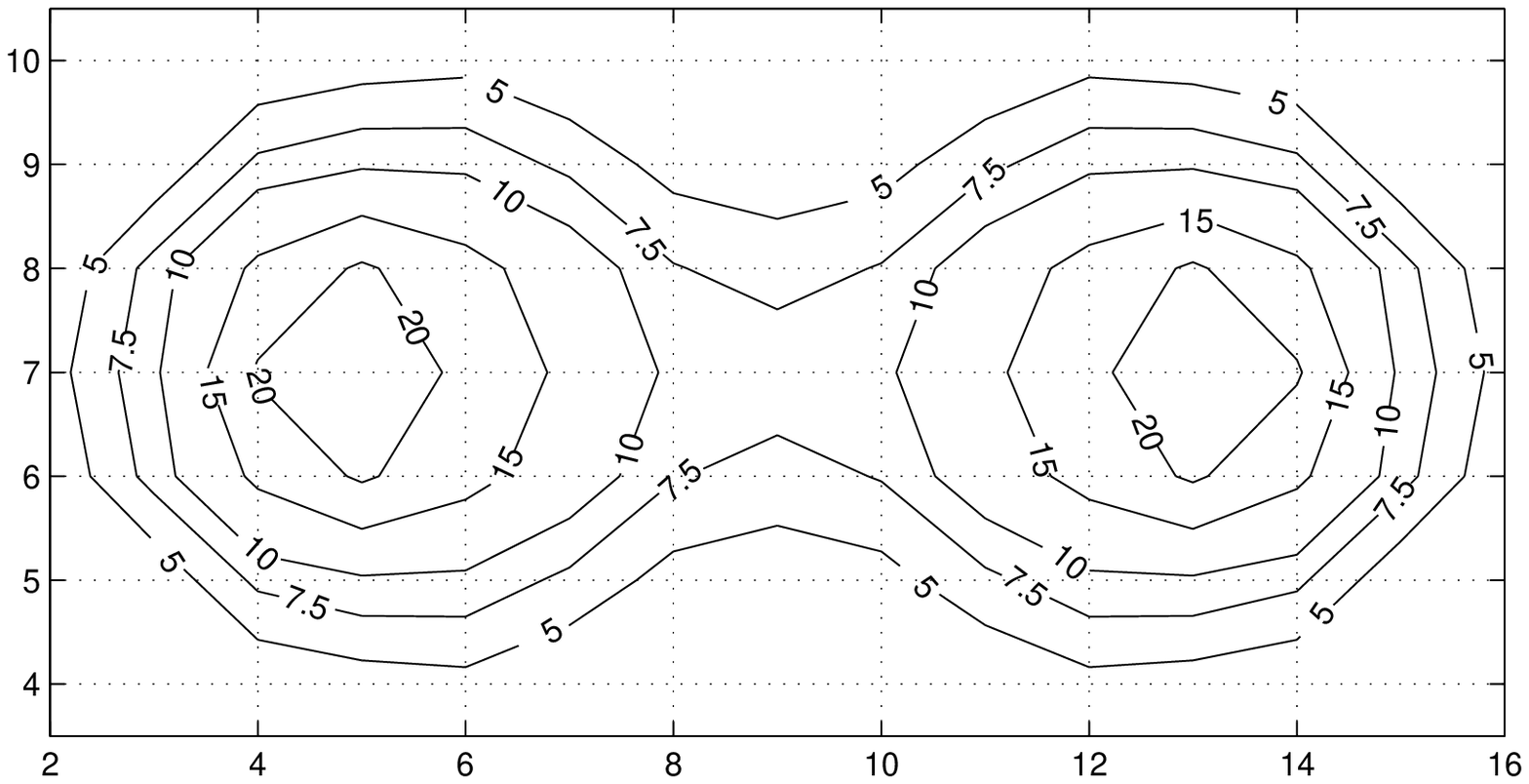} 
 \caption{As in Fig.~\ref{fqb} but in the plane transverse to the 
quarks and in the middle of them.}
 \label{fqsln}
\end{figure}

\begin{figure}[htbp]
\hspace{0cm}\epsfxsize=200pt\epsfbox{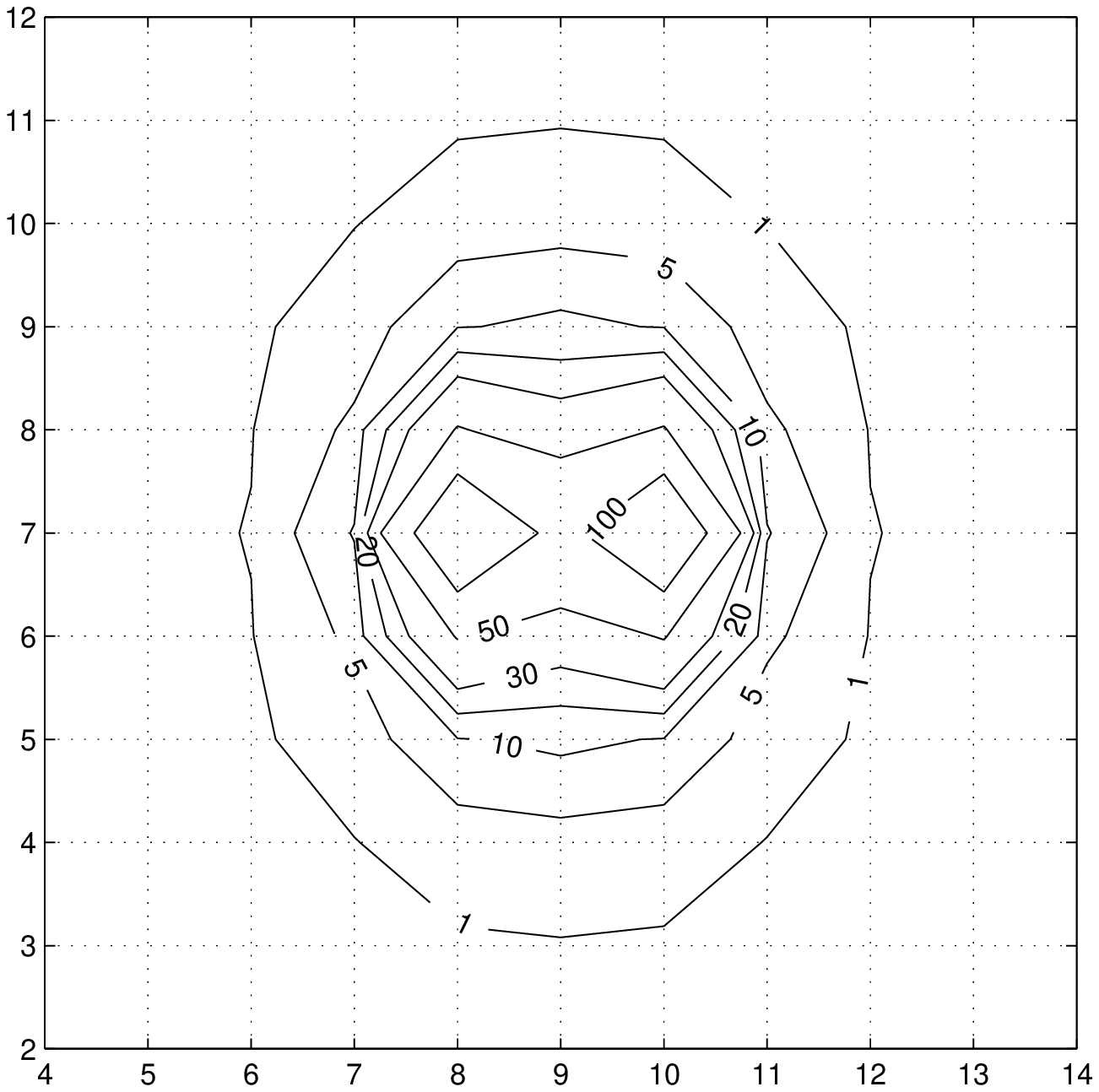}\epsfxsize=200pt\epsfbox{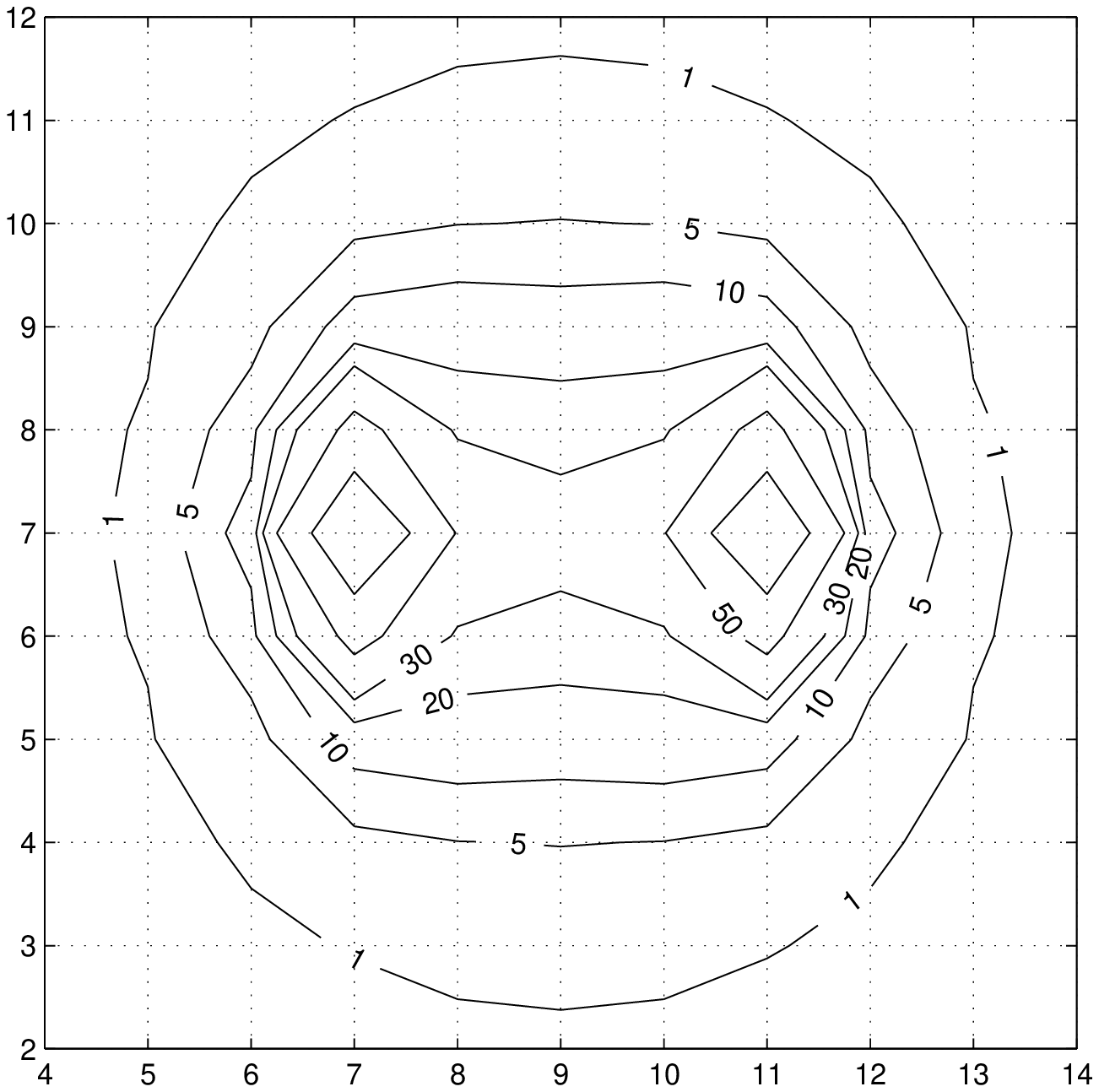} 
\hspace{0cm}\epsfxsize=200pt\epsfbox{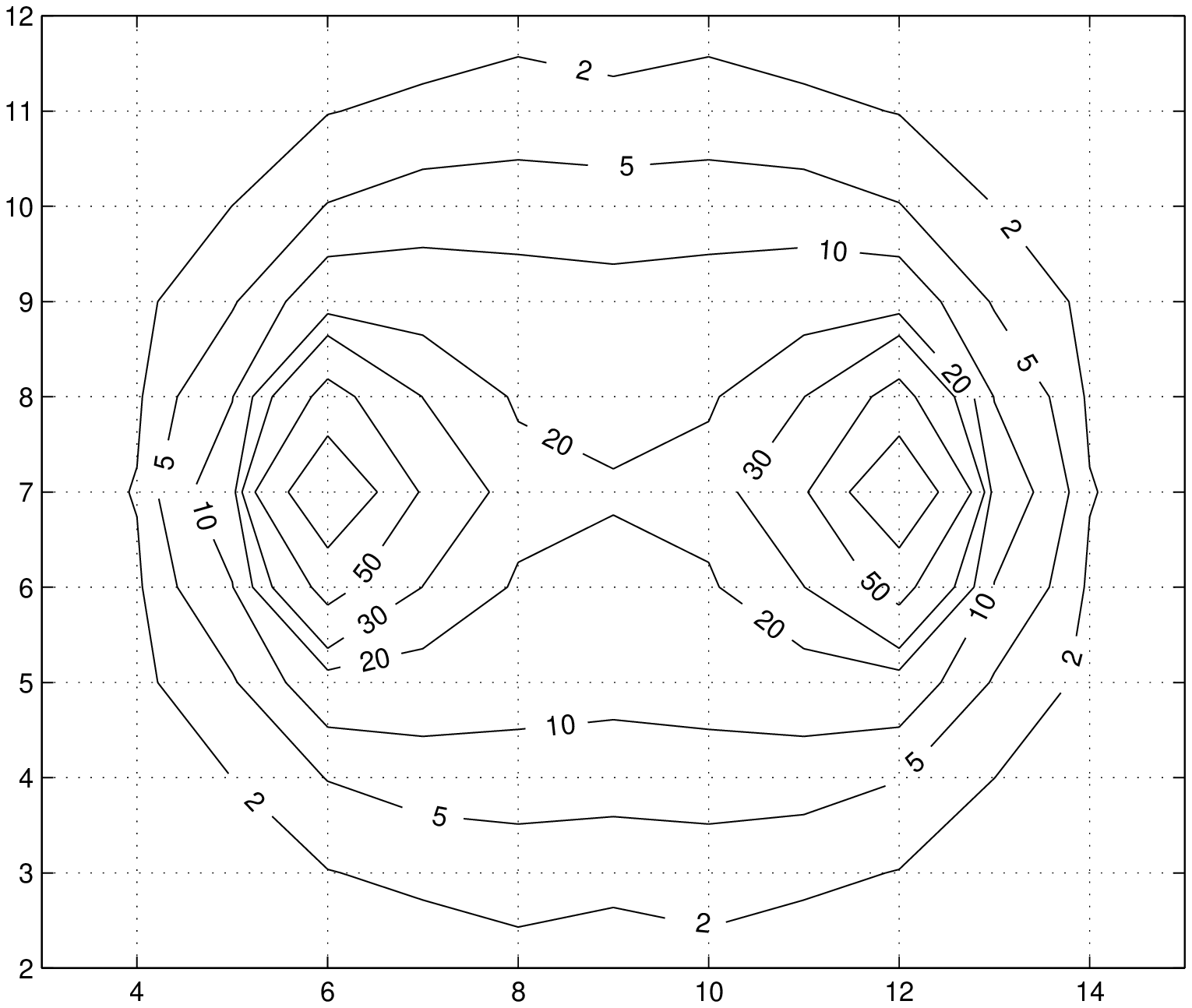}\epsfxsize=200pt\epsfbox{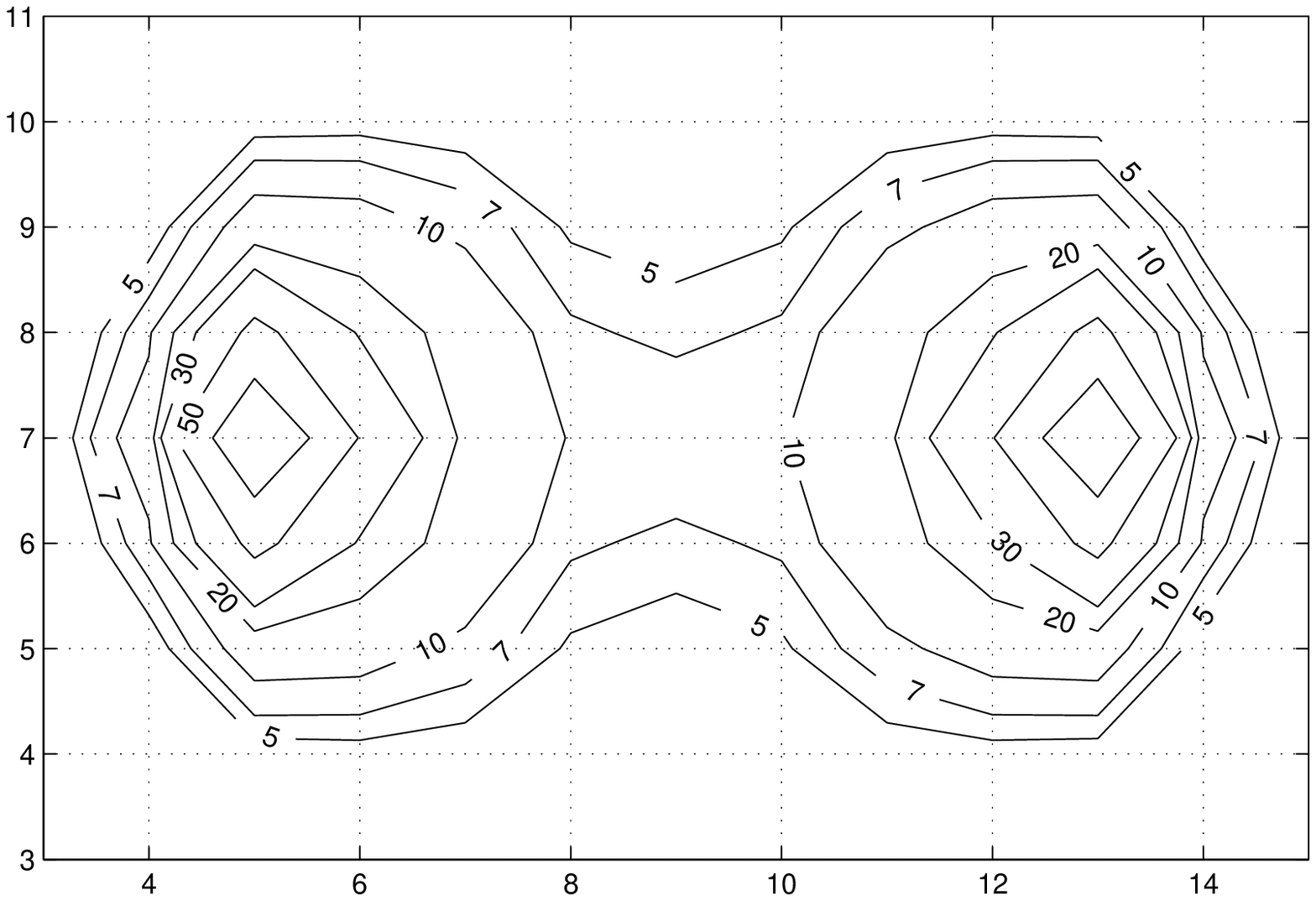} 
 \caption{As in Fig.~\protect\ref{fqb} but in the plane transverse to the 
quarks and cutting diagonally through them. }
 \label{fqsld}
\end{figure}

\begin{figure}[hbtp]
\hspace{0cm}\epsfxsize=200pt\epsfbox{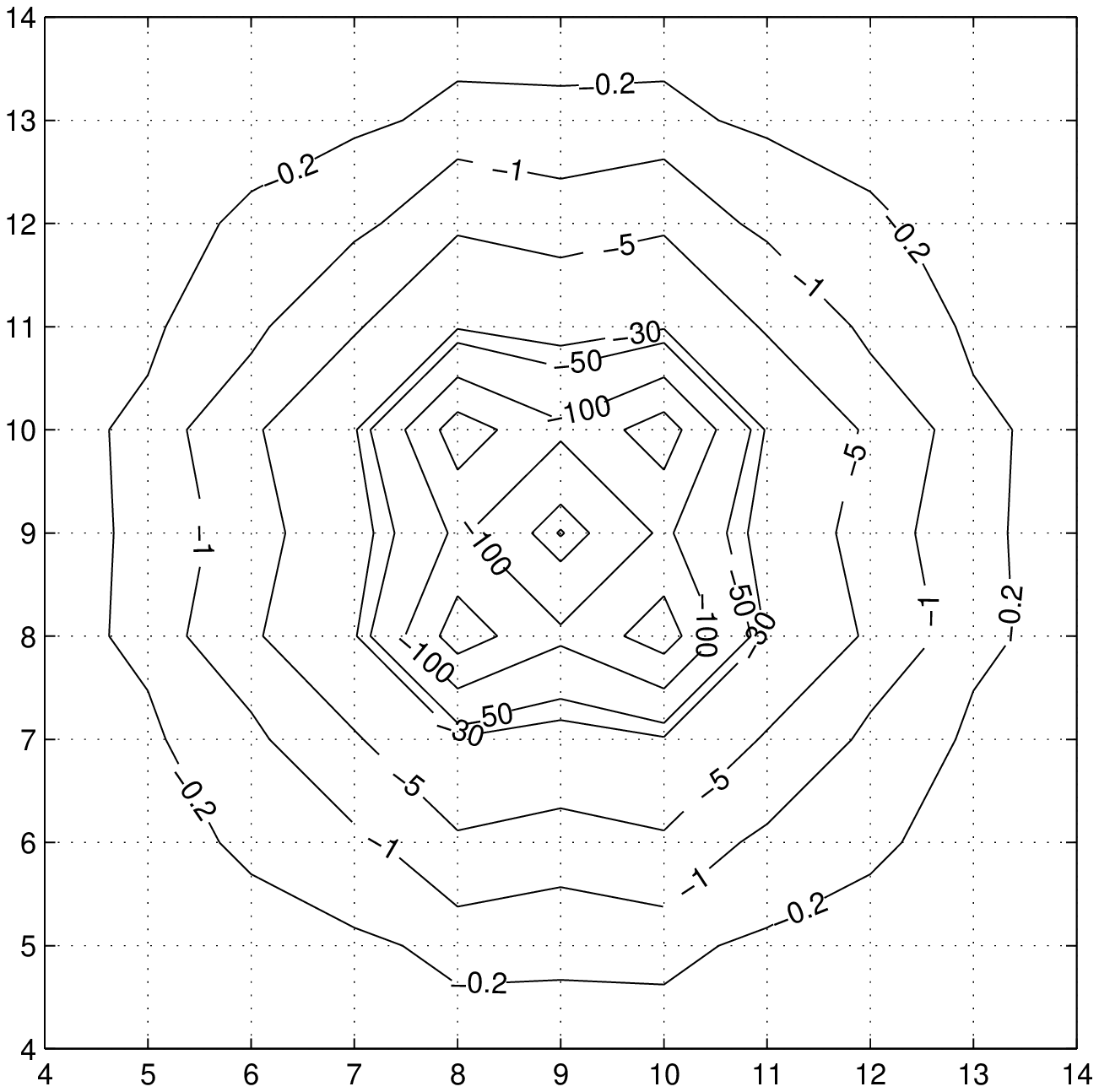}\epsfxsize=200pt\epsfbox{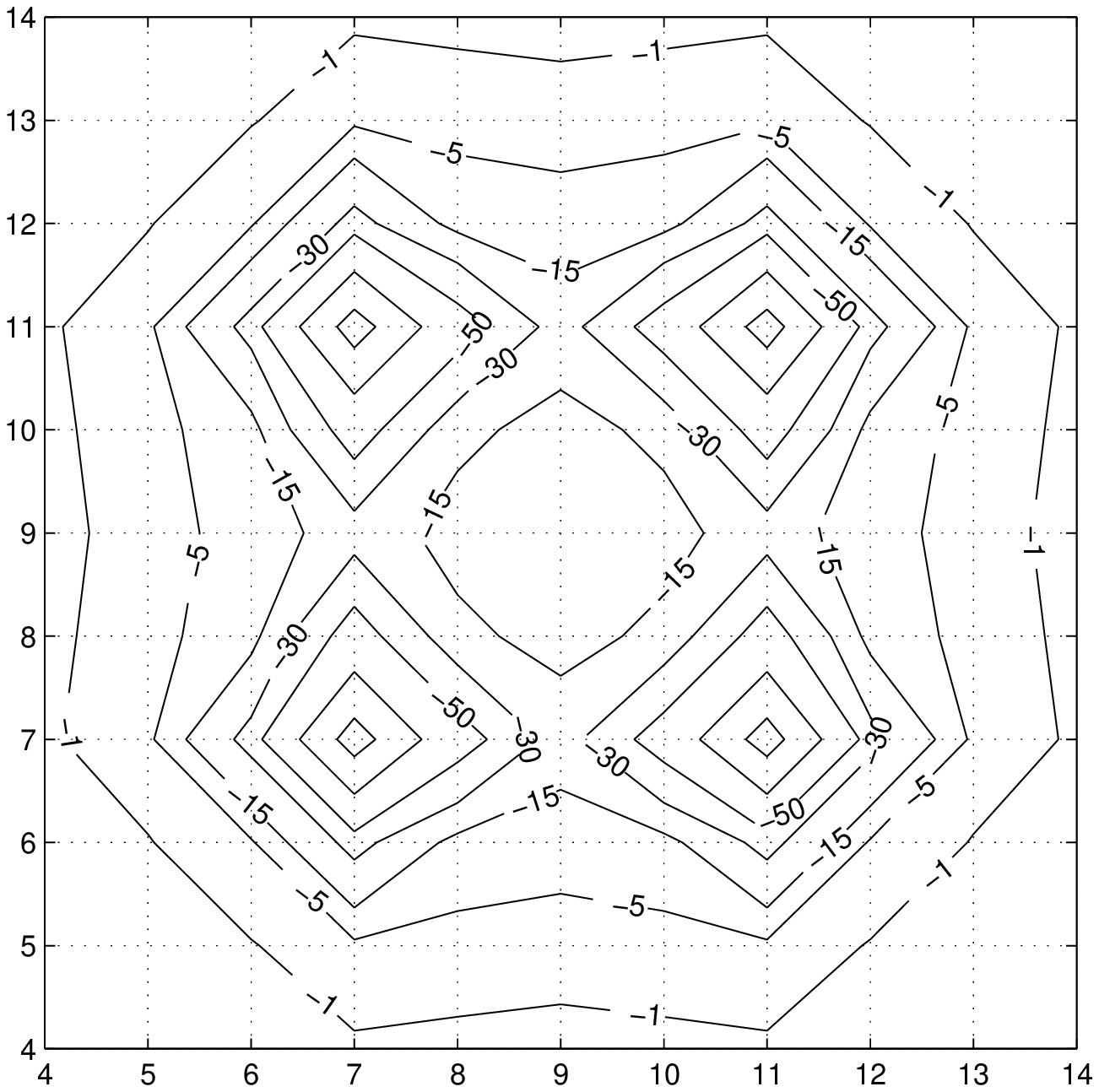} 
\hspace{0cm}\epsfxsize=200pt\epsfbox{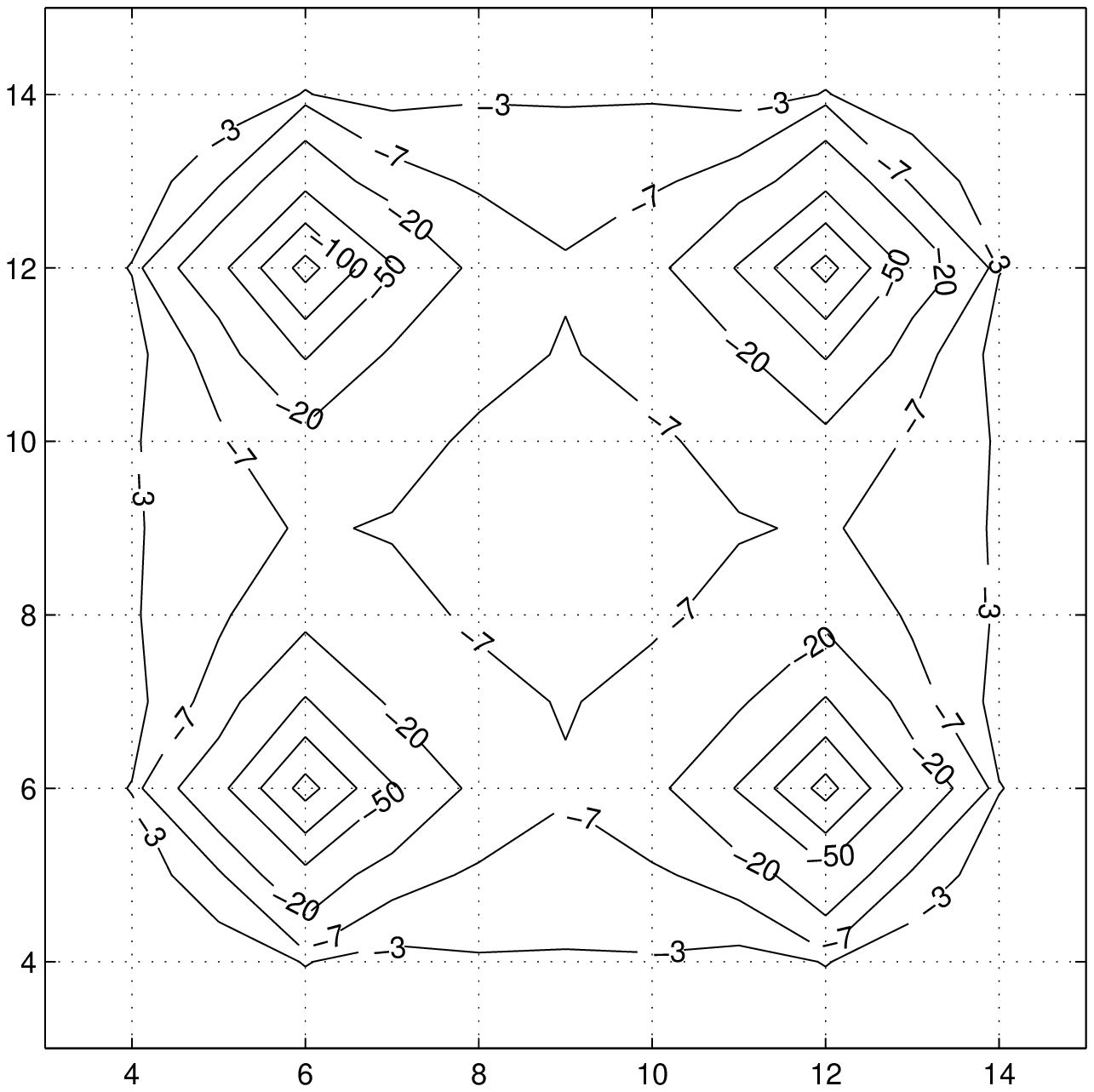}\epsfxsize=200pt\epsfbox{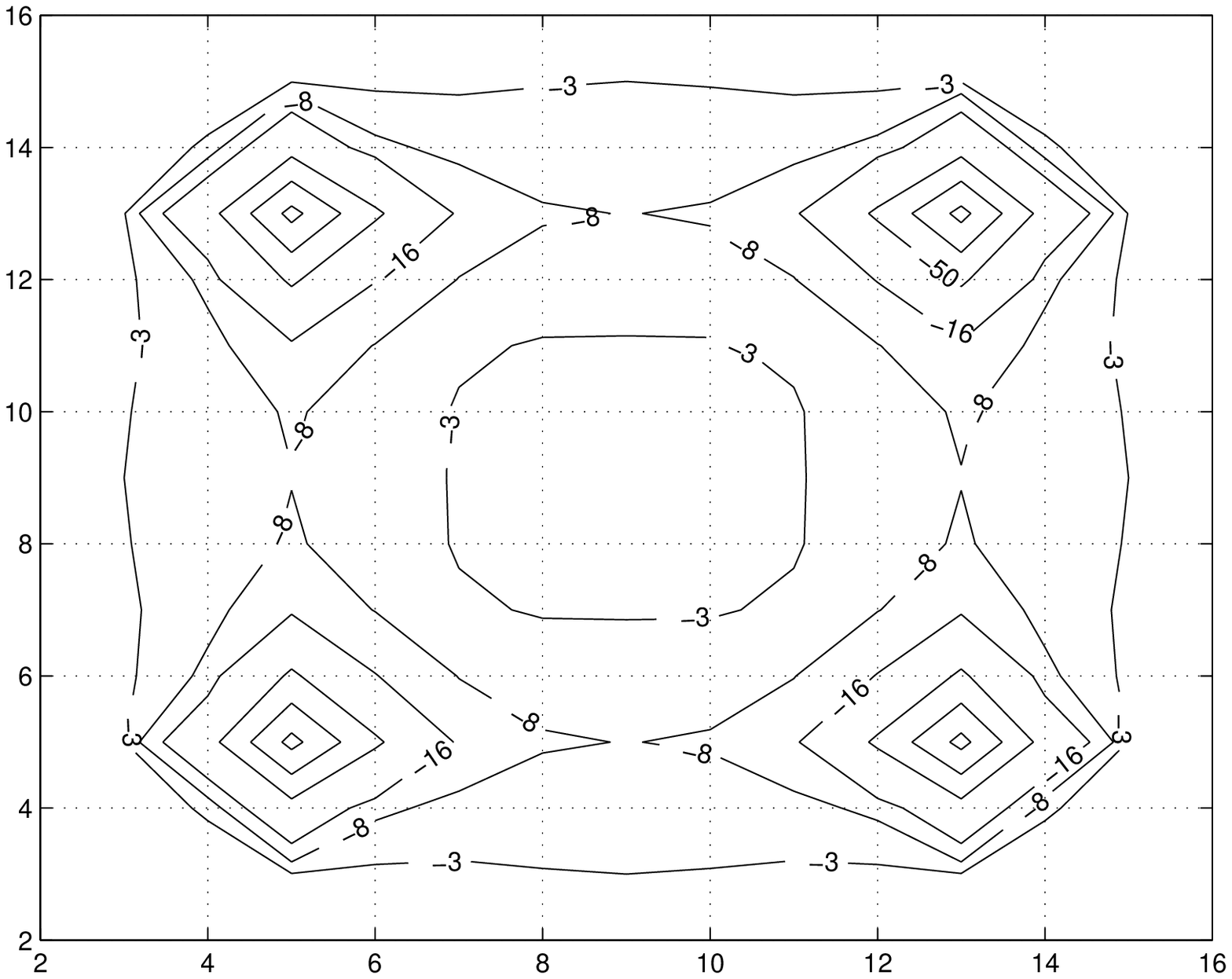} 
\caption{As in Fig.~\protect\ref{fqb} but for the energy.} 
\label{fqb9}
\end{figure}

\begin{figure}[htbp]
\hspace{0cm}\epsfxsize=200pt\epsfbox{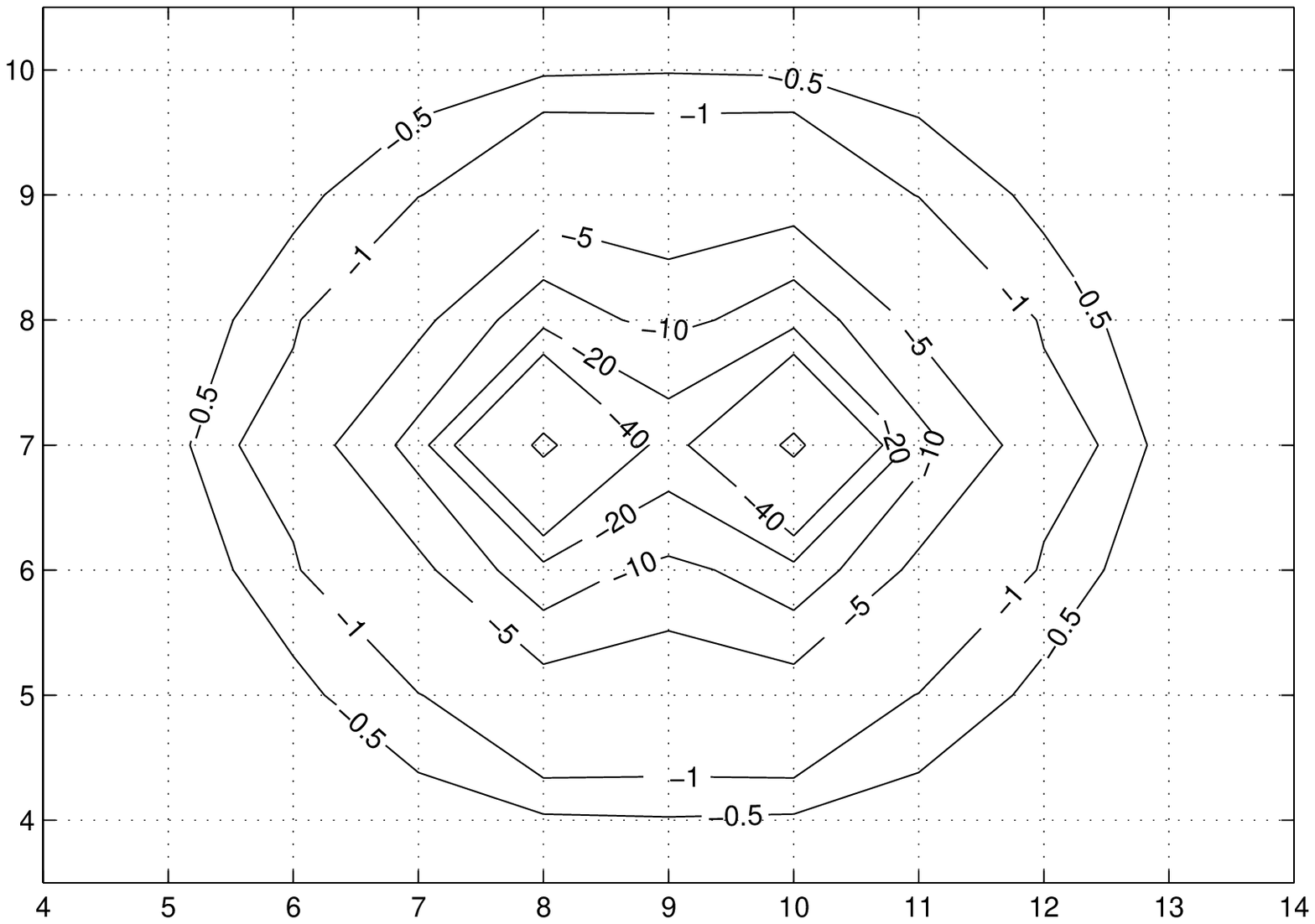}\epsfxsize=200pt\epsfbox{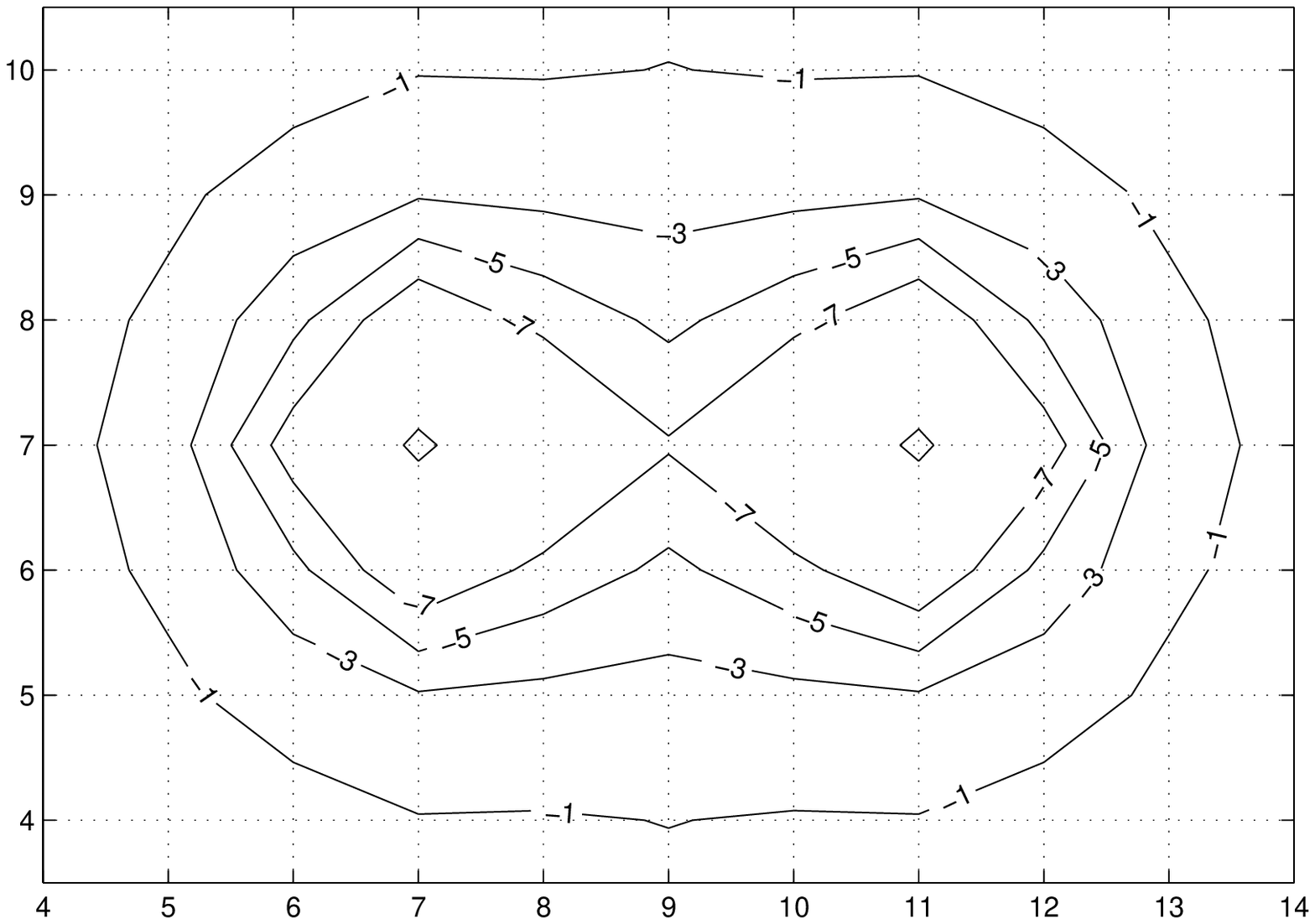} 
\hspace{0cm}\epsfxsize=200pt\epsfbox{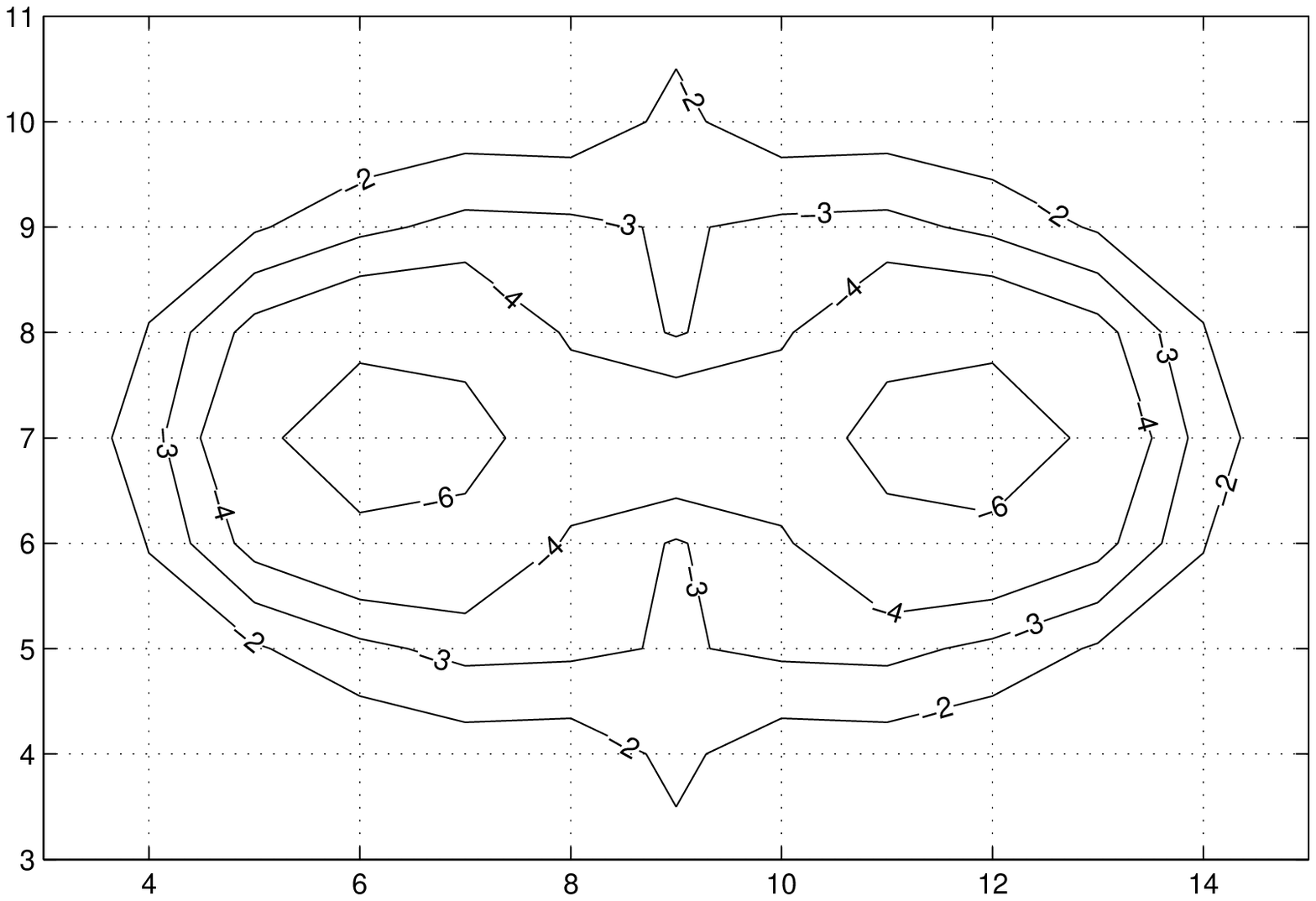}\epsfxsize=200pt\epsfbox{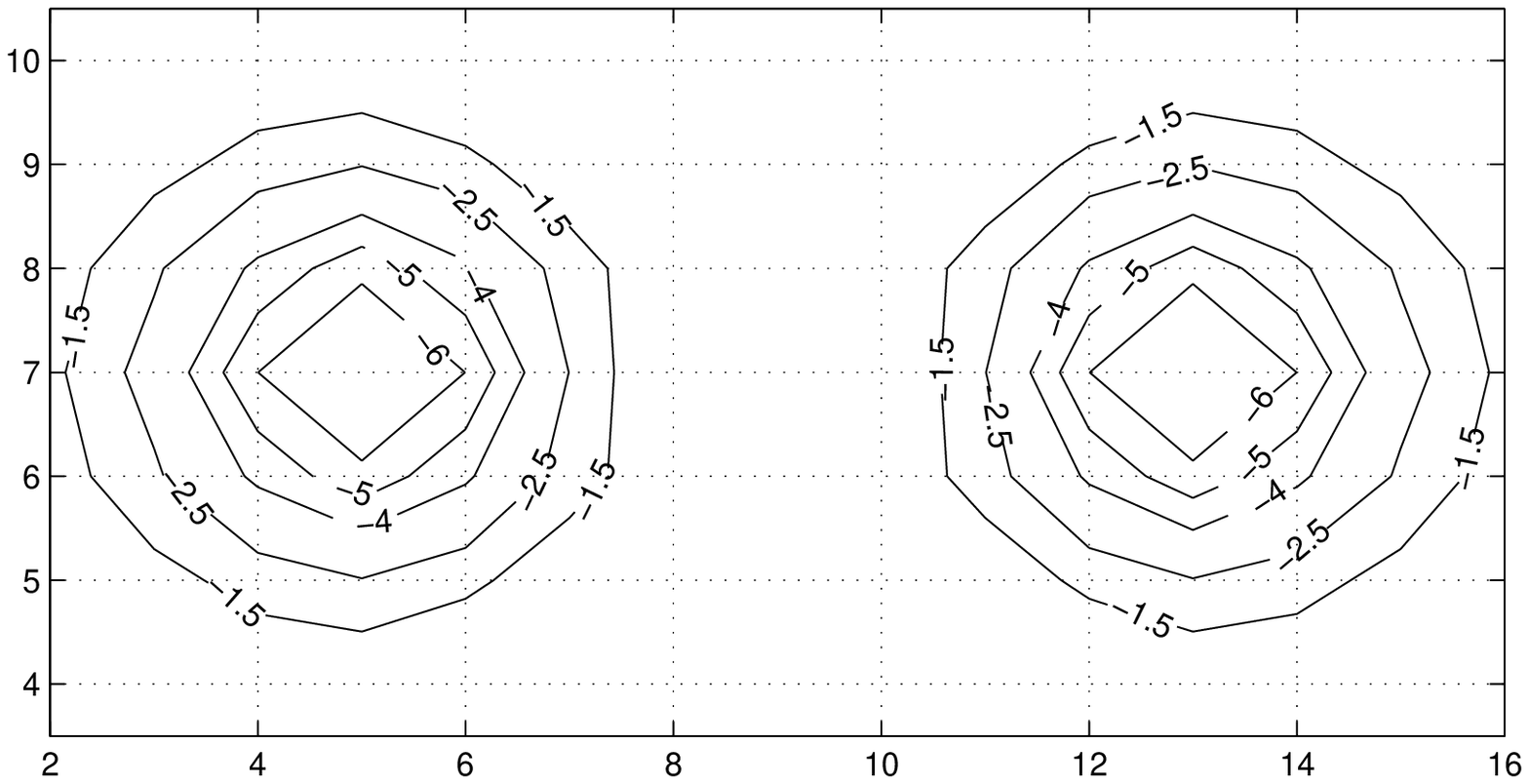} 
 \caption{As in Fig.~\protect\ref{fqsln} but for the energy. }
 \label{fqsln9}
\end{figure}

\begin{figure}[htbp]
\hspace{0cm}\epsfxsize=200pt\epsfbox{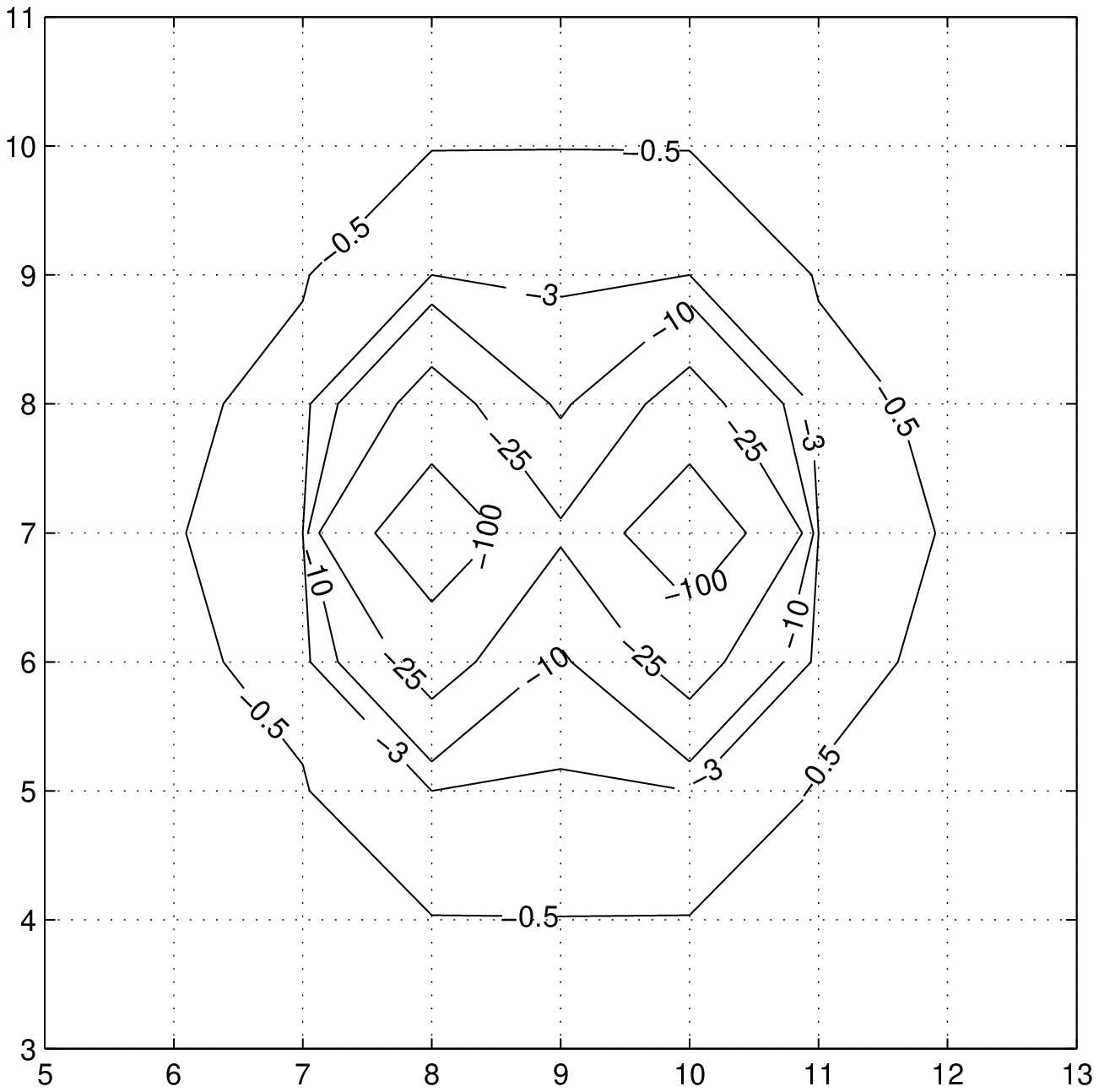}\epsfxsize=200pt\epsfbox{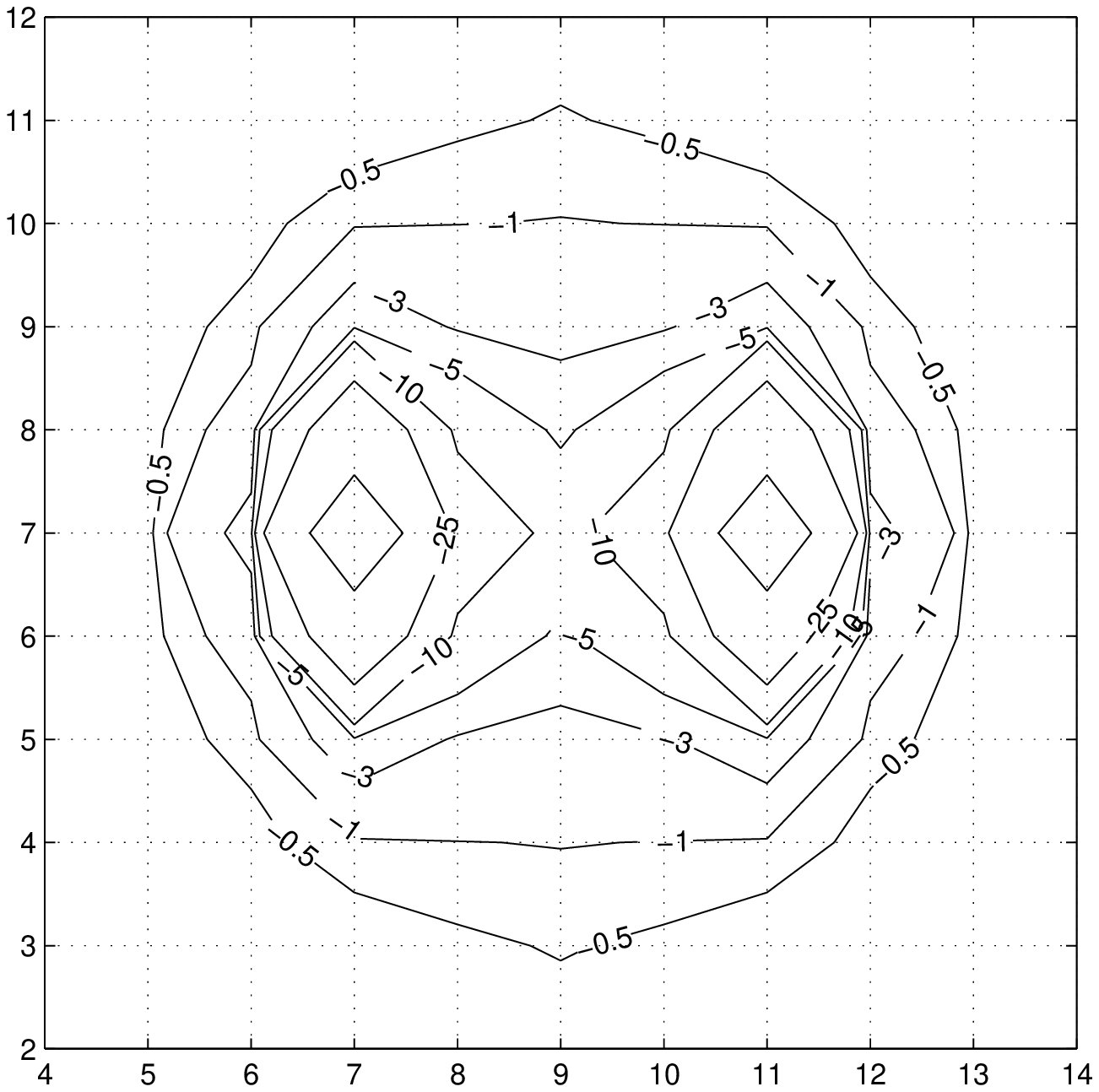} 
\hspace{0cm}\epsfxsize=200pt\epsfbox{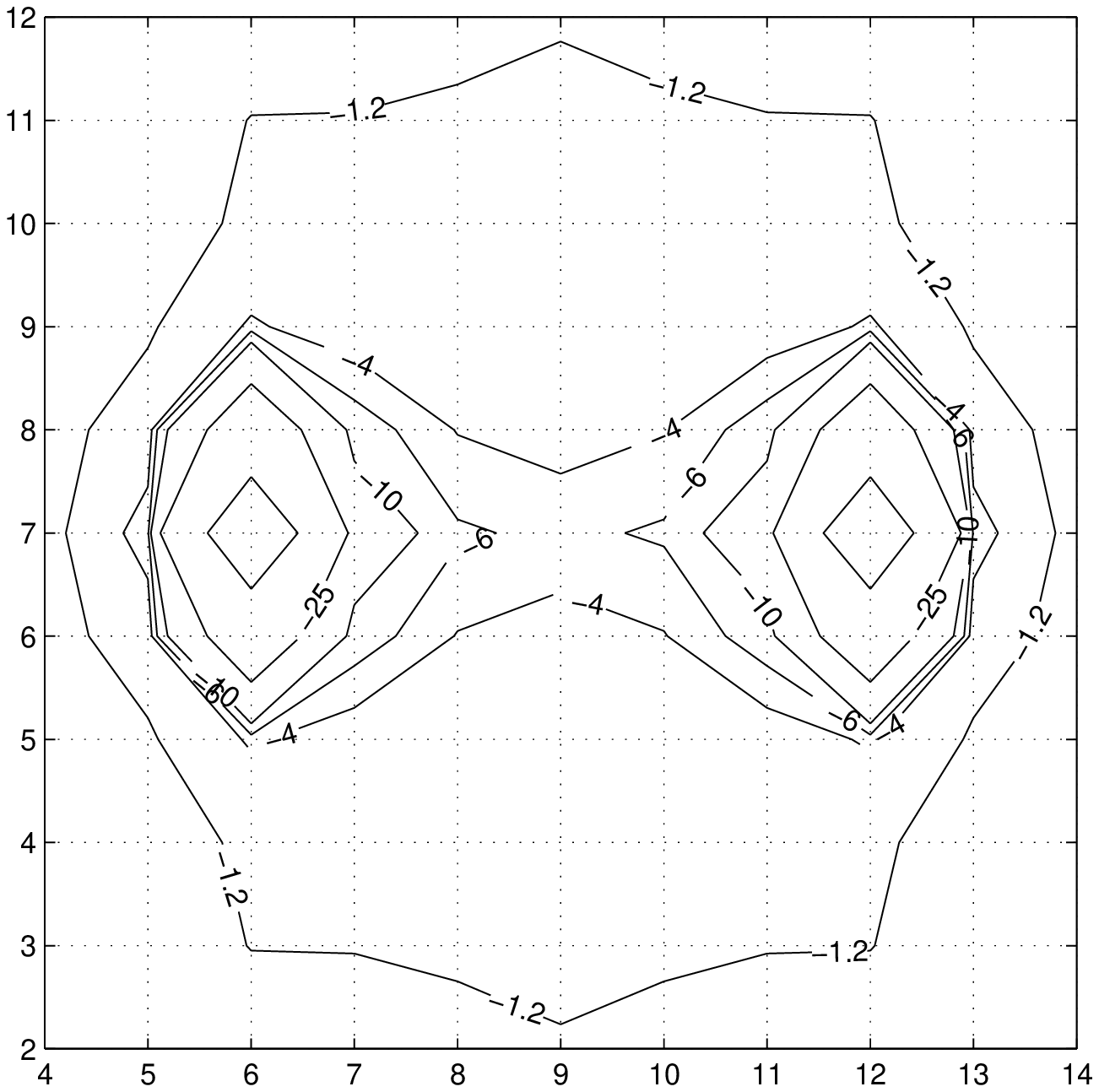}\epsfxsize=200pt\epsfbox{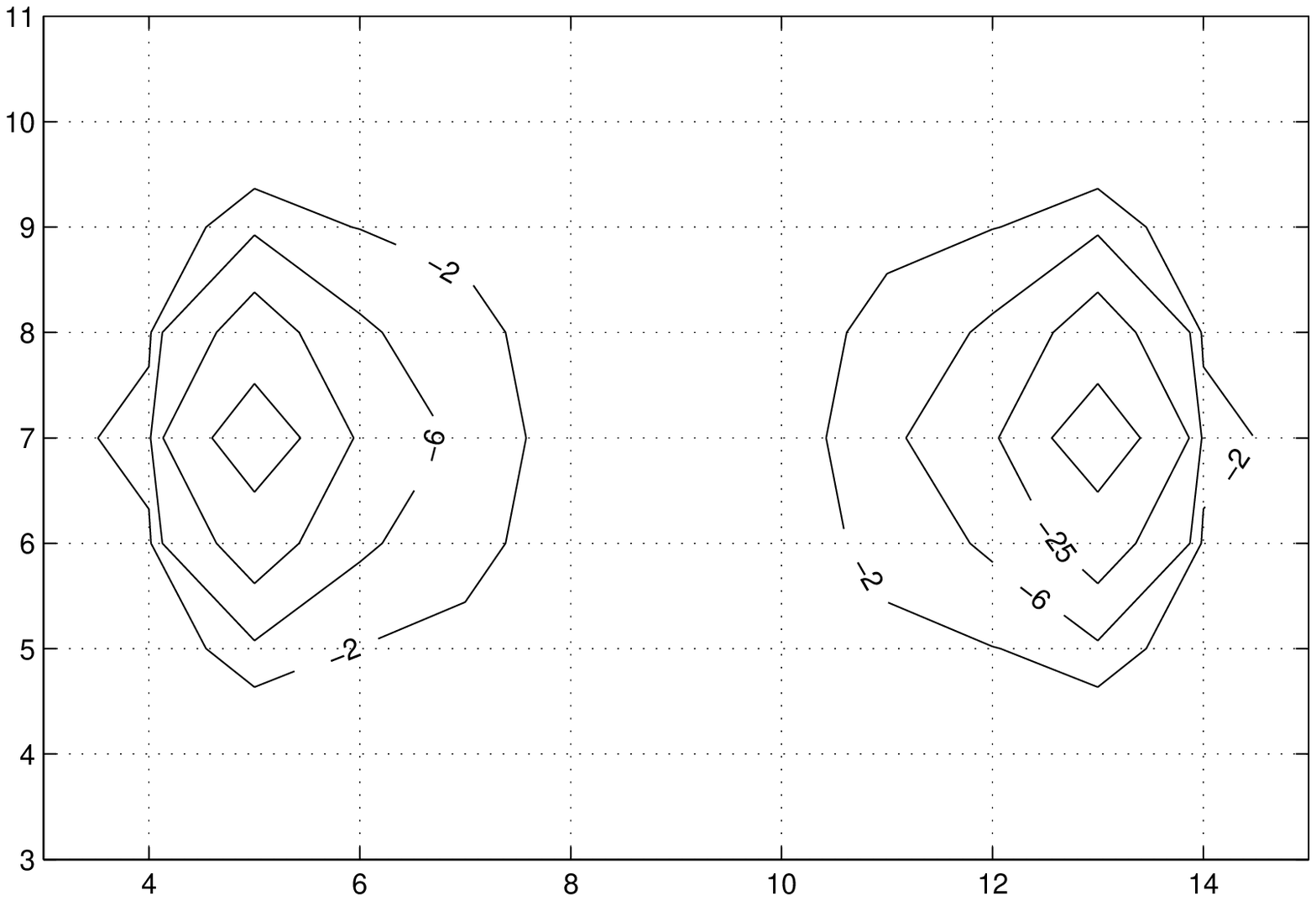} 
 \caption{As in Fig.~\protect\ref{fqsld} but for the energy. }
 \label{fqsld9}
\end{figure}

We cut two-dimensional slices through the four-quark color field 
distributions to illuminate 
details. Let us first concentrate on the four-quark flux-distributions 
before any two-body contributions are subtracted from them.
In Fig.~\ref{fqb} this is carried out for four quarks in the plane on which
they lie. In Fig.~\ref{fqsln} we look at the plane perpendicular to the one
on which the quarks lie and cutting through the
middle of the flux tubes along the sides. In Fig.~\ref{fqsld} the plane
is also perpendicular to the quark plane, but now cuts diagonally through two 
of the quarks. 
Figs.~\ref{fqb9}, \ref{fqsln9}, \ref{fqsld9} show the same slices but for 
the energy
distribution (as in Eq.~\ref{sume}). The $R=8$ data is taken in these three 
figures at $T=2$ because of a lack of signal at $T=3$.

Several points should be noted in these figures:

1) In Figs.~\ref{fqb} and \ref{fqb9} the self-actions and -energies in the 
neighborhoods of the four quarks clearly stand out, with the values at the 
actual positions of the quarks being given in Table~\ref{tselfe}, where they 
are compared with the corresponding two quark cases.

\begin{table}[htb]
 \begin{center}
\begin{tabular}{l|cccc}
          &$R=2$ & 4 & 6 & 8 \\ \hline
4q action  & 0.0659(1)  & 0.0637(2)    & 0.0634(4)   & 0.0638(6) \\
energy     & --0.0708(2) & --0.0680(4) & --0.0677(6) & --0.0684(9)  \\ \hline
2q action  & 0.0652(1)  & 0.0639(1) & 0.0635(1)  & 0.0634(2) \\
energy & --0.0718(1)  & --0.0684(2) & --0.0679(2) & --0.0680(3) 
\end{tabular}
\end{center}
 \caption{Self-energy peaks measured for two and four quarks at $T=2$. 
``Energy'' refers to the combination in Eq.~\protect\ref{sume}. The values
are in lattice units.
\label{tselfe}}
\end{table}

2) As expected, most of the action and energy are contained in the area 
defined by the positions of the four quarks. This effect seems more
pronounced as the sizes of the squares increase.

3) In Figs.~\ref{fqsln} and \ref{fqsln9} the flux tube profiles are seen to 
be distorted from that of two two-quark flux tubes. 
Furthermore, the distortion is such that the contours between the 
sides are more spread out than those outside the square. 
As mentioned in Sect.~\ref{s2q} a similar effect occurs with two 
quarks. This is a consequence of the additional attraction that arises when 
two two-quark flux tubes are brought together. As seen in Fig.~\ref{fqsln}, 
this attraction becomes very weak for $R\ge 8$, since then the flux tubes 
are essentially those of two independent flux tubes i.e. rotational
invariance about their axes has been restored.

4) Figs.~\ref{fqsld} and \ref{fqsld9} show the self-actions 
and -energies at the end of the
diagonals -- the features are similar to those already seen in 
Figs.~\ref{fqsln} and \ref{fqsln9}.

\subsection{Four quarks -- after subtraction}

For a square of side $R$ the total four-quark energies $[E(4)]$ 
corresponding to Figs.~\ref{fqb9}--\ref{fqsld9} can be viewed 
as a combination of two terms:\\
1) The energy $E(AB)$ of two independent two-quark flux tubes of length $R$.
This is simply $0.5[E(A)+E(B)]$ -- due to the symmetry between the 
two partitions $A$ and $B$ depicted in Fig.~\ref{fpair}.

2) The  energy $[B(4)]$ binding the two two-quark systems i.e.
$E(4)=E(AB)+B(4)$.

In practice  $B(4)$ is only a few percent of $E(AB)$ as seen in 
Table~\ref{tafter}.

\begin{table}[htb]
 \begin{center}
\begin{tabular}{l|cccc}
          &$R=2$ & 4 & 6 & 8 \\ \hline
$E(4)$    & 1.069(1) &   1.511(1) &   1.861(4) & 2.242(3) \\
$E(AB)$   & 1.127(1) &   1.566(1) &   1.891(2) & 2.259(3) \\
$B(4)$    &--0.058(1)&  --0.055(3)&  --0.030(1)& --0.017(1) 
\end{tabular}
\end{center}
 \caption{Comparison of four-quark total and binding energies
\label{tafter}}
\end{table}

Since energies are related to the profiles through Eq.~\ref{sume} it is, 
therefore, natural to also view the flux tube energy profile $F(4)$ in 
Figs.~\ref{fqb9}--\ref{fqsld9}
as being a combination of two terms $F(AB)+FB(4)$, where 
$F(AB)=0.5[F(A)+F(B)]$ is the average of the energy profiles for states 
$A$ and $B$ 
in Fig.~\ref{fpair}. In other words, after subtracting 
the energy component of the two-body flux distributions $F(AB)$ of 
the two-body pairings, we get the distribution $FB(4)$, which can be considered
as corresponding to the binding energy of the four quarks. The hope is that the
form of $FB(4)$ will serve as a guide when constructing the type of model to
be discussed in Sect.~\ref{sf} -- a model that only depends on the quark
degrees of freedom. For the action 
there is no clear meaning to this subtraction, and so the action plots in this
subsection are of an exploratory nature and will be compared to the
energy plots to see what similarities exist. 

\begin{figure}[hbtp]
\hspace{0cm}\epsfxsize=200pt\epsfbox{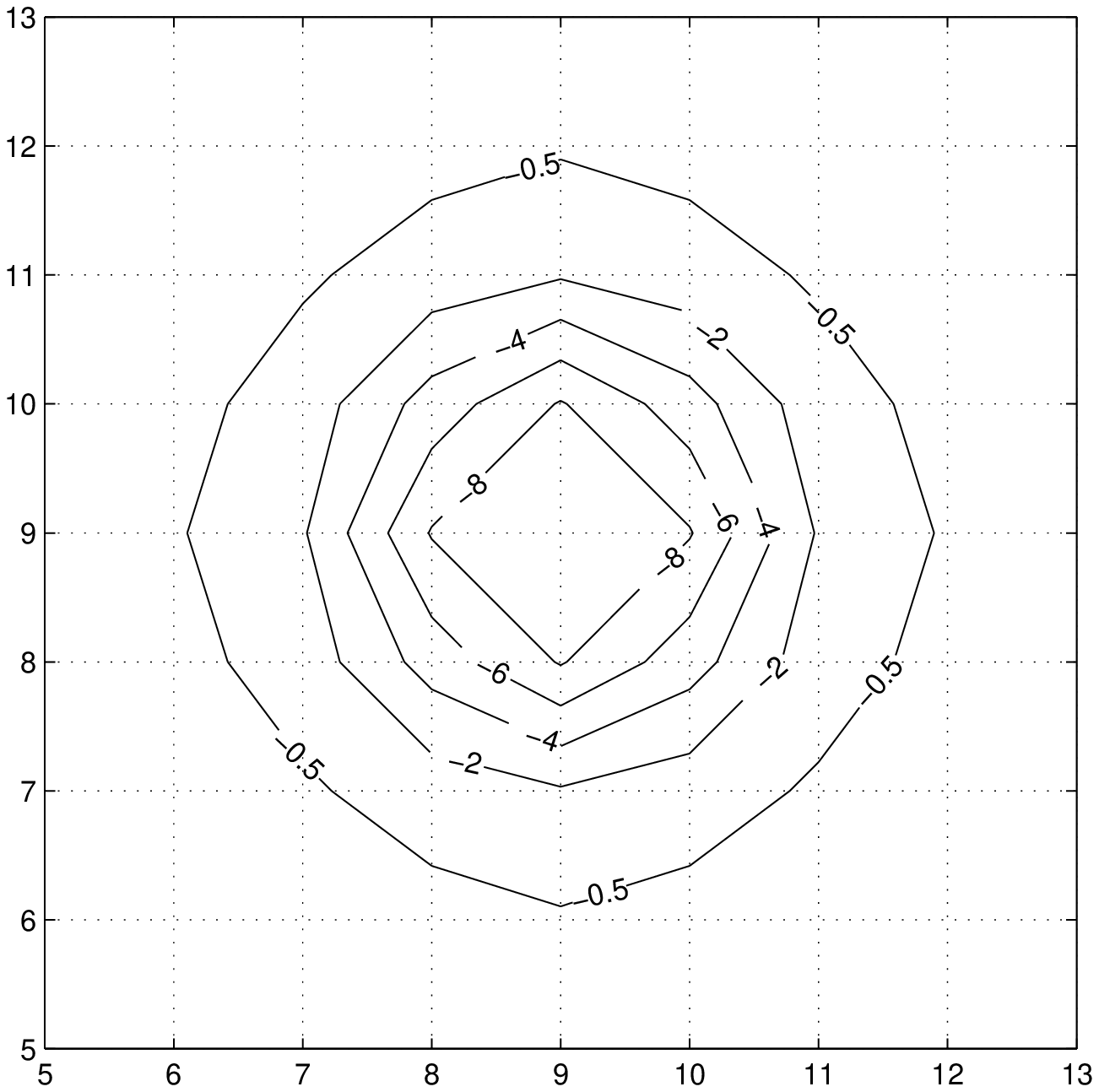}\epsfxsize=200pt\epsfbox{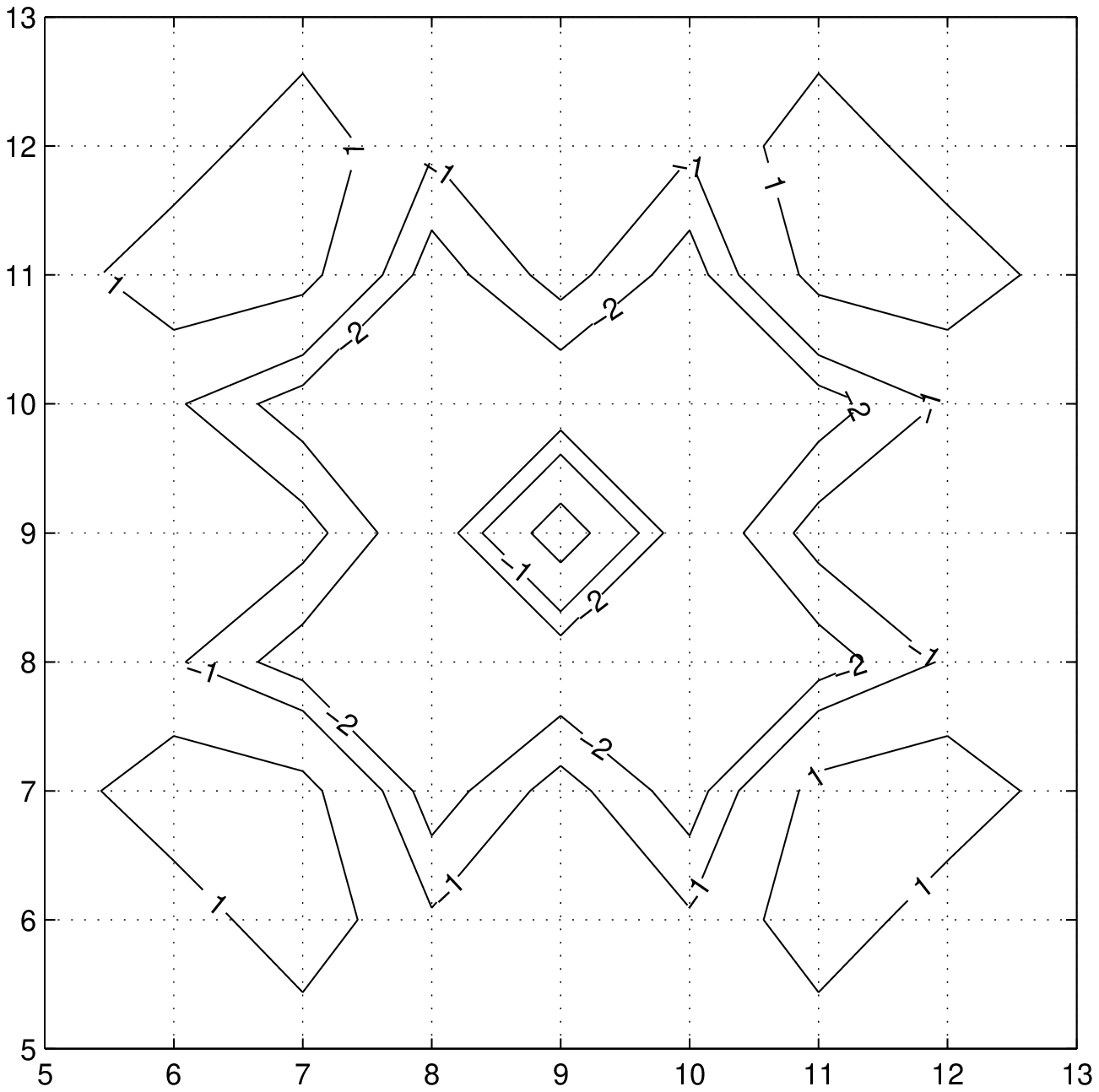} 
\hspace{0cm}\epsfxsize=200pt\epsfbox{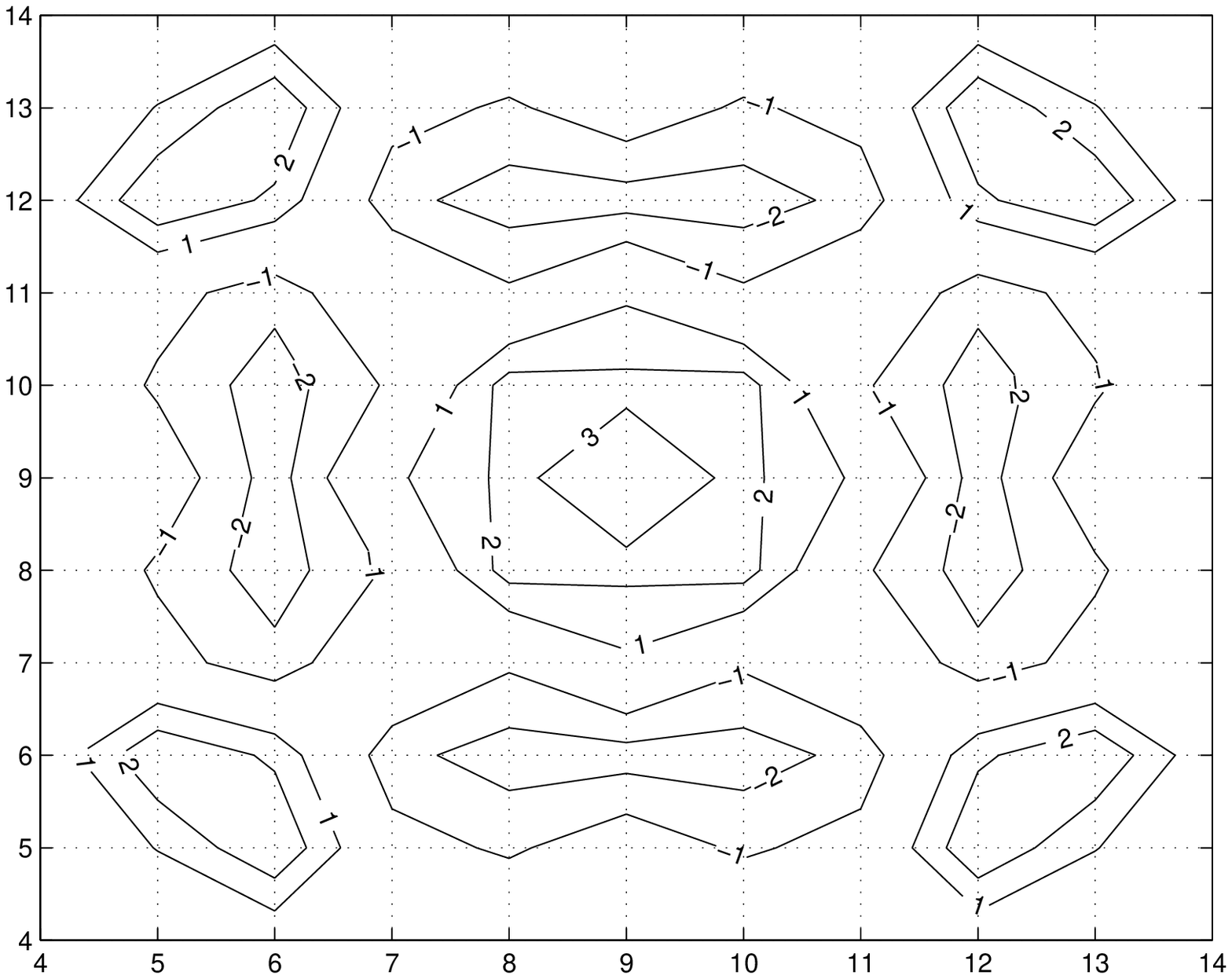}
\caption{As in Fig.~\protect\ref{fqb} but after subtracting two-body tubes along the sides.} \label{fqba}
\end{figure}

\begin{figure}[htbp]
\hspace{0cm}\epsfxsize=200pt\epsfbox{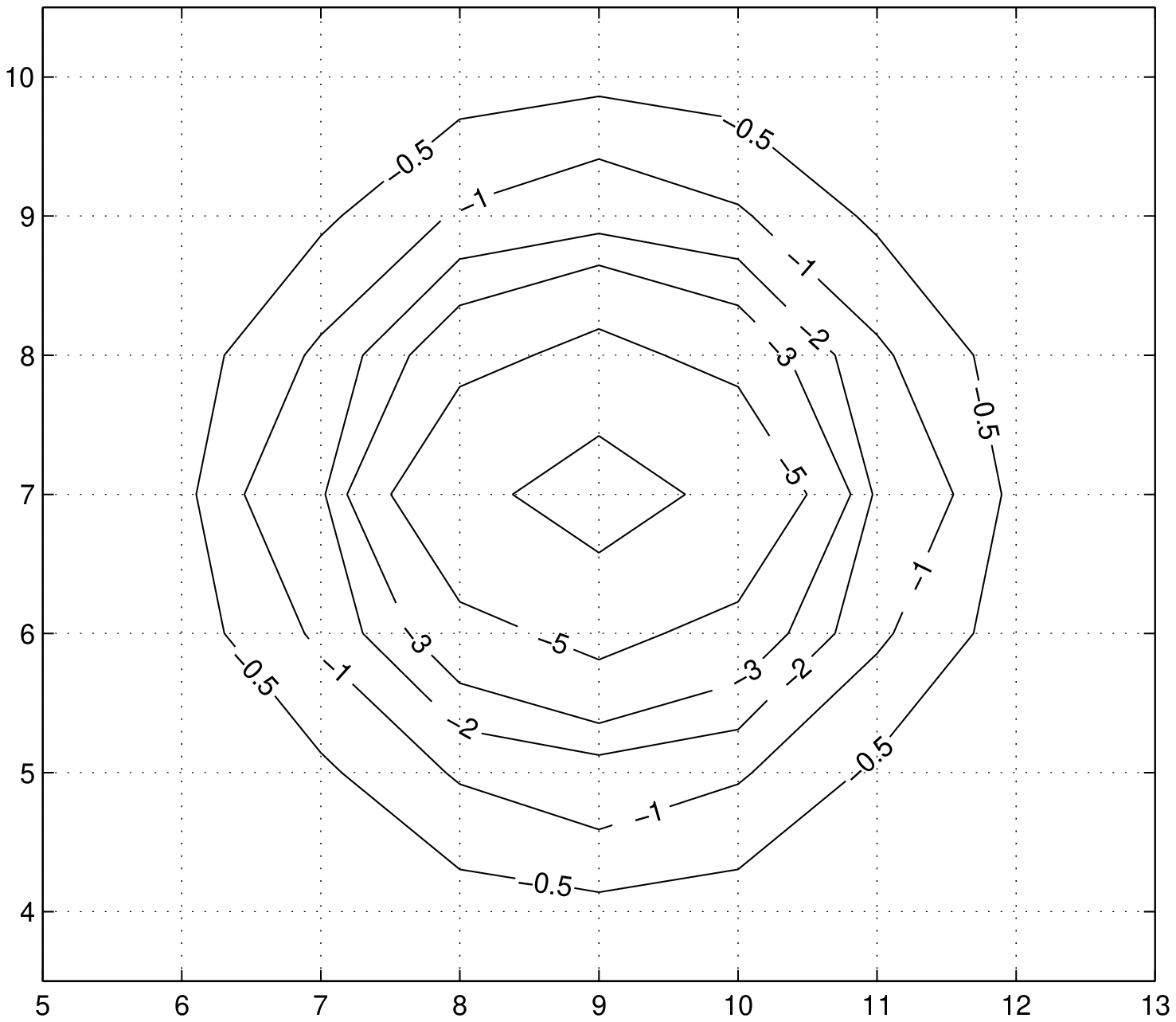}\epsfxsize=200pt\epsfbox{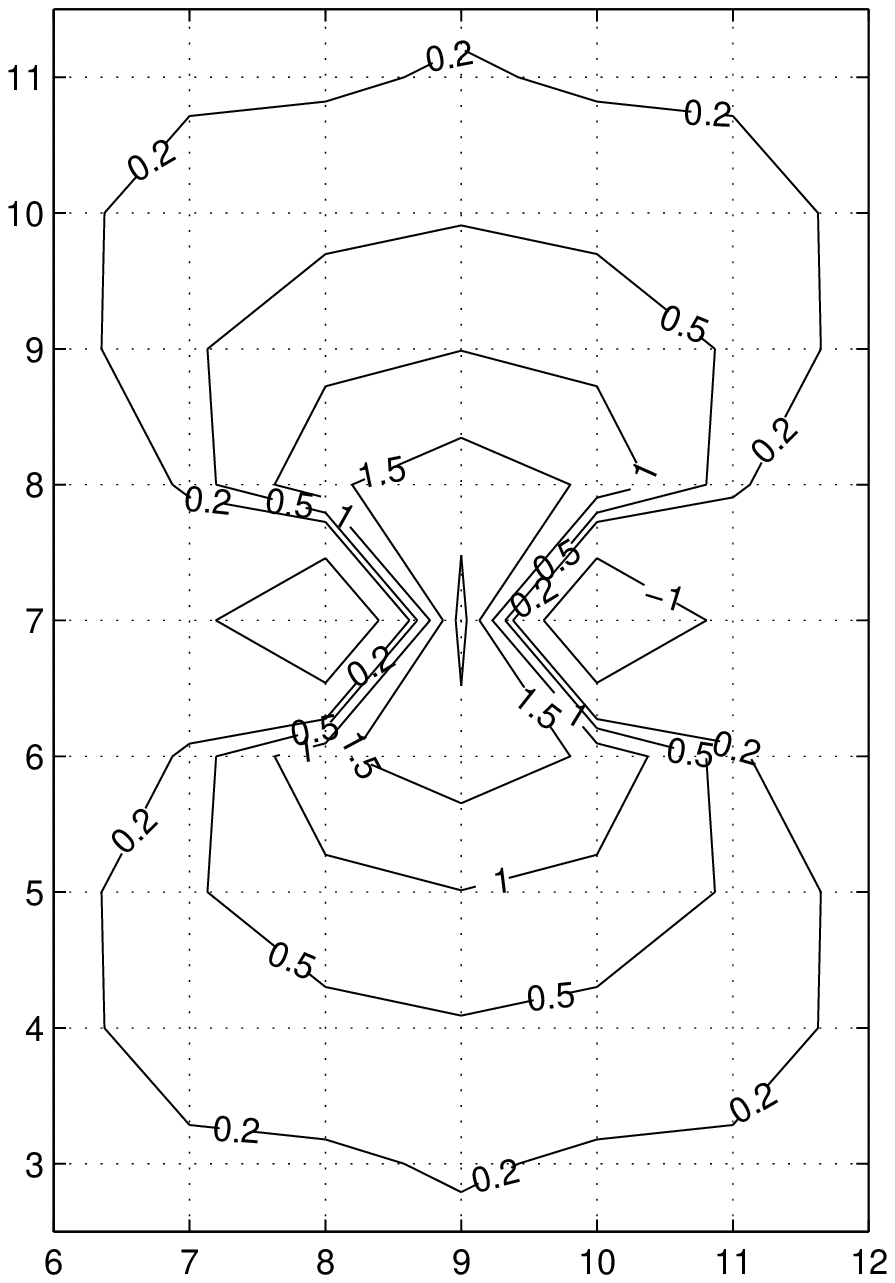} 
\hspace{0cm}\epsfxsize=200pt\epsfbox{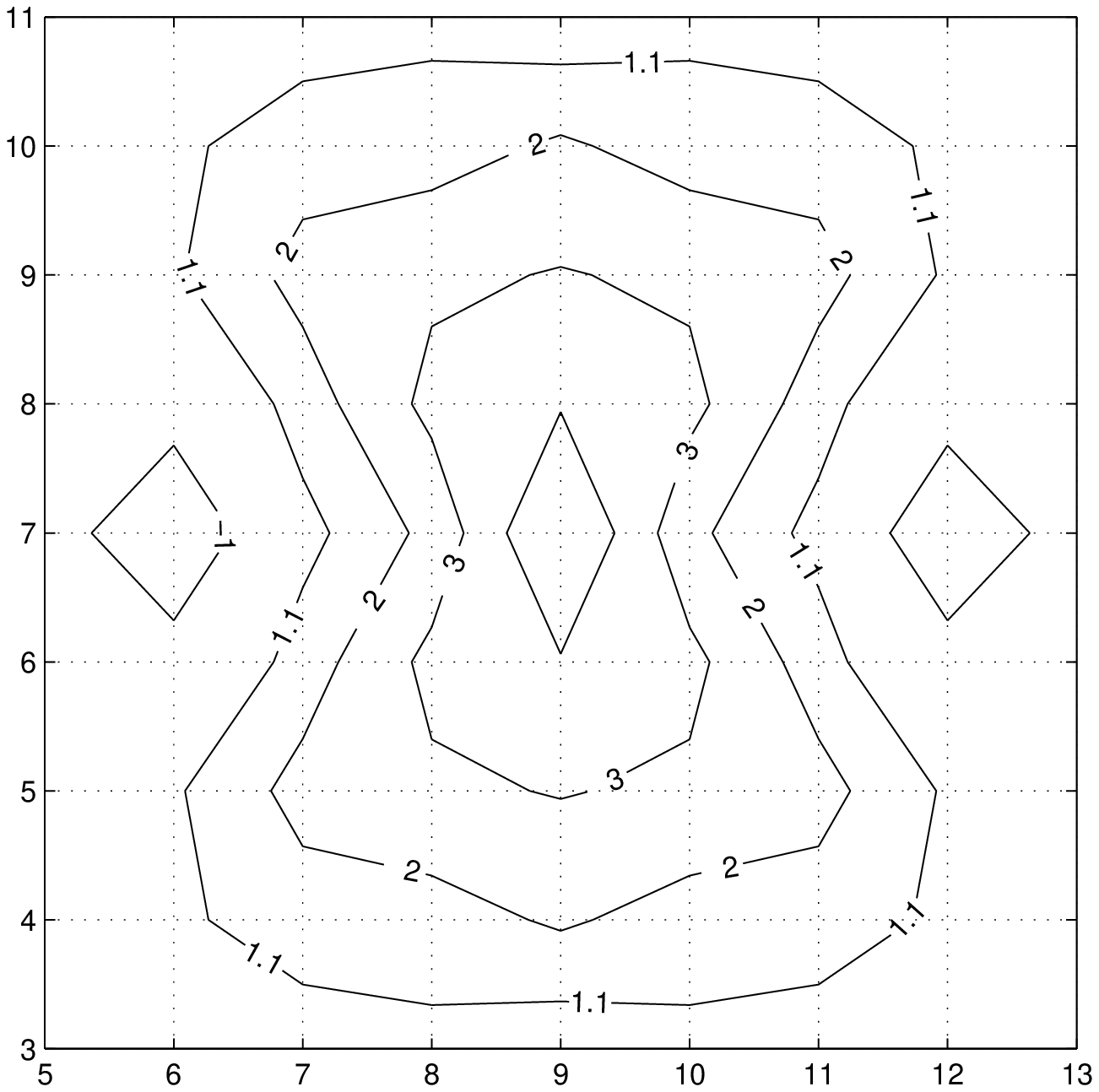}
\caption{As in Fig.~\protect\ref{fqsln} but after subtraction. }
 \label{fqslna}
\end{figure}

\begin{figure}[htbp]
\hspace{0cm}\epsfxsize=200pt\epsfbox{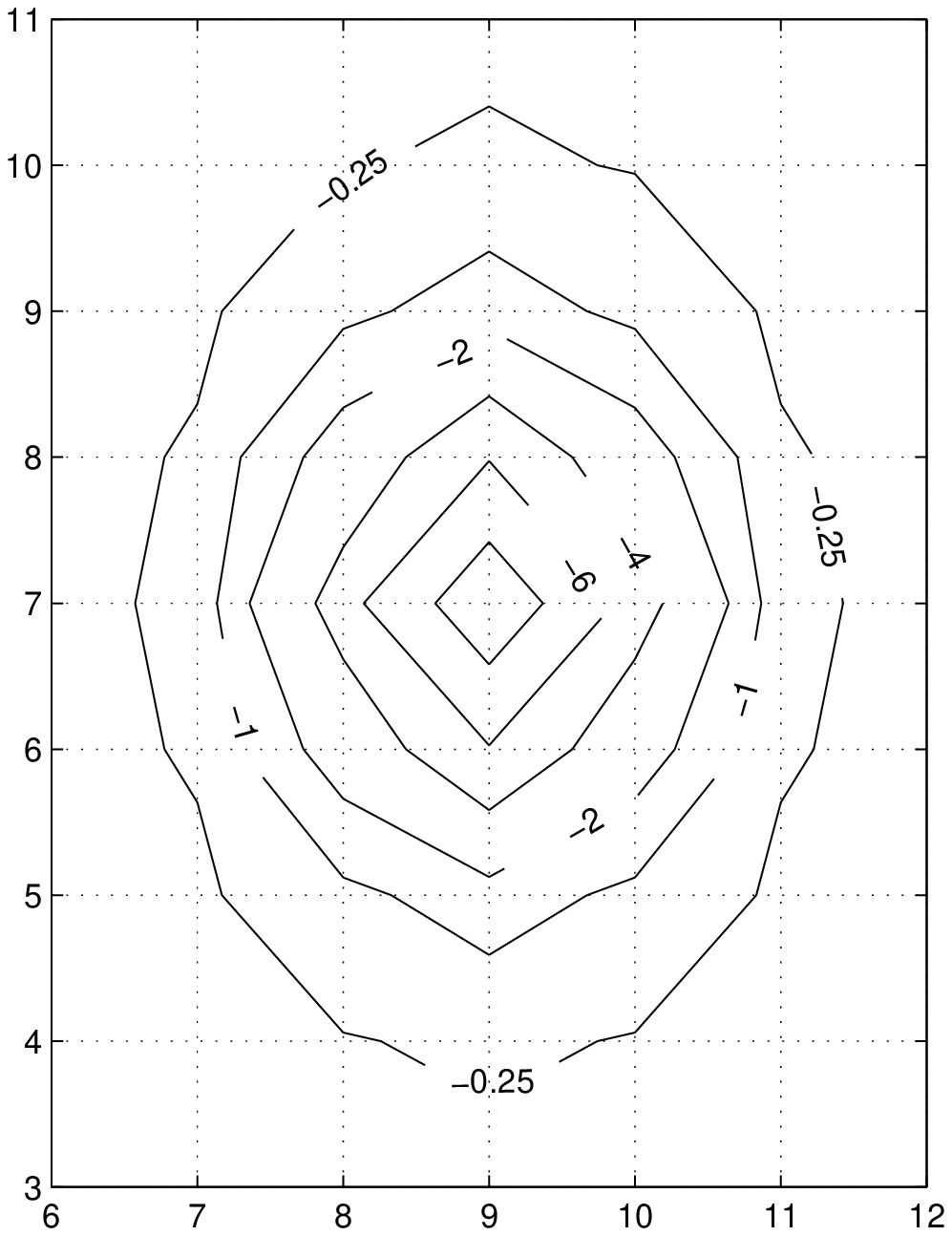}\epsfxsize=200pt\epsfbox{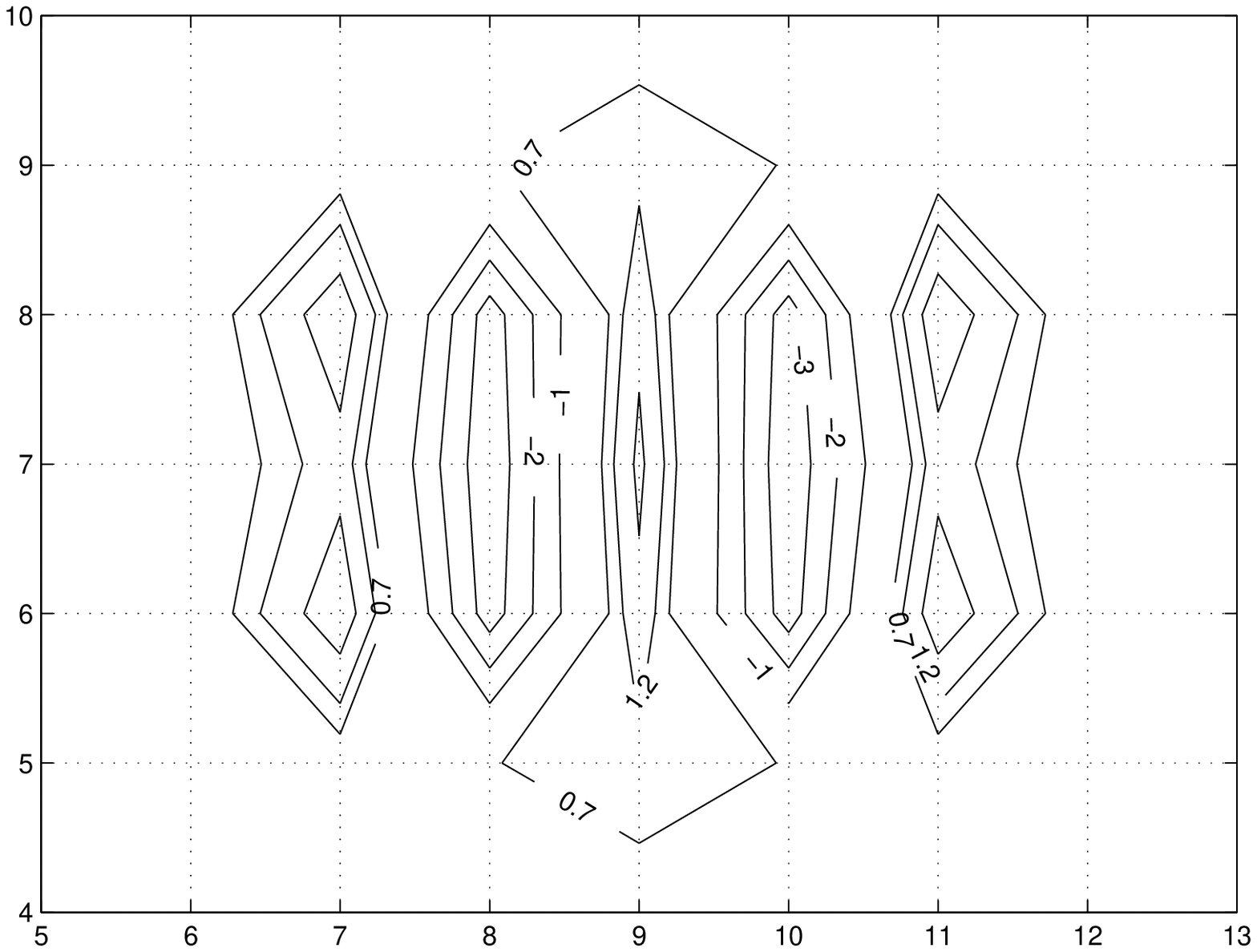} 
\hspace{0cm}\epsfxsize=200pt\epsfbox{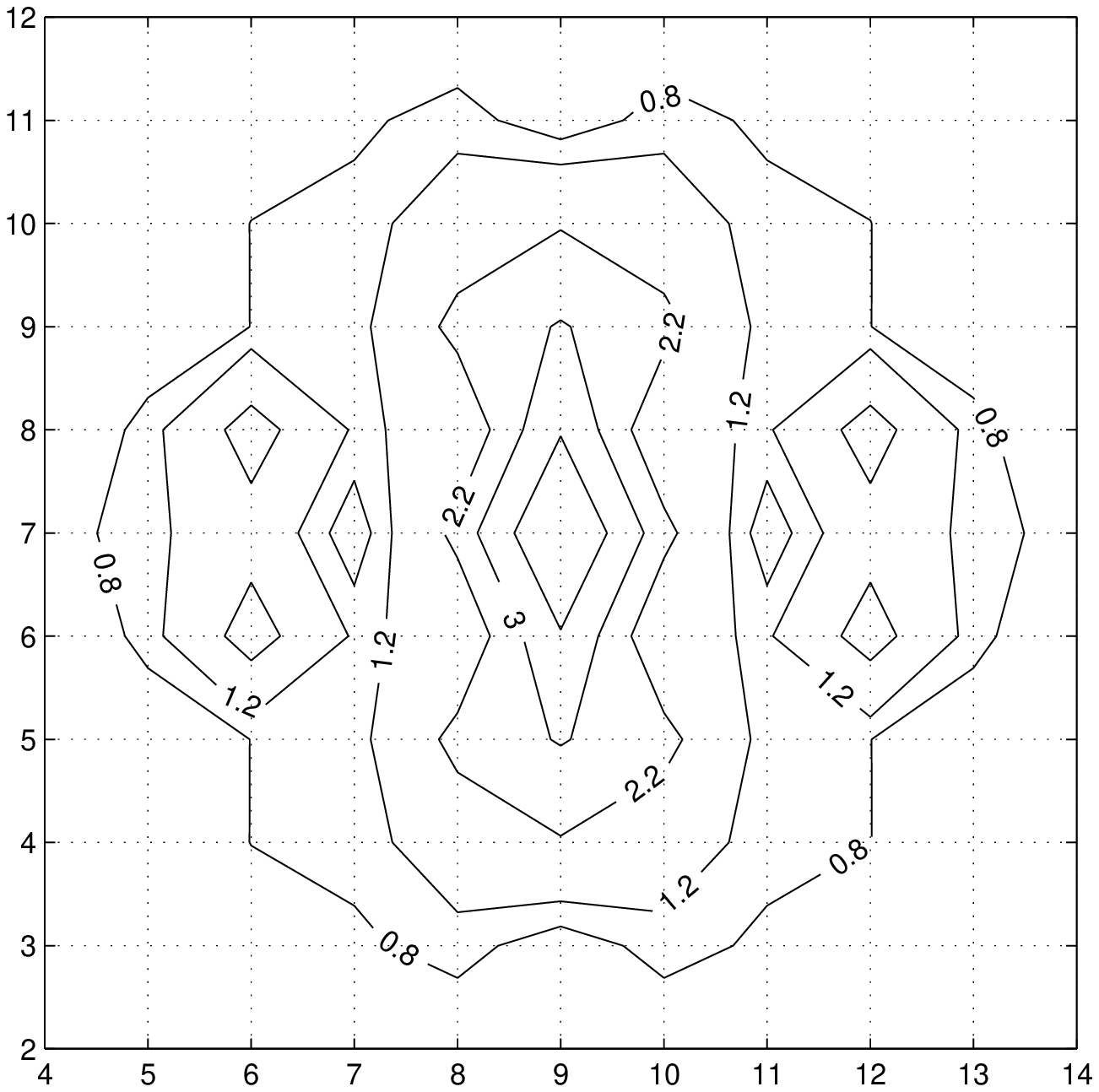}
 \caption{As in Fig.~\protect\ref{fqsld} but after subtraction. }
 \label{fqslda}
\end{figure}

\begin{figure}[hbtp]
\hspace{0cm}\epsfxsize=200pt\epsfbox{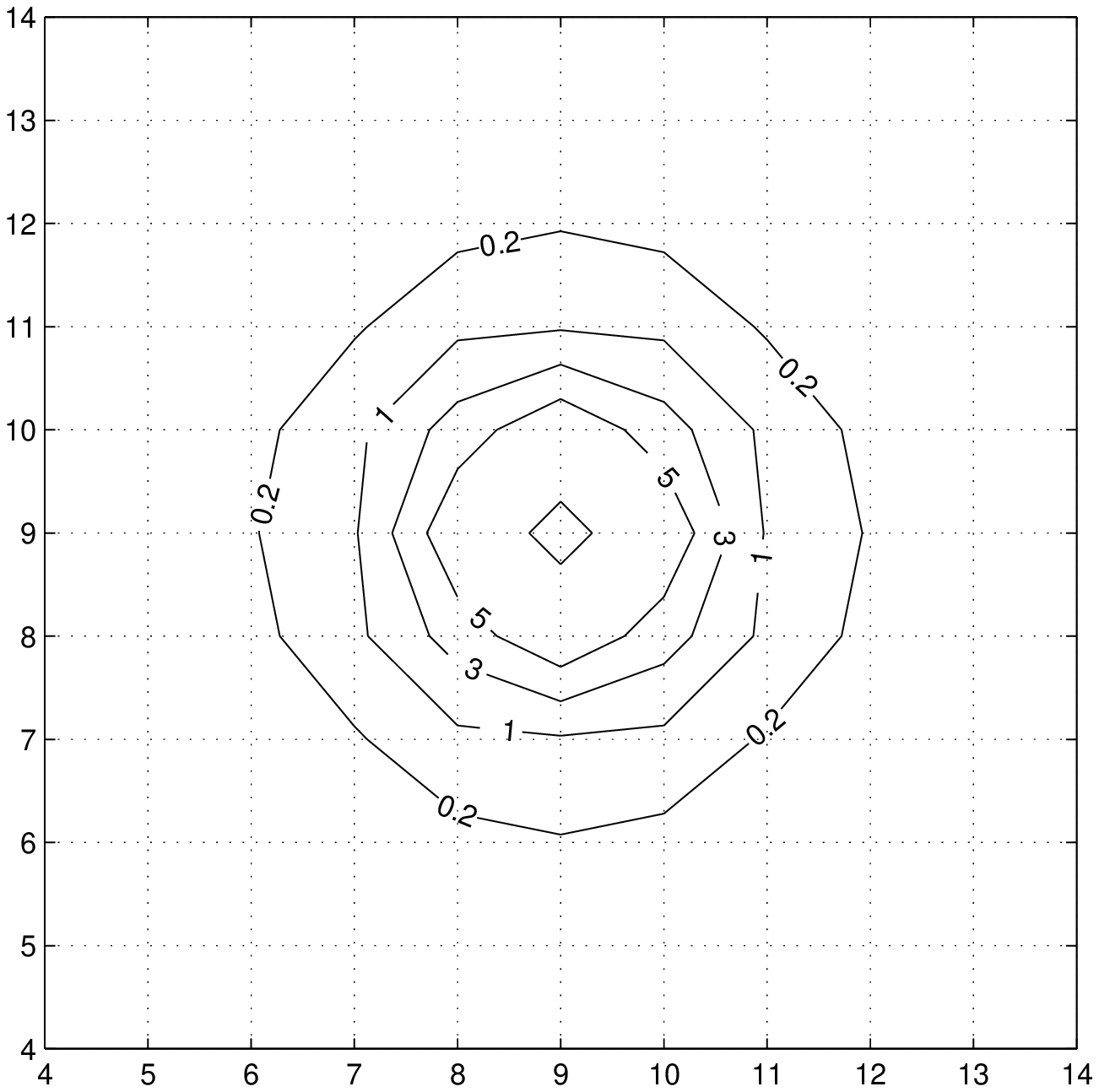}\epsfxsize=200pt\epsfbox{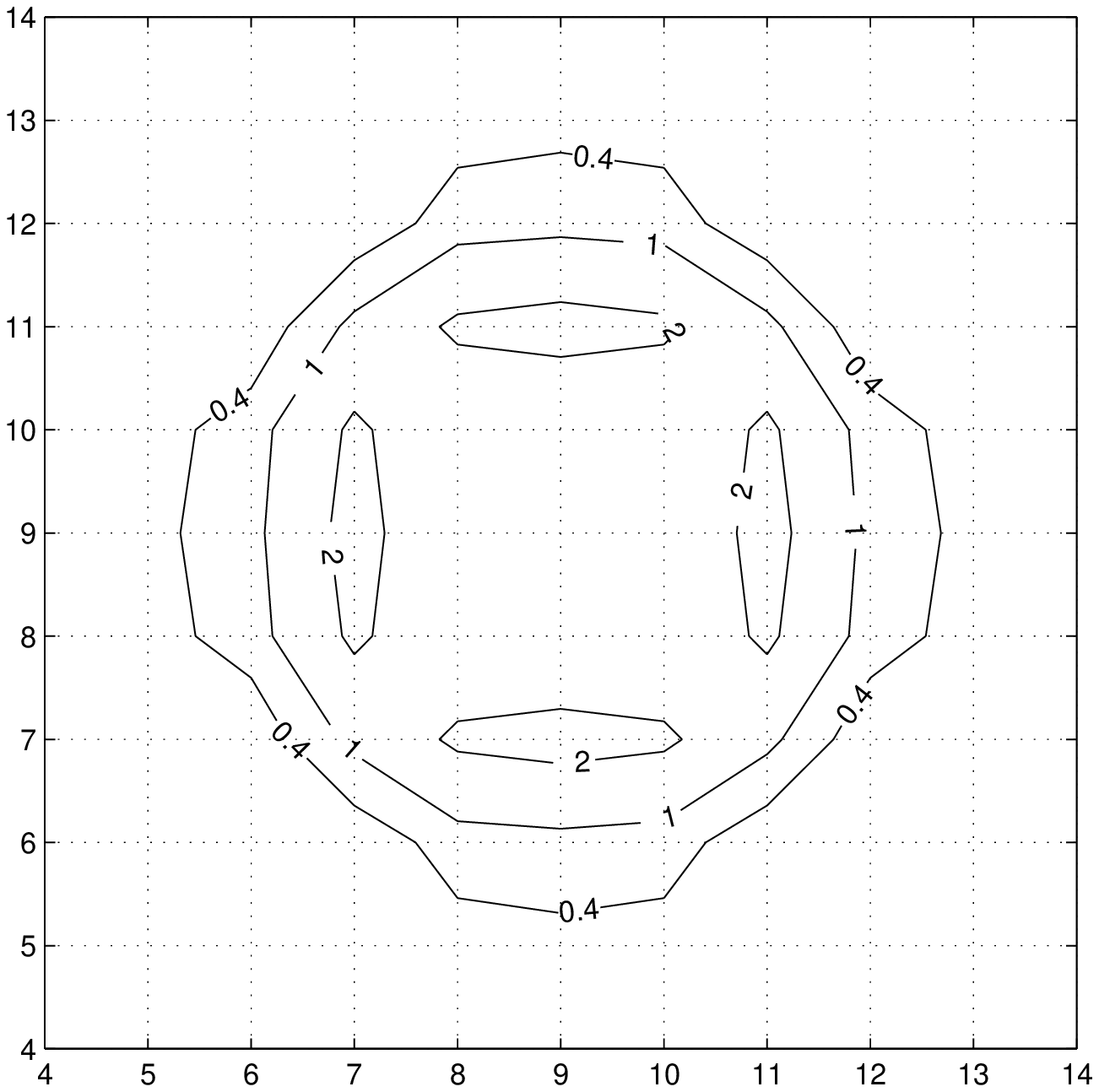}
\hspace{0cm}\epsfxsize=200pt\epsfbox{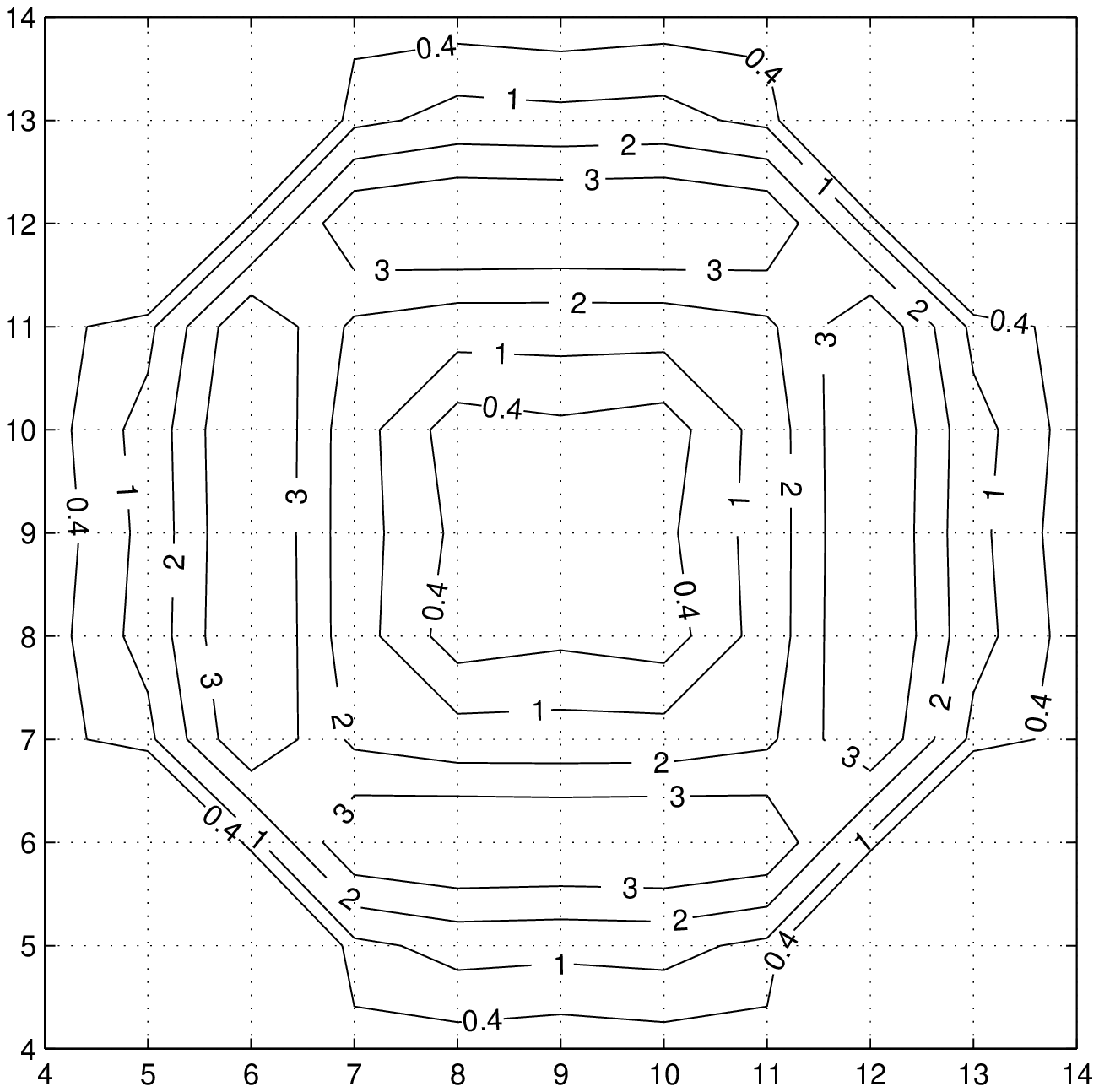}
\caption{As in Fig.~\protect\ref{fqba} but for the energy -- called
$FB(4)$ in the text.} \label{fqba9}
\end{figure}

\begin{figure}[htbp]
\hspace{0cm}\epsfxsize=200pt\epsfbox{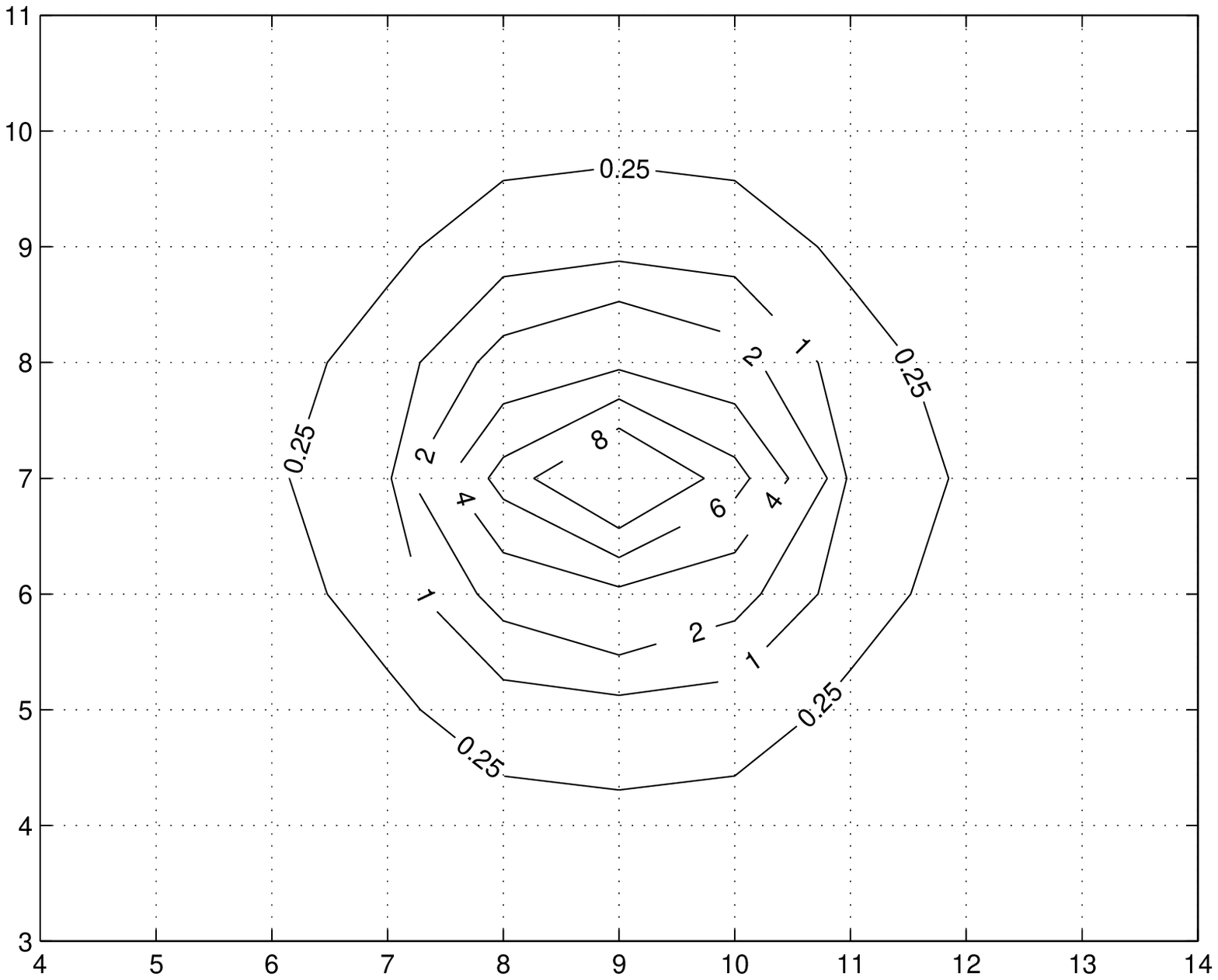}\epsfxsize=200pt\epsfbox{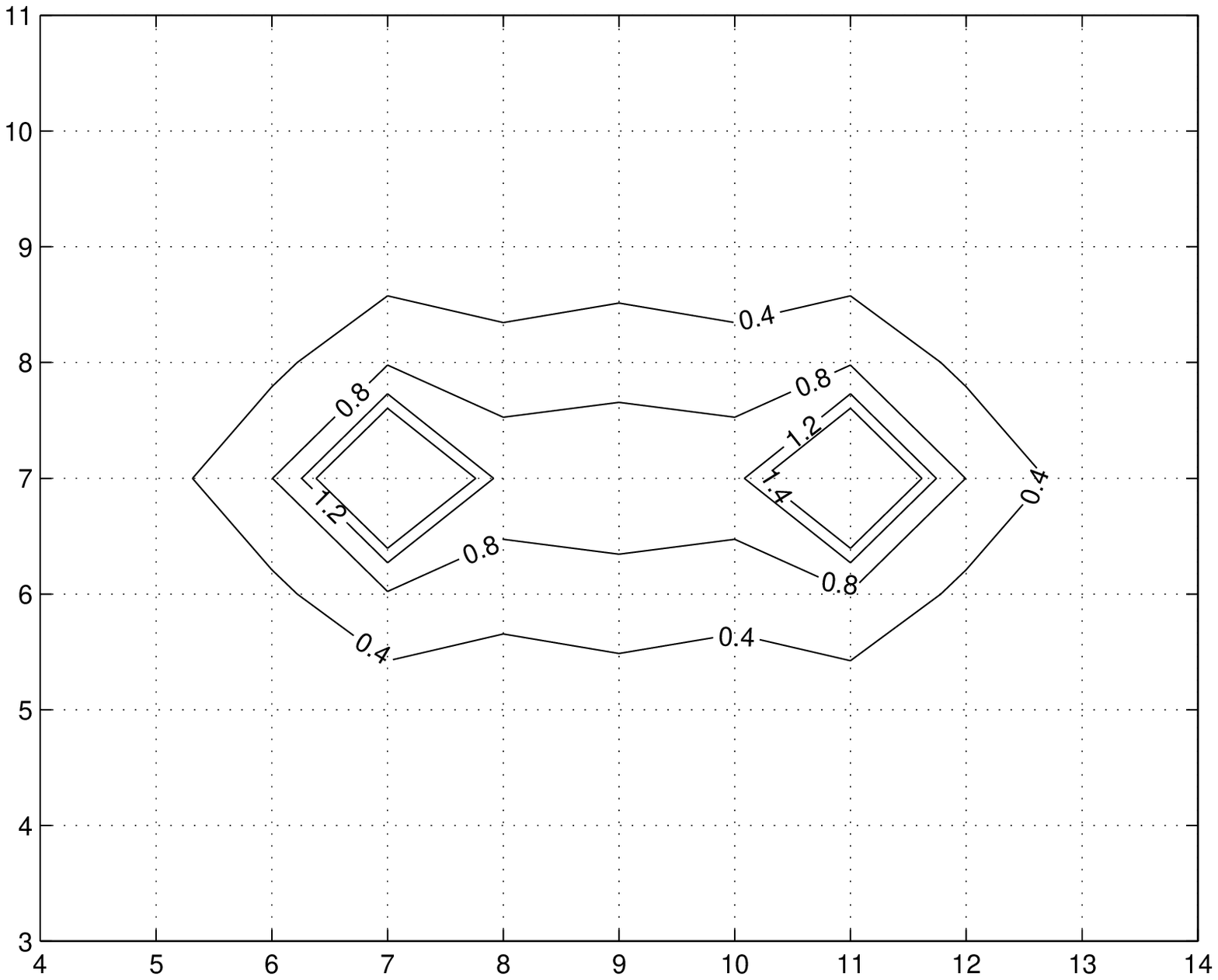} 
\hspace{0cm}\epsfxsize=200pt\epsfbox{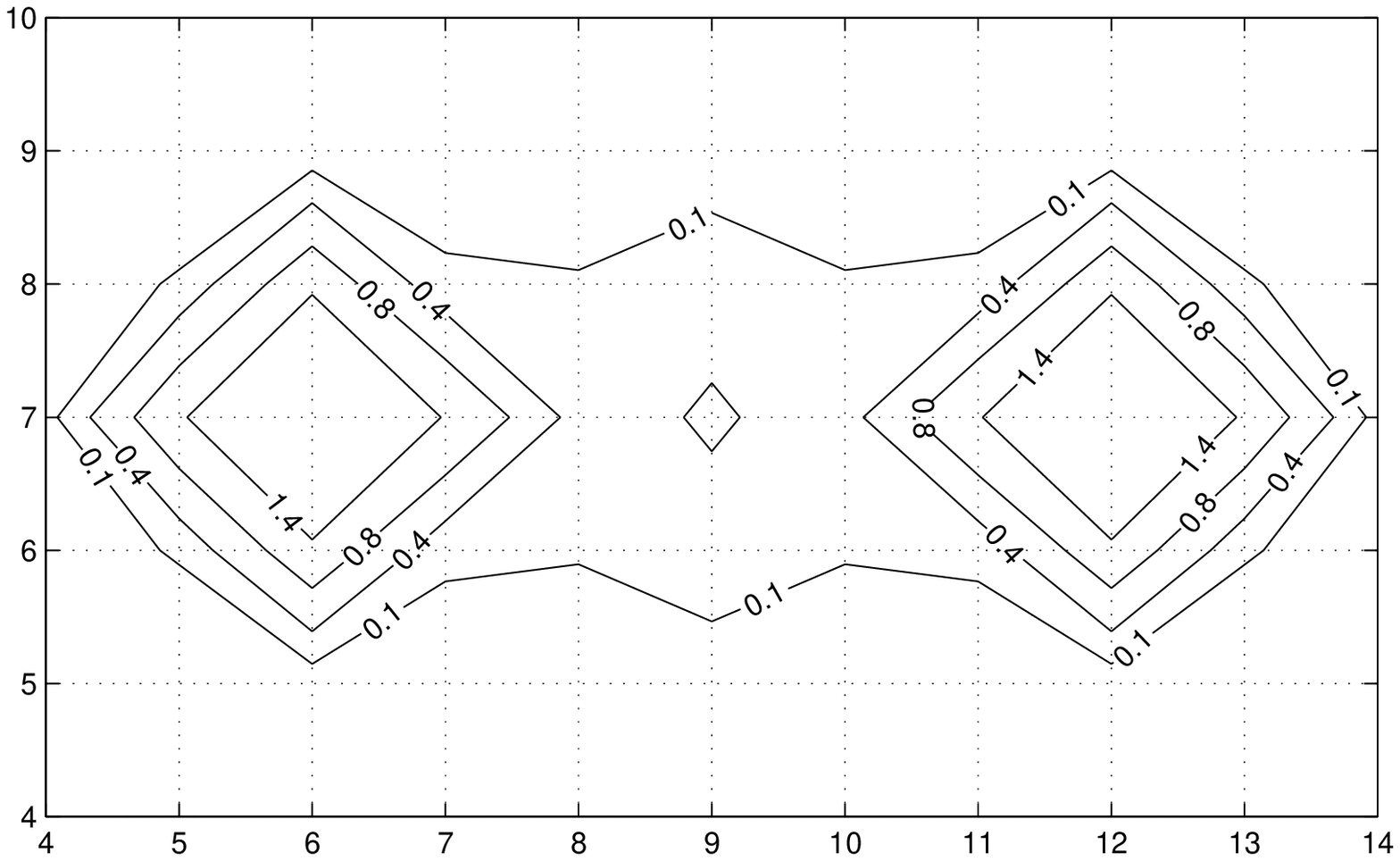}
 \caption{As in Fig.~\protect\ref{fqslna} but for the energy. }
 \label{fqslna9}
\end{figure}

\begin{figure}[htbp]
\hspace{0cm}\epsfxsize=200pt\epsfbox{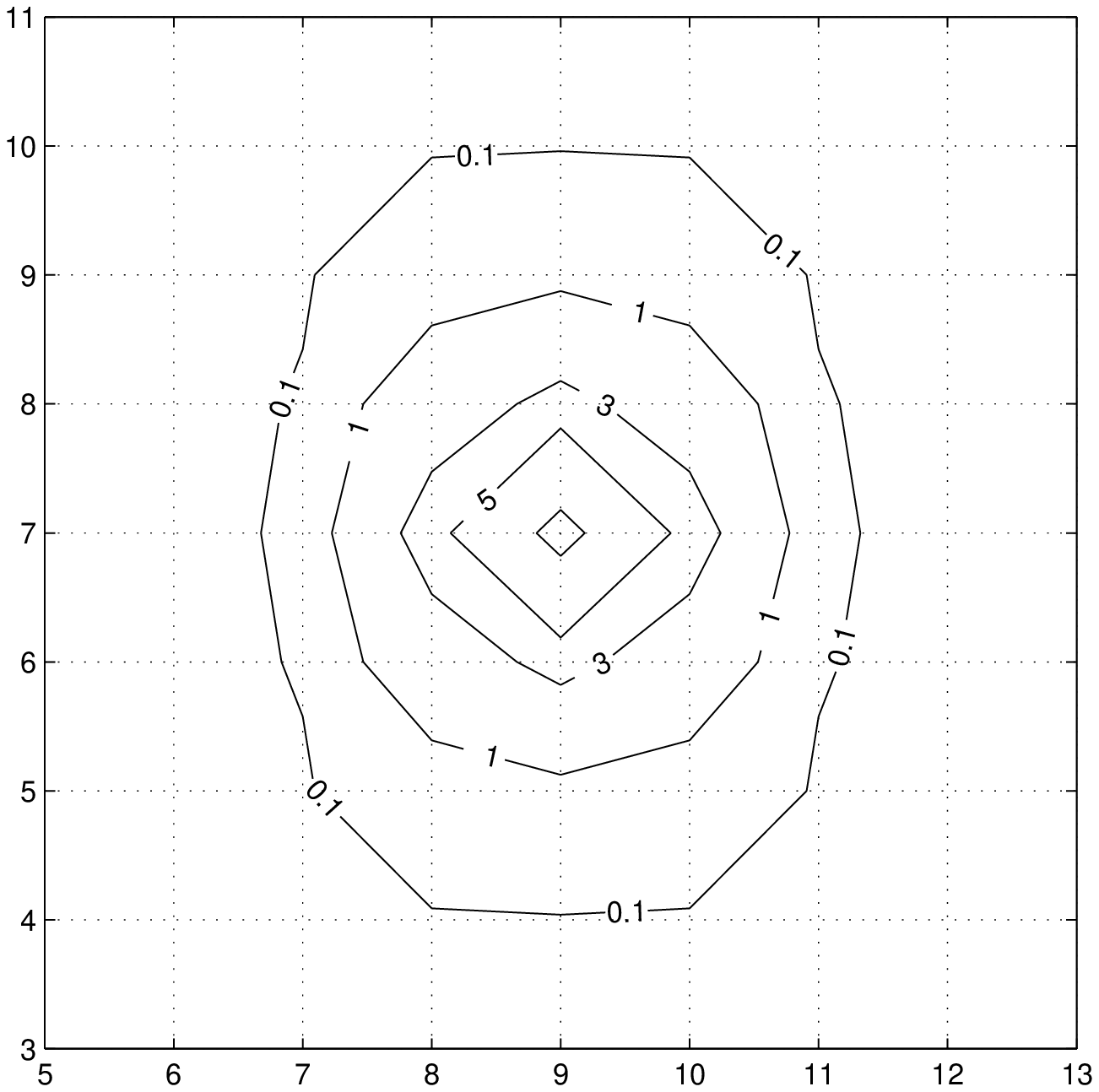}\epsfxsize=200pt\epsfbox{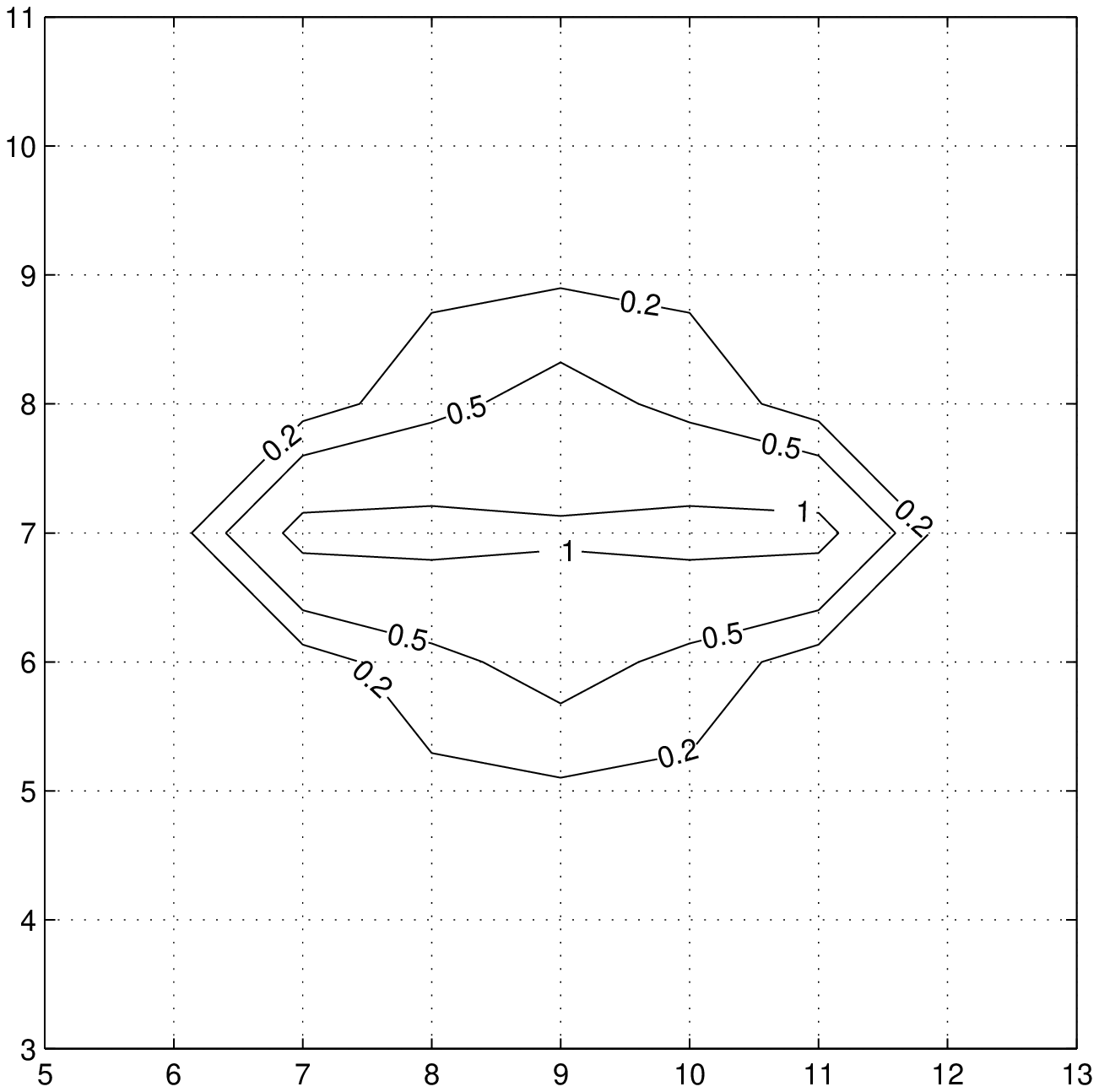} 
\hspace{0cm}\epsfxsize=200pt\epsfbox{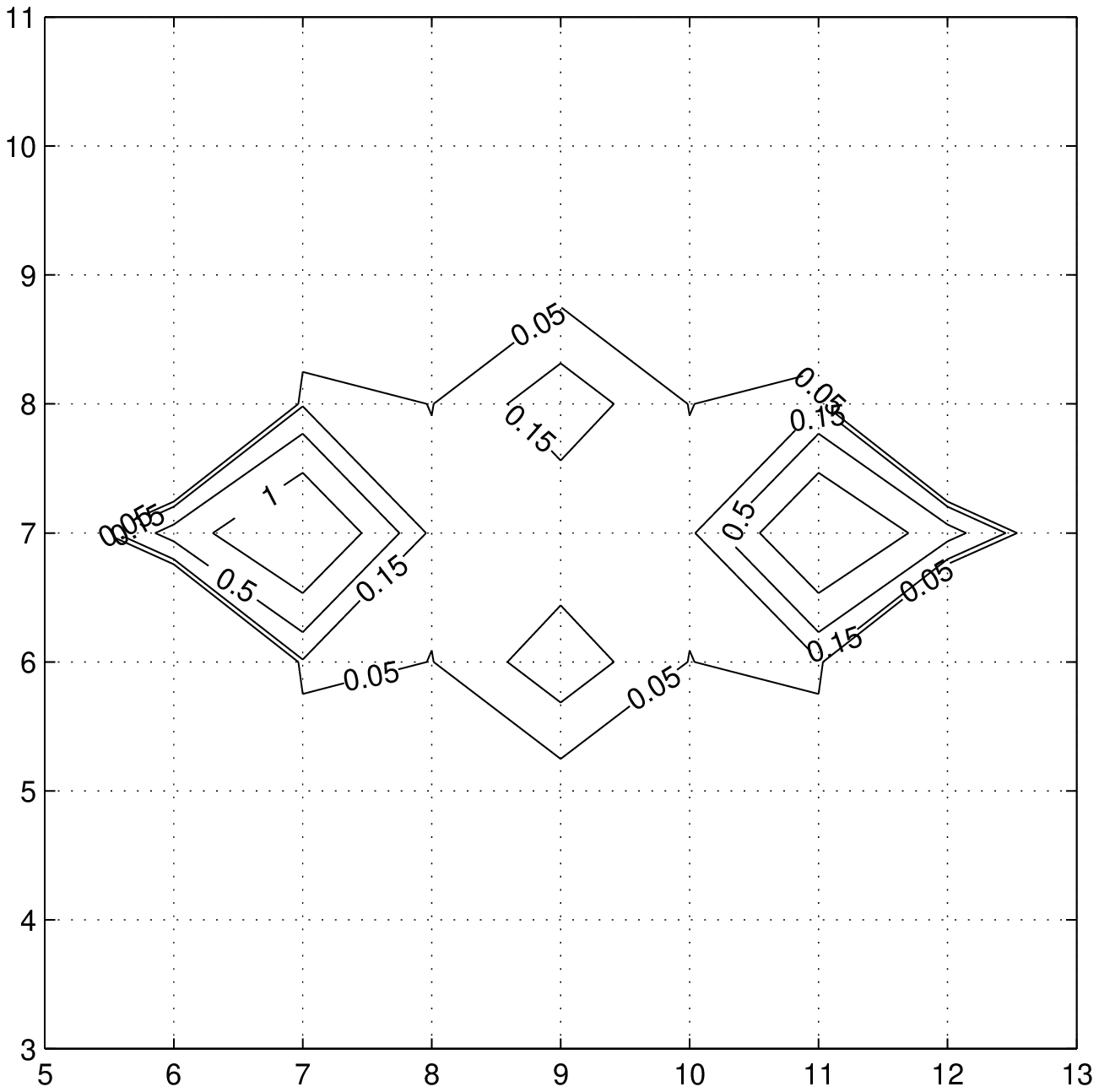}
 \caption{As in Fig.~\protect\ref{fqslda} but for the energy. }
 \label{fqslda9}
\end{figure}

Analogously to the unsubtracted case, Figs.~\ref{fqba}--\ref{fqslda} show the 
action densities for the three slices, while the energy densities are
plotted in Figs.~\ref{fqba9}--\ref{fqslda9}. Being guided 
by the sum rules in Table~\ref{tsumcheck2}, the energy 
densities
are taken at $T=2$ for $R=4,6$ and at $T=3$ for $R=2$.

As expected, there is again a large cancellation between $F(AB)$ and $F(4)$.
However, as seen from Table~\ref{tselfe}, the dominant feature in both 
$F(4)$ and $F(AB)$
--  the self-energies -- are equal to within less than 1\%. For the $R=2$
case the agreement in the table is worse, as the four-quark binding signal 
extends in this small system to the quark positions -- as seen in 
Figs.~\ref{fqba9}--\ref{fqslda9}.
Therefore, 
the residual profile $FB(4)$ is expected to have a realistic signal not 
dominated by self-energy cancellation errors. This is seen in 
Fig.~\ref{fqba9}, where 
there is no particular structure at the positions of the four quarks. 
Elsewhere, $F(4)$ and $F(AB)$ cancel to leave  $|FB(4)|\approx |F(AB)|/10$ 
over the area defined by the four quarks. In spite of this delicate 
cancellation, $FB(4)$ is seen to be everywhere {\em positive} for all the 
$R$'s considered. Therefore, due to our sign convention, $FB(4)$ represents
a negative energy density -- as expected for a bound state. 

It is of interest to see in detail the contributions to $FB(4)$ from
the five terms $F(4)$ and $F(AB)=0.5[F(12)+F(34)+F(13)+F(24)]$. These
are given in Table~\ref{tfbreakup} for three different points in the plane
of the quarks. Here it is seen that at point (b) -- in the middle of the line
connecting quarks 1 and 2 -- the cancellation is a 
complicated procedure
with the resulting attraction arising from the combined effect of flux tubes 
(12), (13) and (24). The effect from (12) alone is not enough to overcome
the signal in the four-quark distribution.

\begin{table}[htb]
 \begin{center}
\begin{tabular}{l|c@{\hspace{1mm}}c@{\hspace{1mm}}c@{\hspace{1mm}}c@{\hspace{1mm}}c@{\hspace{1mm}}c}
Position  &$F(4)$  & $0.5F(12)$ & $0.5F(34)$ & $0.5F(13)$ & $0.5F(24)$ & $FB(4)$\\ \hline
(a)         & --0.00262(4) & --0.00076(1) & --0.00076(1) & --0.00076(1) & --0.00076(1) & 0.00043(3)  \\ 
(b)         & --0.00834(3) & --0.00778(3) & --0.00002(1) & --0.00076(1) & --0.00076(1) & 0.00098(2) \\
(c)         & --0.06797(34) & --0.03420(8) & --0.00002(1) & --0.03420(8) & --0.00002(1) & 0.00047(26) \\
\end{tabular}
\end{center}
 \caption{The contributions to $FB(4)$ at three points in the plane of the square with side $R=4$ (here the $T=2$ data is used). The positions are (a) -- the center of the square, (b) -- the middle of (12), (c) -- at quark 1.
\label{tfbreakup}}
\end{table}

In Figs.~\ref{fqba9}--\ref{fqslda9}, $FB(4)$ has a roughly spherical
shape for $R=2$, with the shape getting more elongated 
when viewed from the side of the quark plane as in Fig.~\ref{fqslna9}. For 
$R=4$ we observe a clear region of
binding in between the quarks. In Fig.~\ref{fqba9} b) it has the shape of a 
regular
octagon bounded by the quarks, that extends outside the quark square in between
two nearest neighbor quarks. In the latter region we can see maxima
(with errors of 2--20\%) that resemble the two-body flux-tubes. 
This is understandable as in the four-quark distributions before
subtraction we observed that the fields were pulled towards the centerpoint, 
leaving a smaller contribution in the middle of the sides of the square. 
Therefore,
when the two-body distributions are subtracted we are left with larger
positive (binding in our sign convention) contributions at these points. 
The region inside the quark square has a constant density and thickness, except
when viewed diagonally in Fig.~\ref{fqslda9} b), where the maxima at the sides
do not contribute. A qualitatively similar situation is observed for $R=6$,
with more contribution in the maxima at the sides and less in between
them. An area of constant thickness in between the sides of the square 
can still be observed in Fig.~\ref{fqslna9} c). The drop in action density 
right
at the center observed in Figs.~\ref{fqba9} c), \ref{fqslna9} c) can be at 
least 
partly attributed to the poor performance of our variational basis for this 
$R$ value at the center
point as discussed in Sect.~\ref{sloper}. With a better basis we would 
expect the hole in the center of Fig.~\ref{fqslna9} c) to disappear and 
Fig.~\ref{fqba9} c)
to have more contribution in the place of the valley in the center.

The exploratory plots for the action in Figs.~\ref{fqba}--\ref{fqslda} show a 
more complicated structure than the corresponding ones for the energy. For 
$R=4,6$ there is an area of negative (``binding'') density around the center,
where the distribution has a positive sign. Positive contributions are also
found outside the corners of the square. For $R=6$ the negative area
is broken into four separate pieces. These complicated action distributions 
are in sharp contrast to the
the simple connected regularity of the binding distributions in the energy 
case. This fits in with the clear physical interpretation of the energy
distributions in this subtracted case, unlike the ones for the action.

\subsection{First excited state}

The first excited state of four quarks is not bound, and its wavefunction is 
close to $(|A\rangle-|B\rangle)/\sqrt{2}$ both when two or three basis 
states are considered. 

\begin{figure}[hbtp]
\hspace{0cm}\epsfxsize=200pt\epsfbox{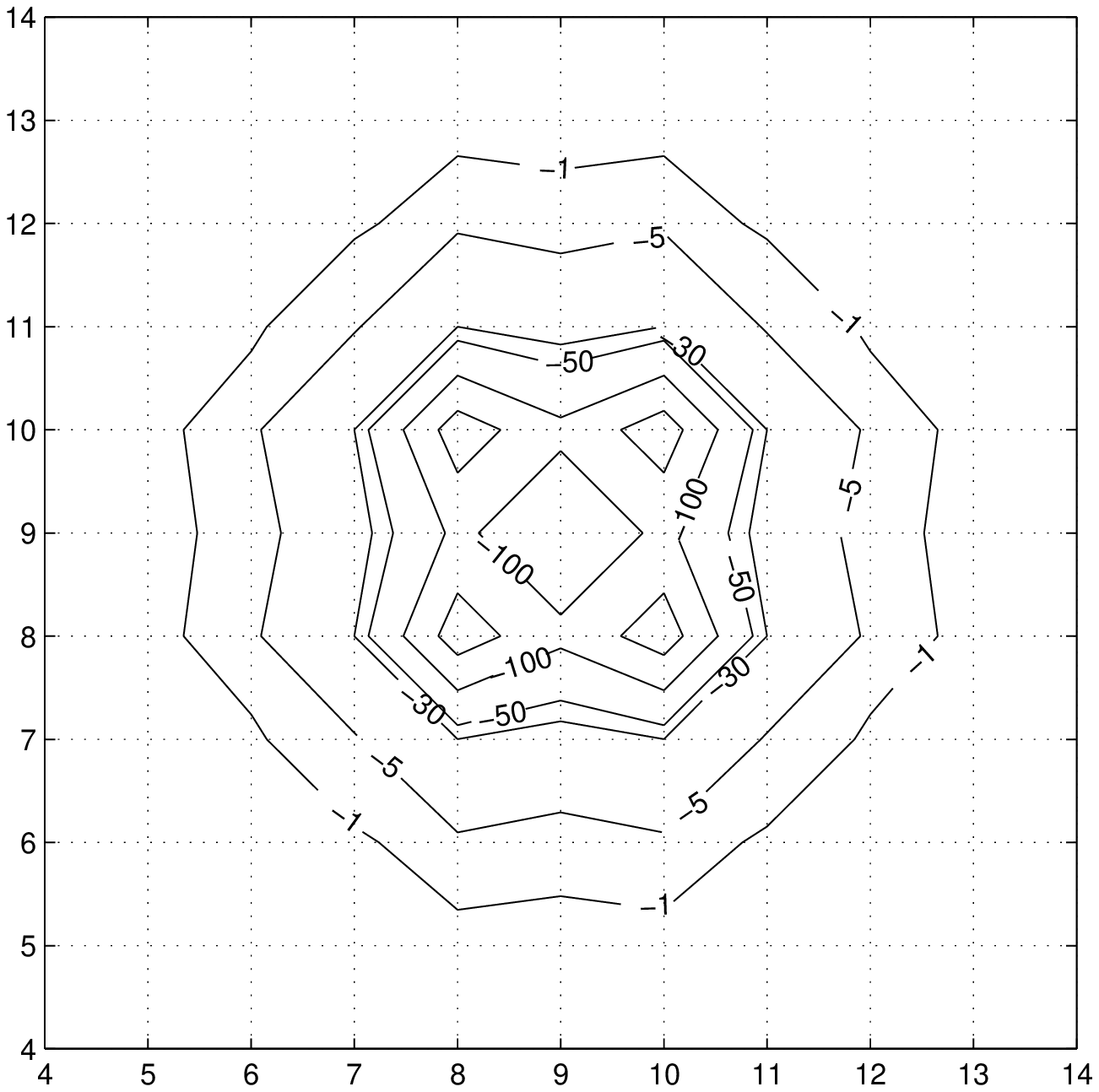}\epsfxsize=200pt\epsfbox{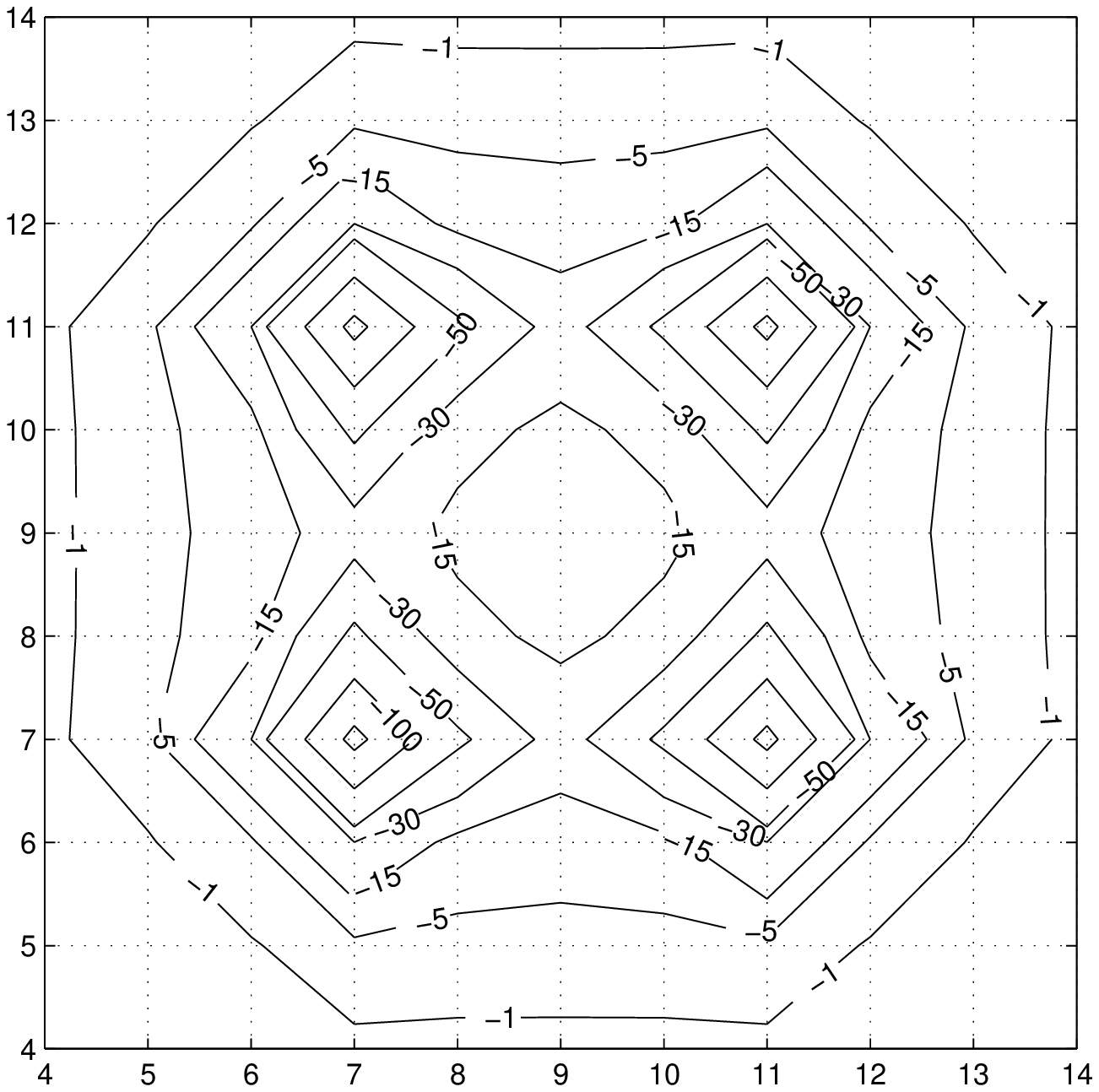} 
\caption{As in Fig.~\protect\ref{fqb9} but for the first excited state and 
$R=2,4$.} 
\label{feb9}
\end{figure}

\begin{figure}[hbtp]
\hspace{0cm}\epsfxsize=200pt\epsfbox{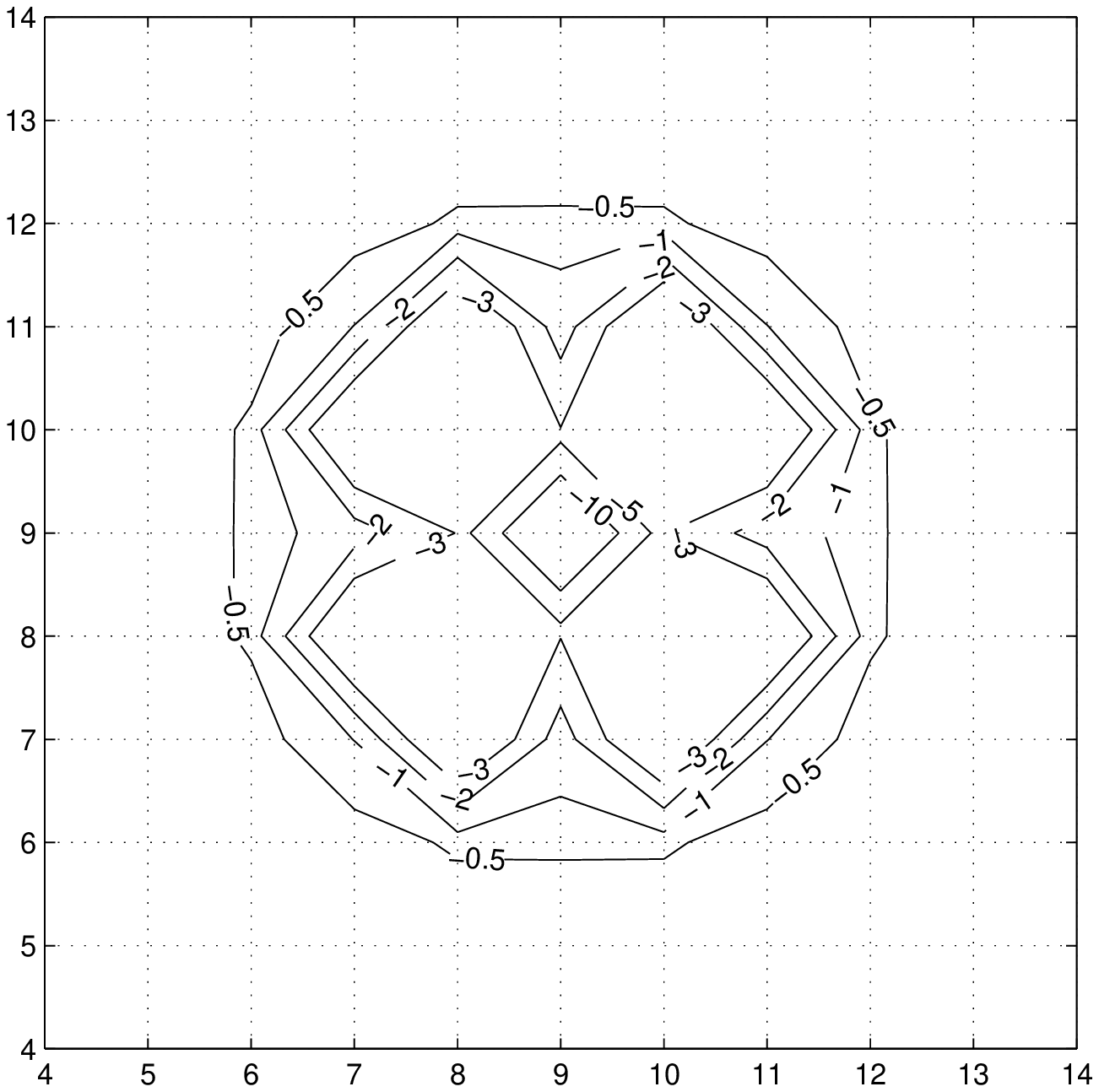}\epsfxsize=200pt\epsfbox{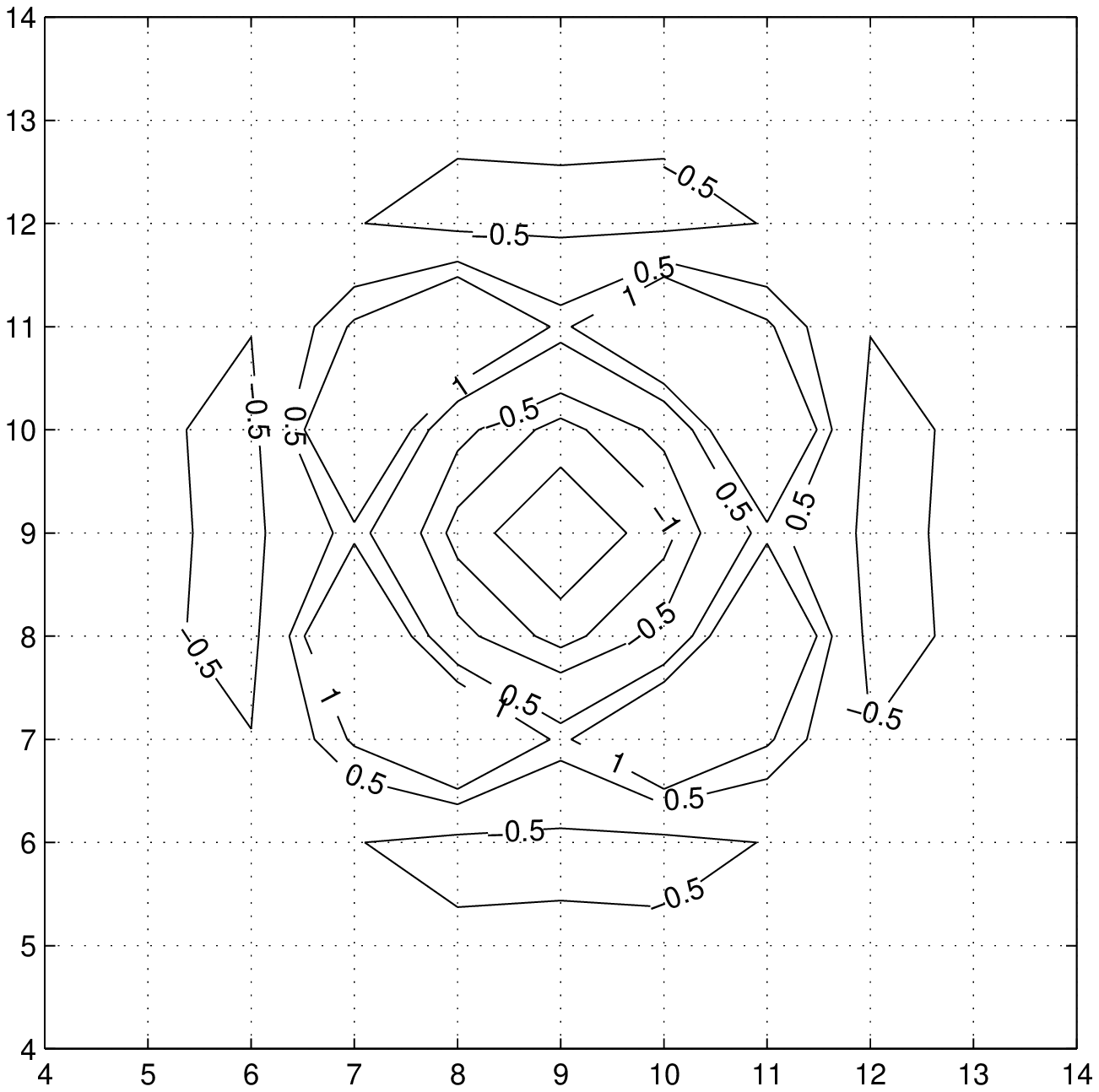}
\caption{As in Fig.~\protect\ref{fqba9} but for the first excited state.} \label{fea9}
\end{figure}

\begin{figure}[htbp]
\hspace{0cm}\epsfxsize=200pt\epsfbox{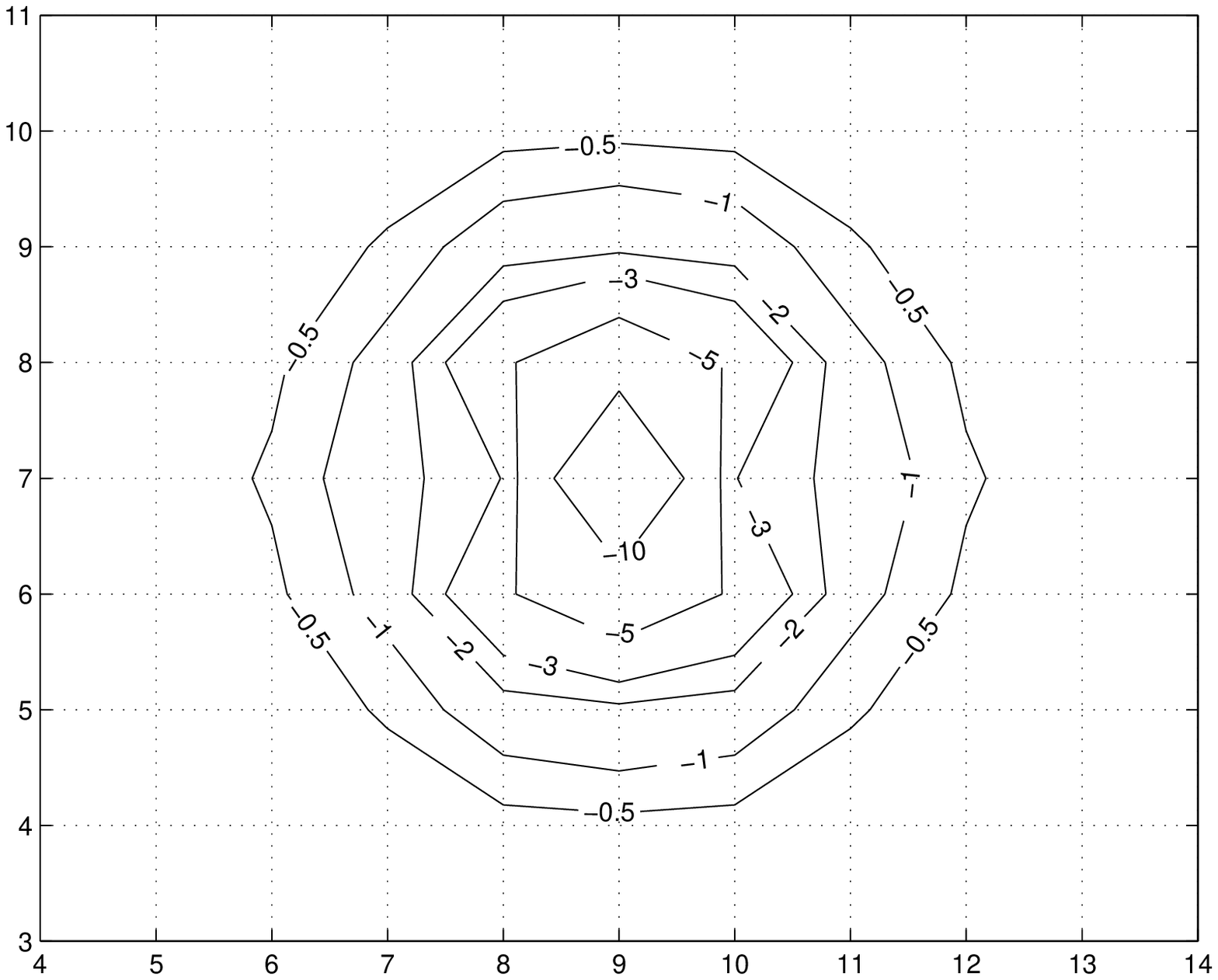}\epsfxsize=200pt\epsfbox{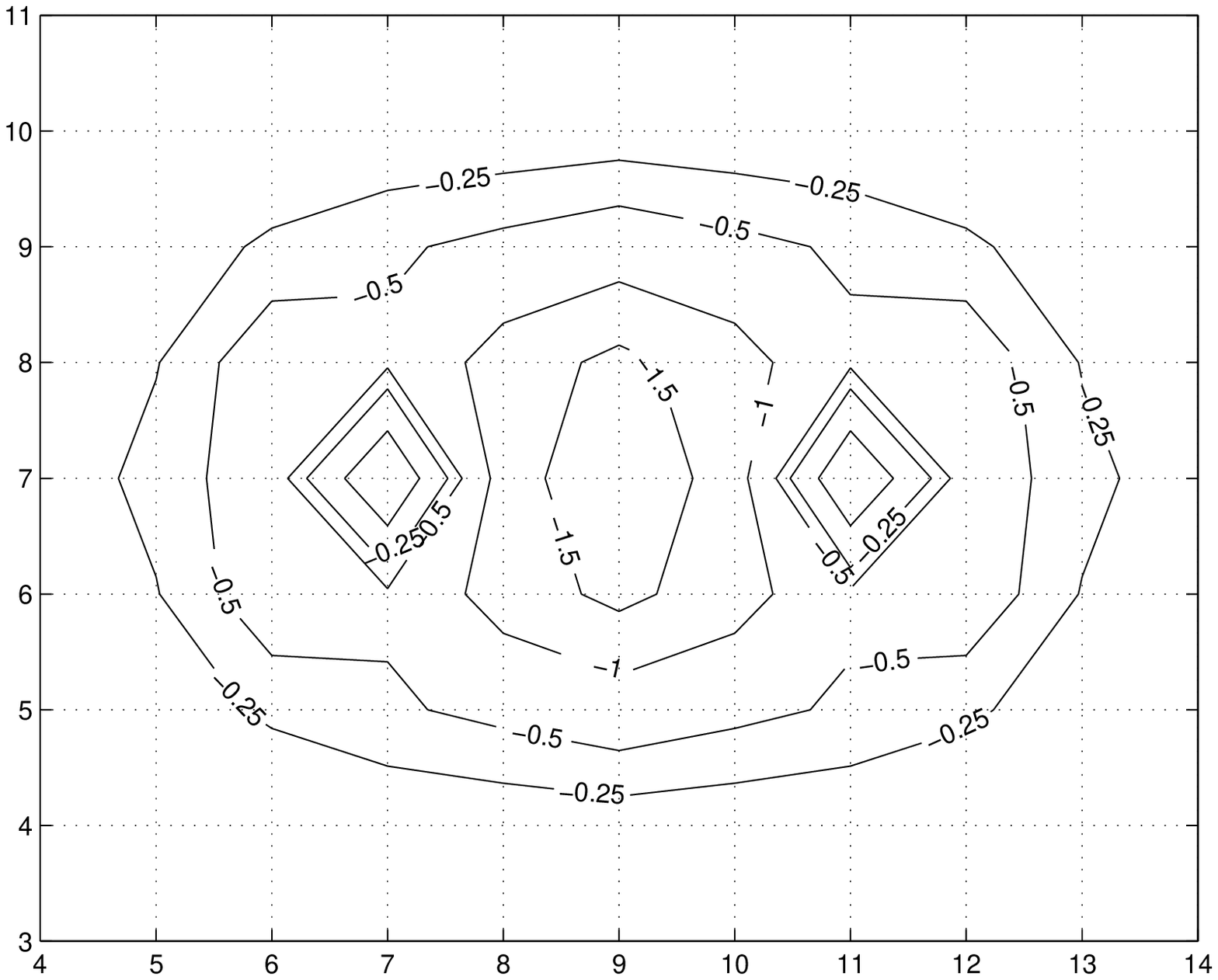} 
 \caption{As in Fig.~\protect\ref{fqslna9} but for the first excited state. }
 \label{feslna9}
\end{figure}

\begin{figure}[htbp]
\hspace{0cm}\epsfxsize=200pt\epsfbox{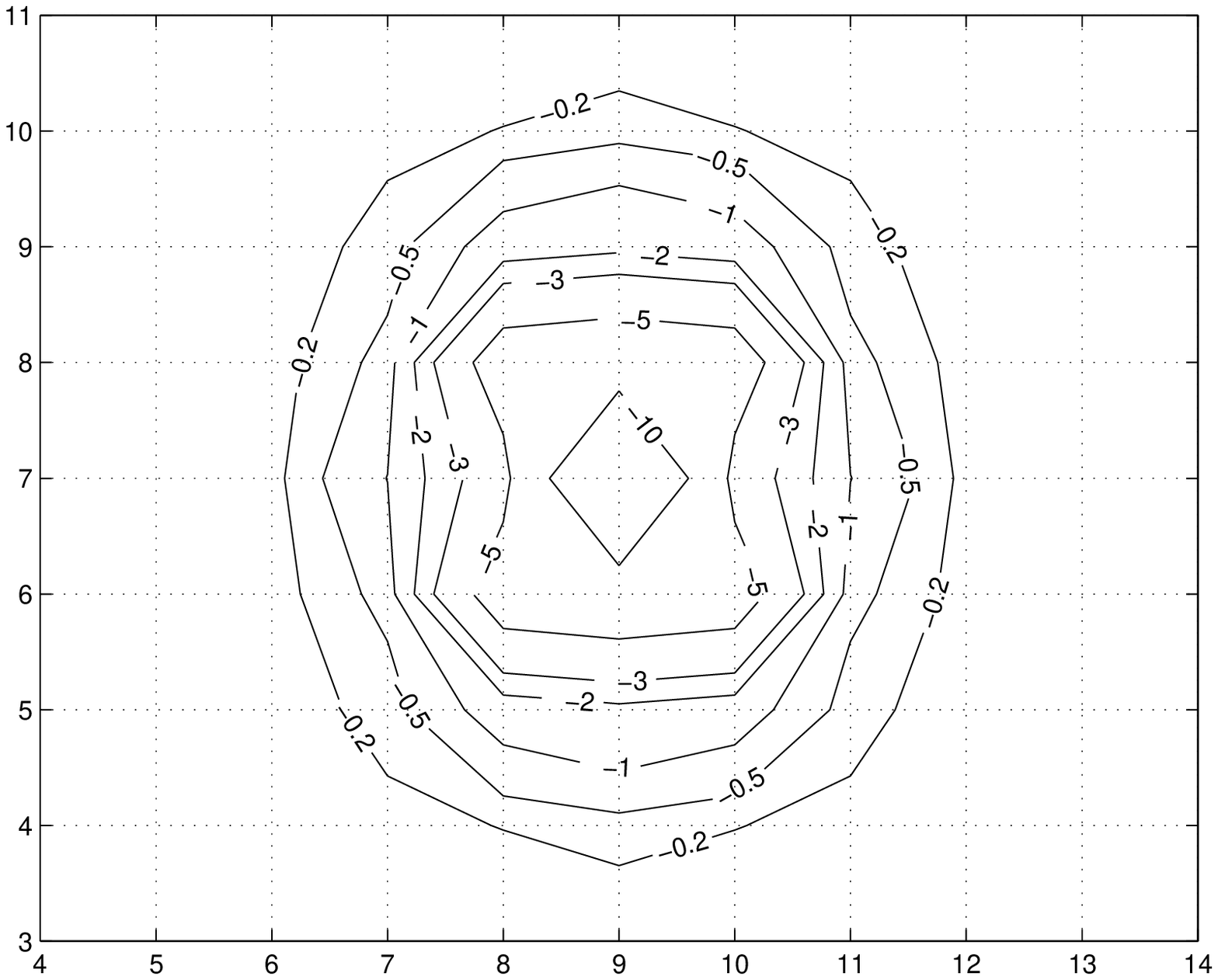}\epsfxsize=200pt\epsfbox{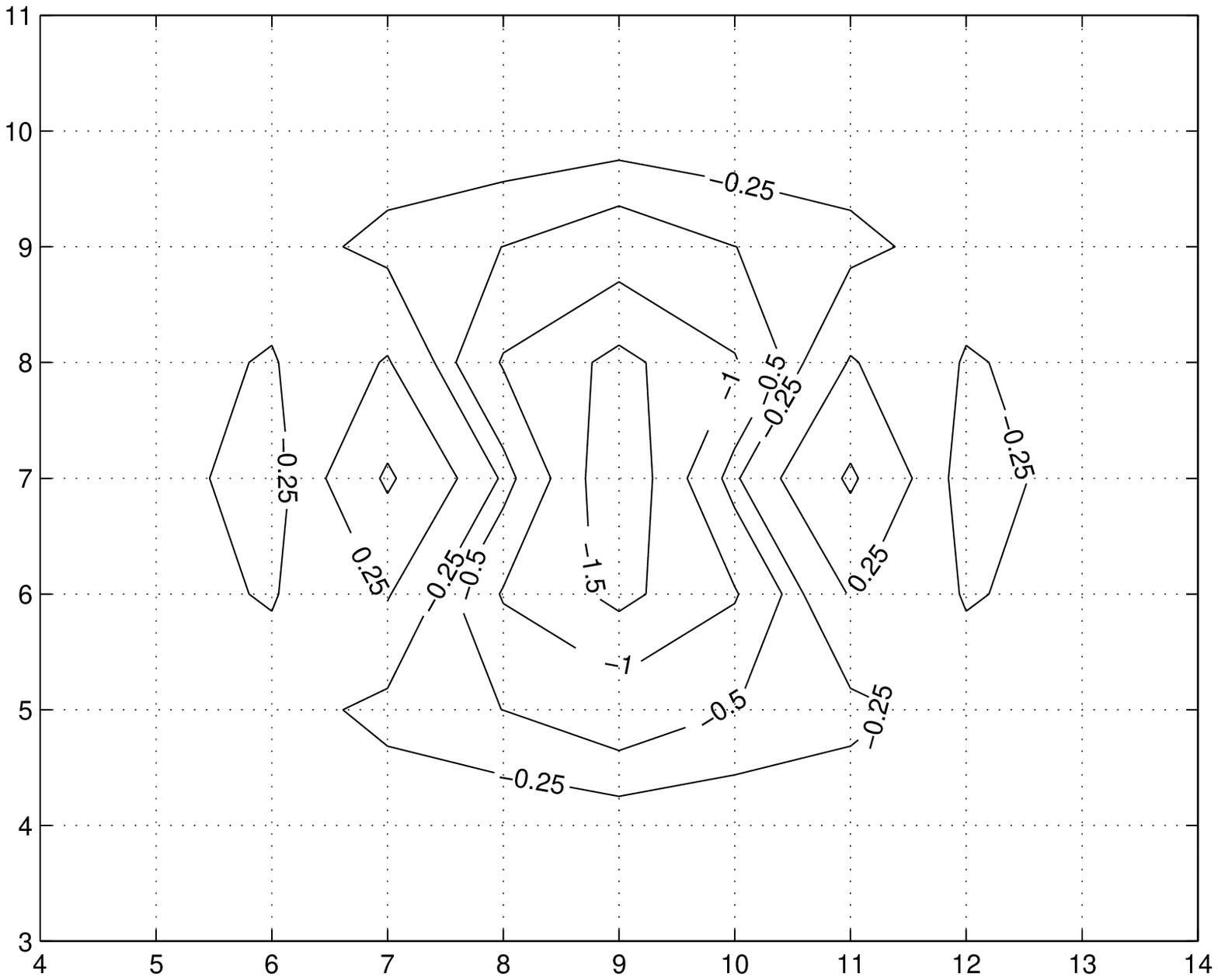} 
 \caption{As in Fig.~\protect\ref{fqslda9} but for the first excited state. }
 \label{feslda9}
\end{figure}

The energy distribution of this state is presented in 
Fig.~\ref{feb9} for $R=2,4$. These are taken at $T=3,2$ respectively, being
again guided by Tables~\ref{tsumcheck} and \ref{tsumcheck2}. Very little 
difference compared with the ground state pictures in Fig.~\ref{fqb9} can be 
seen.
However, after the ground-state two-body flux tubes are subtracted, a 
very different picture emerges -- as seen in Figs.~\ref{fea9}--\ref{feslda9}. 
The large negative  contributions (due to our sign convention) in these three 
figures are evidence of the unbound nature of the state. Comparison
of Figs.~\ref{fea9} and Fig.~\ref{fqba9} shows clearly the different symmetry
in this case; for the ground state a roughly spherical distribution is found
with concentrated areas at the sides of the square,
whereas for the excited state the distribution is concentrated in the
corners of the square near the quarks and decreased at the middle of the 
sides, showing a cloverleaf-shaped structure.
For $R=4$ the negative distribution is concentrated in the center with 
remnants outside the sides of the square, with a positive ``cloverleaf''
in between these regions, indicating a node in the wavefunction of the
excited state. In Figs.~\ref{feslna9} and \ref{feslda9} the distribution
can be seen to have a larger extent outside the quark plane than the ground
state. 

\subsection{Chromomagnetic fields}

As discussed in Sect. \ref{scolmeas}, we measure separately the spatial 
components of chromoelectric and -magnetic fields. However, full information 
on the {\em direction} of these fields is not available, as the measured 
quantities
correspond to the squares of the components. Therefore the pictures in this 
section have been created by inserting the signs of the components by hand.

\begin{figure}[htbp]
\hspace{0cm}\epsfxsize=200pt\epsfbox{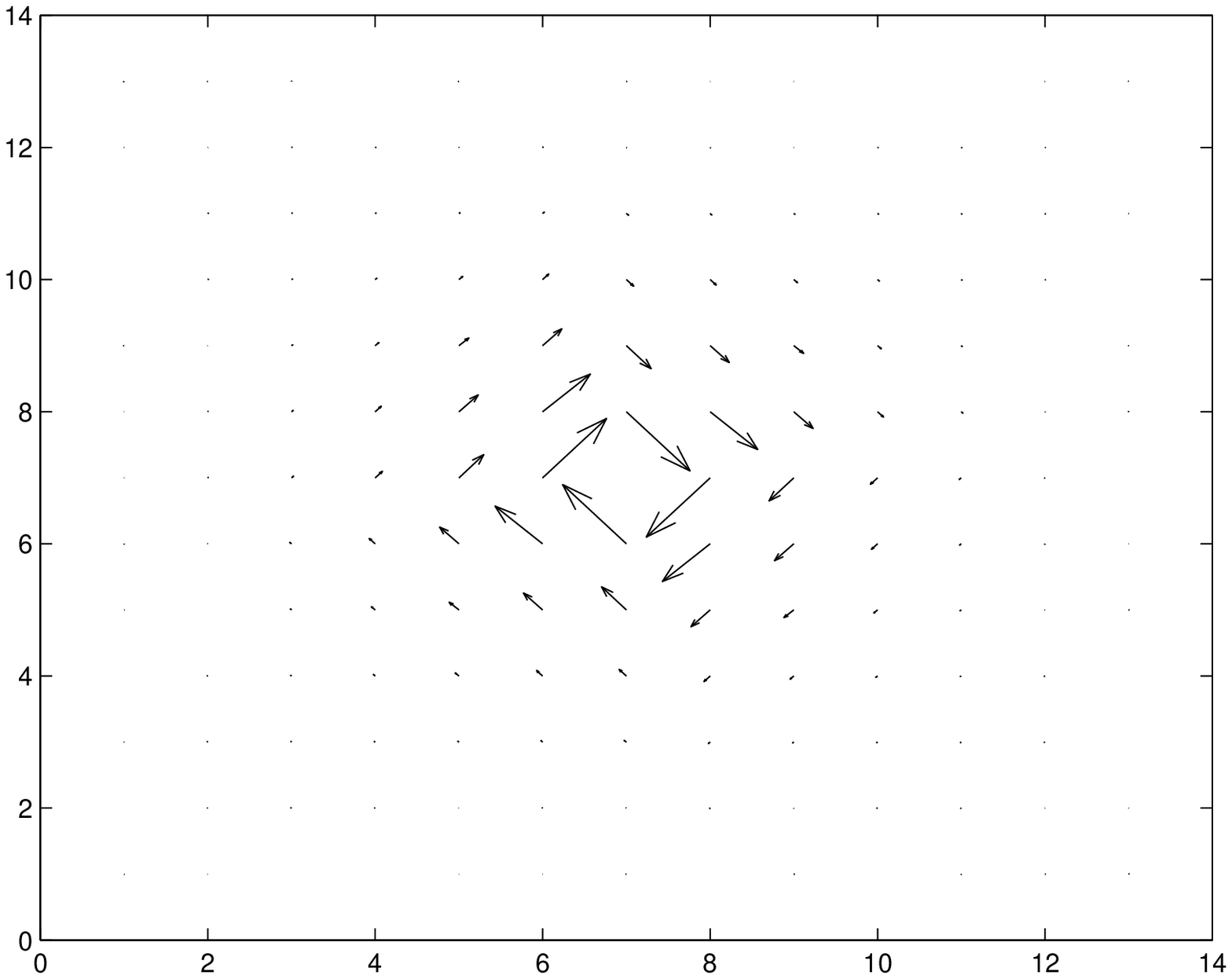}\epsfxsize=200pt\epsfbox{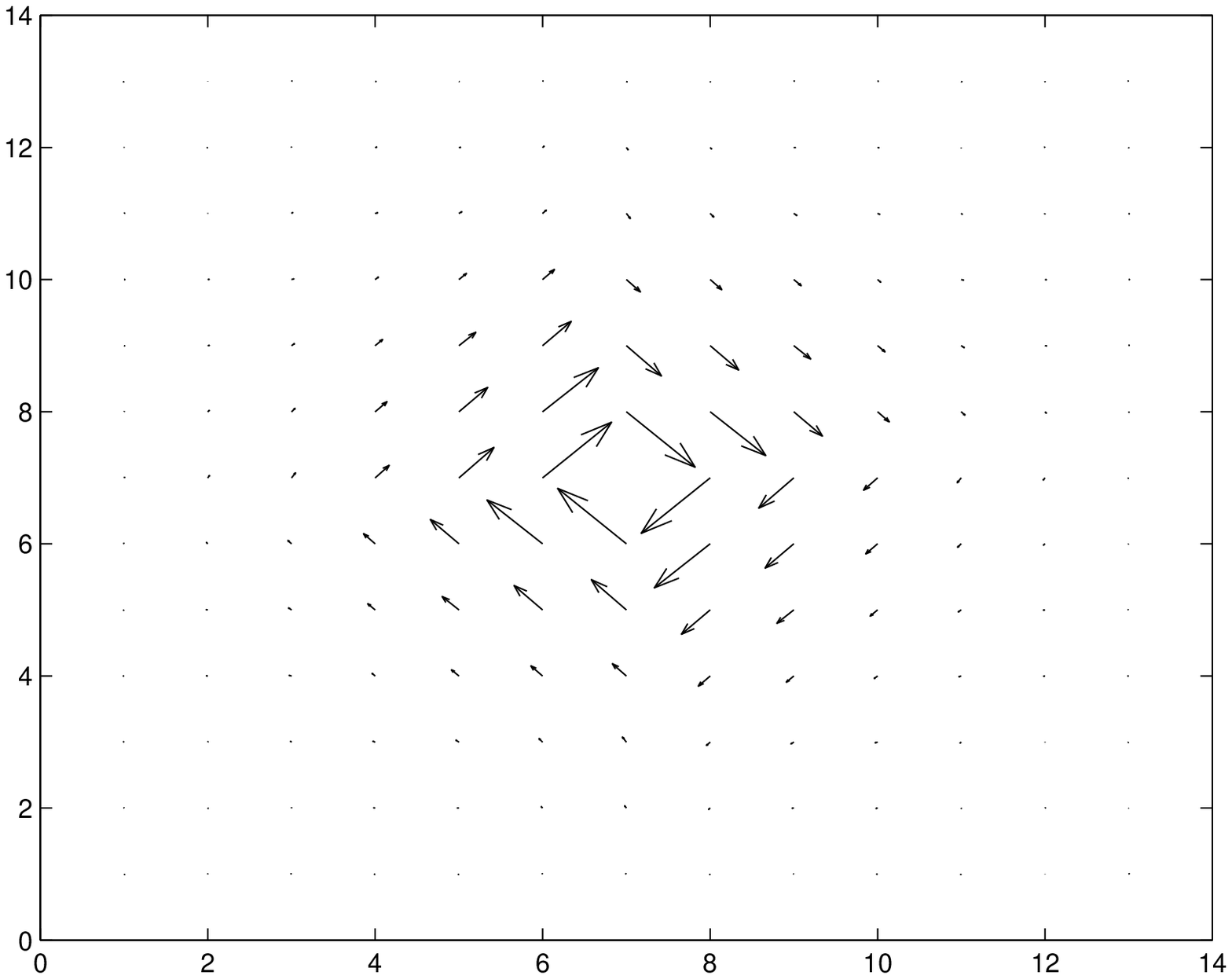} 
 \caption{Magnetic field in the transverse plane in the middle of two quarks at separation a) $R=4$ and b) $R=8$. }
 \label{fonmt}
\end{figure}

\begin{figure}[htbp]
\hspace{0cm}\epsfxsize=200pt\epsfbox{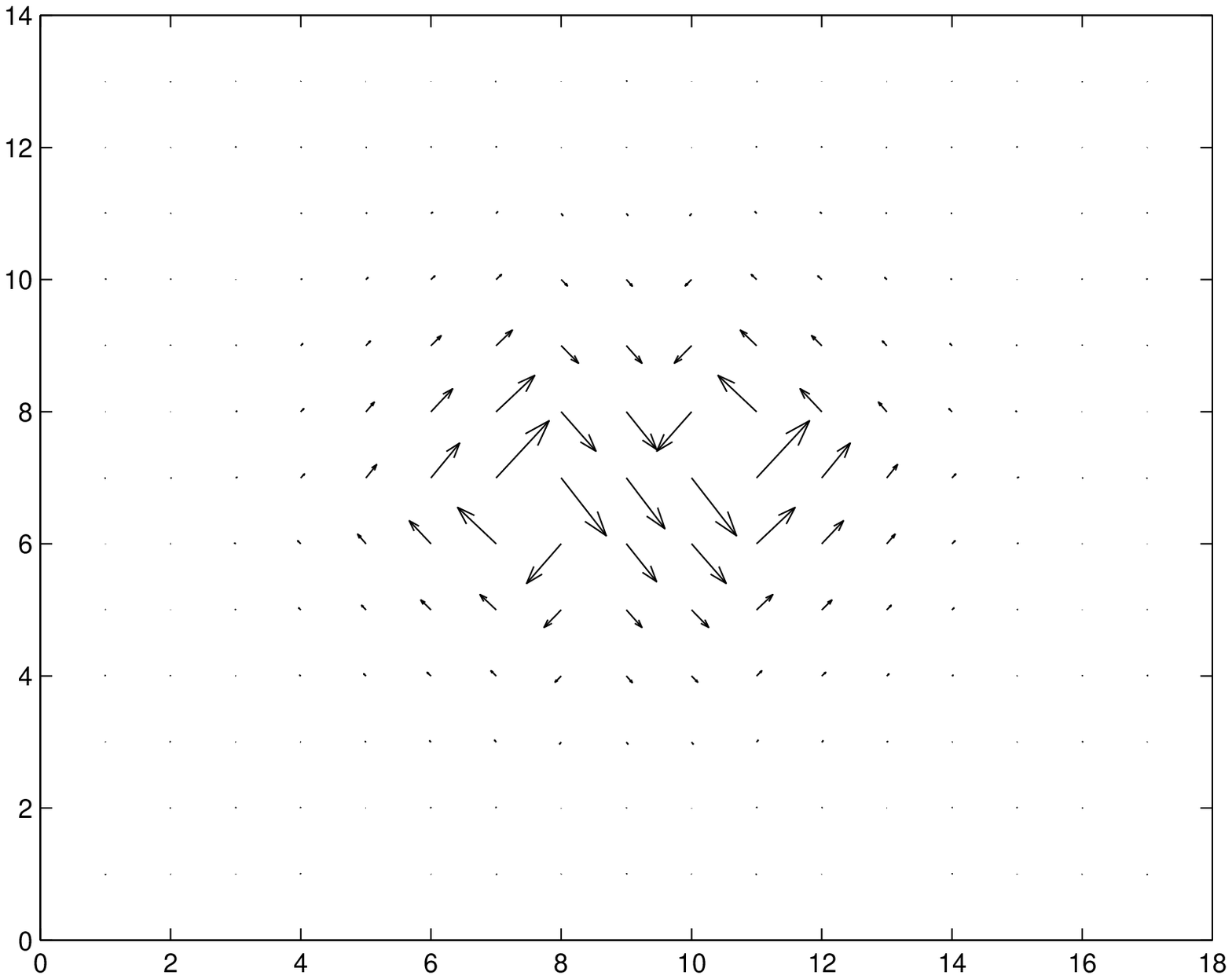}\epsfxsize=200pt\epsfbox{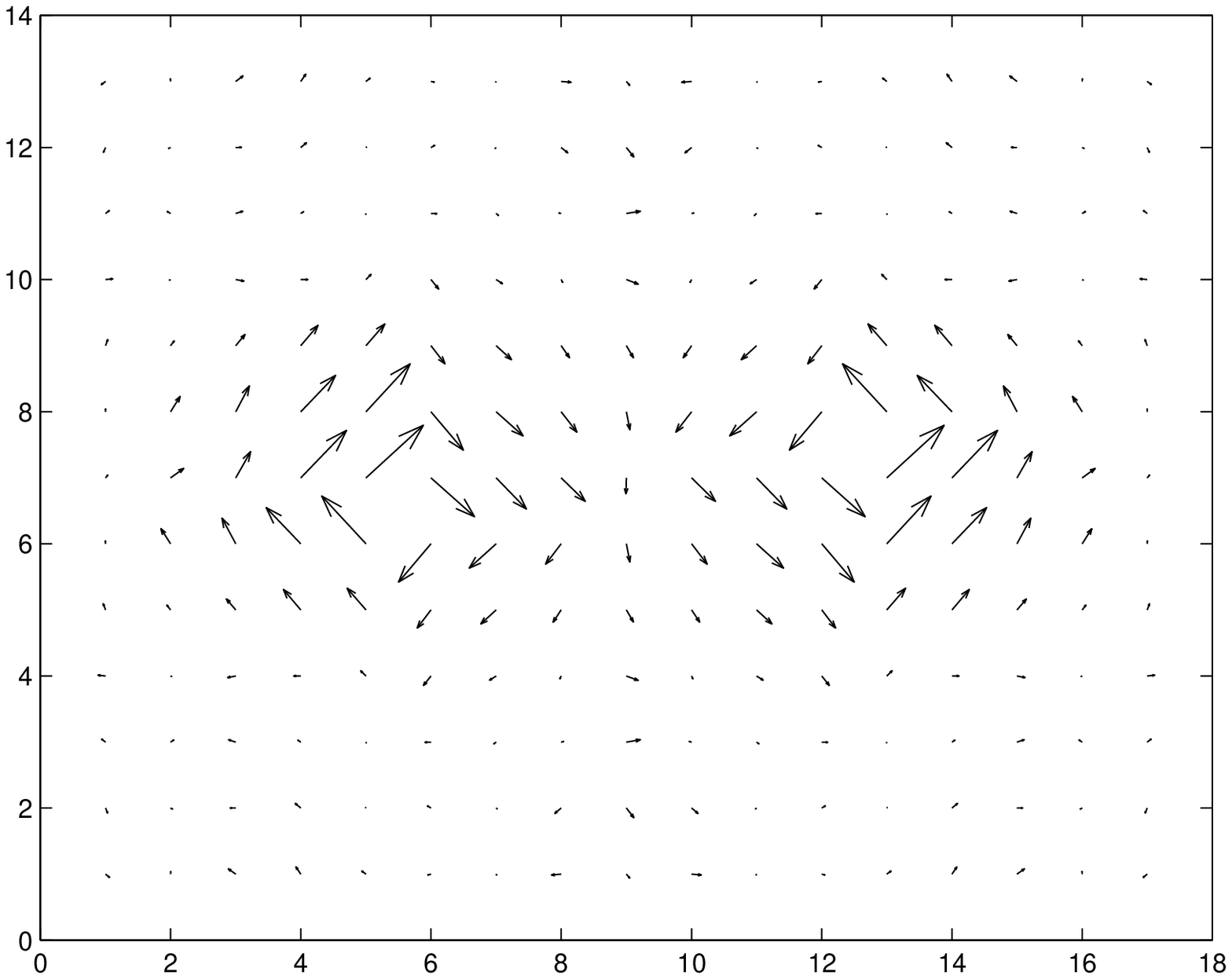} 
 \caption{Magnetic field in the transverse plane in the middle of four quarks at separation a) $R=4$ and b) $R=8$. }
 \label{fslnm}
\end{figure}

In Fig.~\ref{fonmt} for two quarks the magnetic field in the plane 
perpendicular to the 
interquark axis and in the middle of the quarks is shown. The signs have been 
chosen so that the magnetic field rotates around a flux-tube. In 
Fig.~\ref{fslnm} the 
magnetic field is shown for the four-quark case in the plane perpendicular
to the quark plane and cutting through the middle as in Fig.~\ref{fqsln}. 
Comparison between 
Figs.~\ref{fonmt} and \ref{fslnm} shows how the two-quark fields get 
distorted in the four-quark case -- an effect already seen in earlier figures.
The field in the middle of the four quarks can be seen to have a direction
more perpendicular to the quark plane in the $R=8$ case than for $R=4$. 

\section{Overlap of fluxes and a model for the energies \label{sf}}

A simple version of the binding energy model in Ref.~\cite{gmp:93} using 
only two basis states (A,B) 
reproduces well the observed ground and excited state binding energies of 
four quarks at the corners
of a square. Therefore it is interesting to see how the observed 
flux distributions corresponding to the binding energy relate to the model. 

The two-state version of the model gives the energies as eigenvalues $E(4)$ of
\begin{equation}
\label{VN}
\left[{\bf V}-E(4) {\bf N}\right]=0,
\end{equation}  
where
\begin{equation}
\label{NV}
{\bf N}=\left(\begin{array}{ll}
1&-f/N \\
-f/N&1\end{array}\right)\ \ {\rm and}\ \ {\bf V}=\left(\begin{array}{cc}
v_{13}+v_{24} & \frac{f}{N}V_{AB}\\
\frac{f}{N}V_{BA}&v_{14}+v_{23}\end{array}\right), 
\end{equation}
and $v_{ij}$ are the static two-body potentials between quarks $i$ and 
$j$. The matrix element $V_{AB}$ ($=V_{BA}$) comes from the perturbative 
expression
\begin{equation}
V_{ij}=-\frac{1}{N^2-1}{\bf T_i} \cdot {\bf T_j} v_{ij}=
-\left(v_{13} +v_{24} +v_{14}+v_{23} - v_{12}-v_{34}\right), \label{evij}
\end{equation}
where for a color singlet state $[ij]^0$ the normalization is chosen to give\\
$\langle [ij]^0|V_{ij}|[ij]^0\rangle =v_{ij}$. The four-quark 
binding energies $B(4)$ are obtained by subtracting the 
internal energy of the basis state with the lowest energy, e.g. 
$$
B(4) = E(4)-(v_{13}+v_{24}).
$$
In our case we take them from Table~\ref{tafter}.

A central 
element in this model is the phenomenological factor $f$ appearing in the 
overlap of the basis states $\langle A|B\rangle  = -f/N$ for $SU(N_c)$. This 
factor is a 
function of  the spatial coordinates of all four quarks, making the 
off-diagonal 
elements of ${\bf V}$ in Eq.~\ref{NV} {\em four-body} potentials. It 
attempts to take into account the decrease of overlap from the weak coupling 
limit, where $\langle A|B\rangle =-1/N$, to the strong coupling limit where 
$\langle A|B\rangle=0$.
Perturbation theory to $O(\alpha^2)$ also produces the two-state model of Eqs. 
\ref{VN}--\ref{evij} with $f=1$ \cite{lan:95}. 
A working parameterization for $f$ is
\begin{equation}
f = e^{-k_A b_S A-k_P \sqrt{b_S} P}. \label{ef}
\end{equation}
Here $b_S$ is the string tension and 
$k_A, k_P$ parameters multiplying, respectively, the minimal area and its 
perimeter bounded by the four quarks. In a fit to energies of square and 
tilted
rectangle geometries at $\beta=2.4$ in Ref.~\cite{pen:96b} the values of 
these parameters were $k_A=0.38(4), \; k_P=0.087(10)$. In a continuum 
extrapolation the $k_A$ increased to a value close to one and $k_P$ 
approached zero. Also
in Ref.~\cite{gre:98} a fit to many additional geometries, but with 
$k_P$ fixed at zero, yielded $k_A=0.57(1)$. 

With this model, the ground state energy for a square geometry is  
\begin{equation}
B(4) = \frac{f}{1+f/2} (v_s-v_d), \label{efen}
\end{equation}
giving 
\begin{equation}
f=\frac{B(4)}{v_s-v_d}(1-\frac{1}{2}\frac{B(4)}{v_s-v_d})^{-1}. \label{eff}
\end{equation}
Here $v_s, v_d$ are the two-body potentials $v_{ij}$ between quarks on one side
and in the opposite corners of the square, respectively. This leads to the 
values of $f$ shown in Table~\ref{tf}. The expression in Eq.~\ref{efen}
can now be rewritten in terms of the sums over the corresponding field 
distributions $\sum F$ in Eq.~\ref{sume} as
\begin{equation}
\sum FB(4)=\frac{f}{1+f/2} \sum \frac{1}{2} [F(AB)-F(C)] \label{efsum}
\end{equation}
Of course, if the sum rules were satisfied exactly, then this equation 
would add nothing new to our knowledge of $f$. However, as seen in 
Table~\ref{tsumcheck2}
the errors on some of the sums can be quite large. 
Therefore, if Eq.~\ref{efsum} is used to extract $f$ values, the resultant 
numbers 
are found to be only meaningful for the $R=2$ case -- as seen in 
Table~\ref{tf}.

\begin{figure}[h]
\hspace{0cm}\epsfxsize=220pt\epsfbox{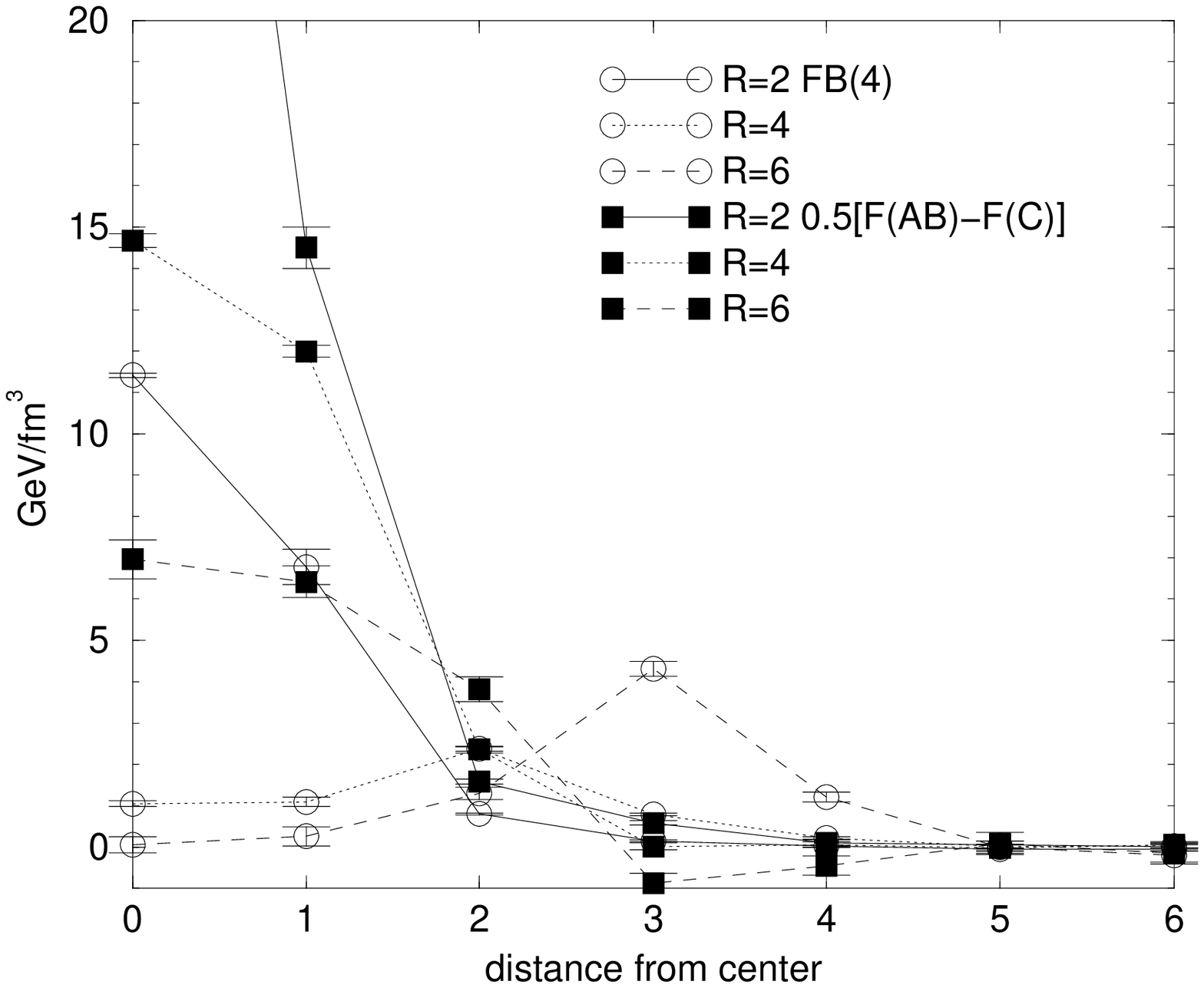} \epsfxsize=220pt\epsfbox{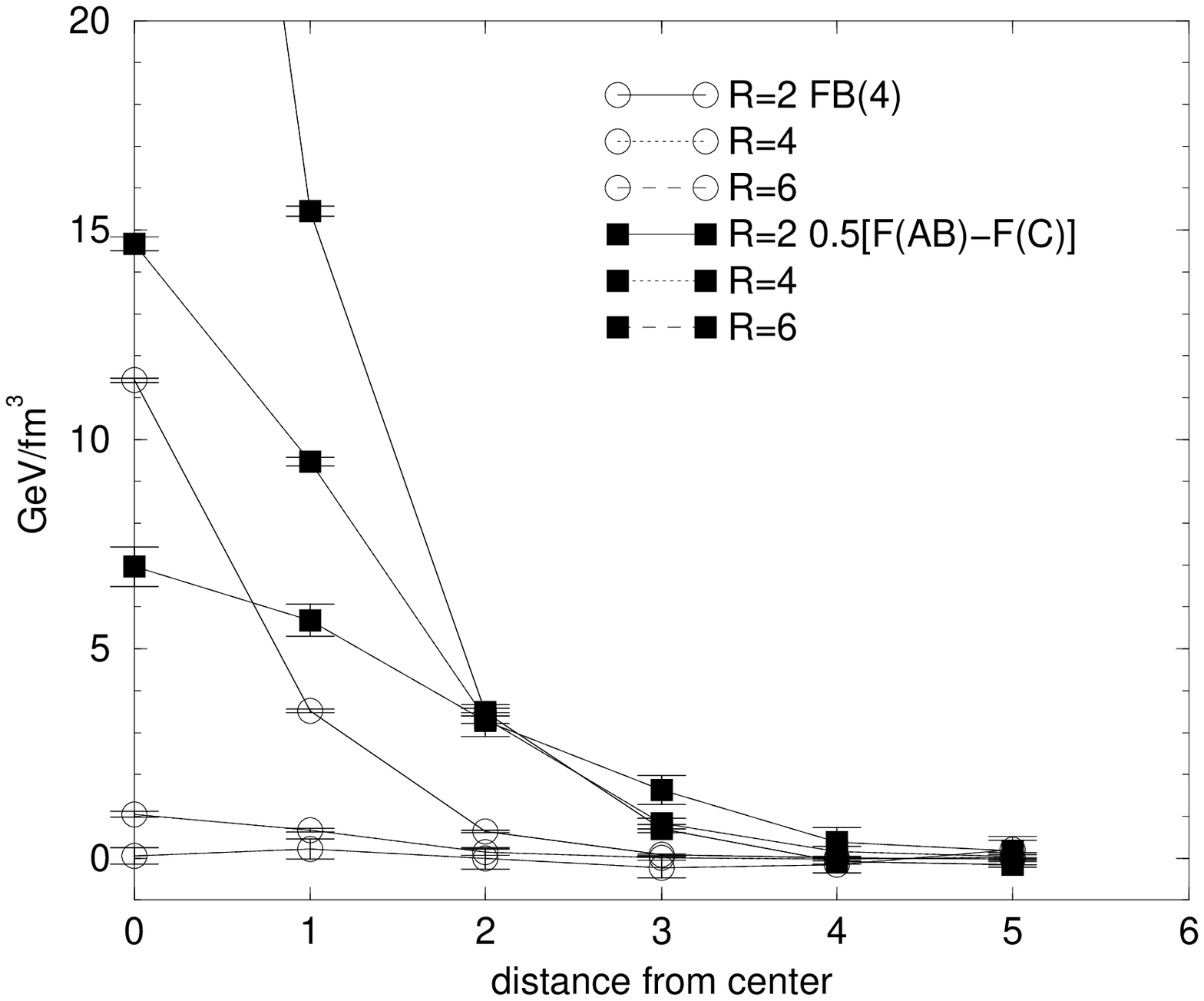} 
\caption{ $FB(4)$ and $0.5[F(AB)-F(C)]$ away from the center point -- a) on the quark plane for $R=2,4,6\;$ -- b) moving up from the quark plane. The $R=2$ data is 
taken at $T=3$ and the $R=4,6$ data at $T=2$.}
 \label{fa}
\end{figure}

\begin{figure}[h]
\hspace{0cm}\epsfxsize=220pt\epsfbox{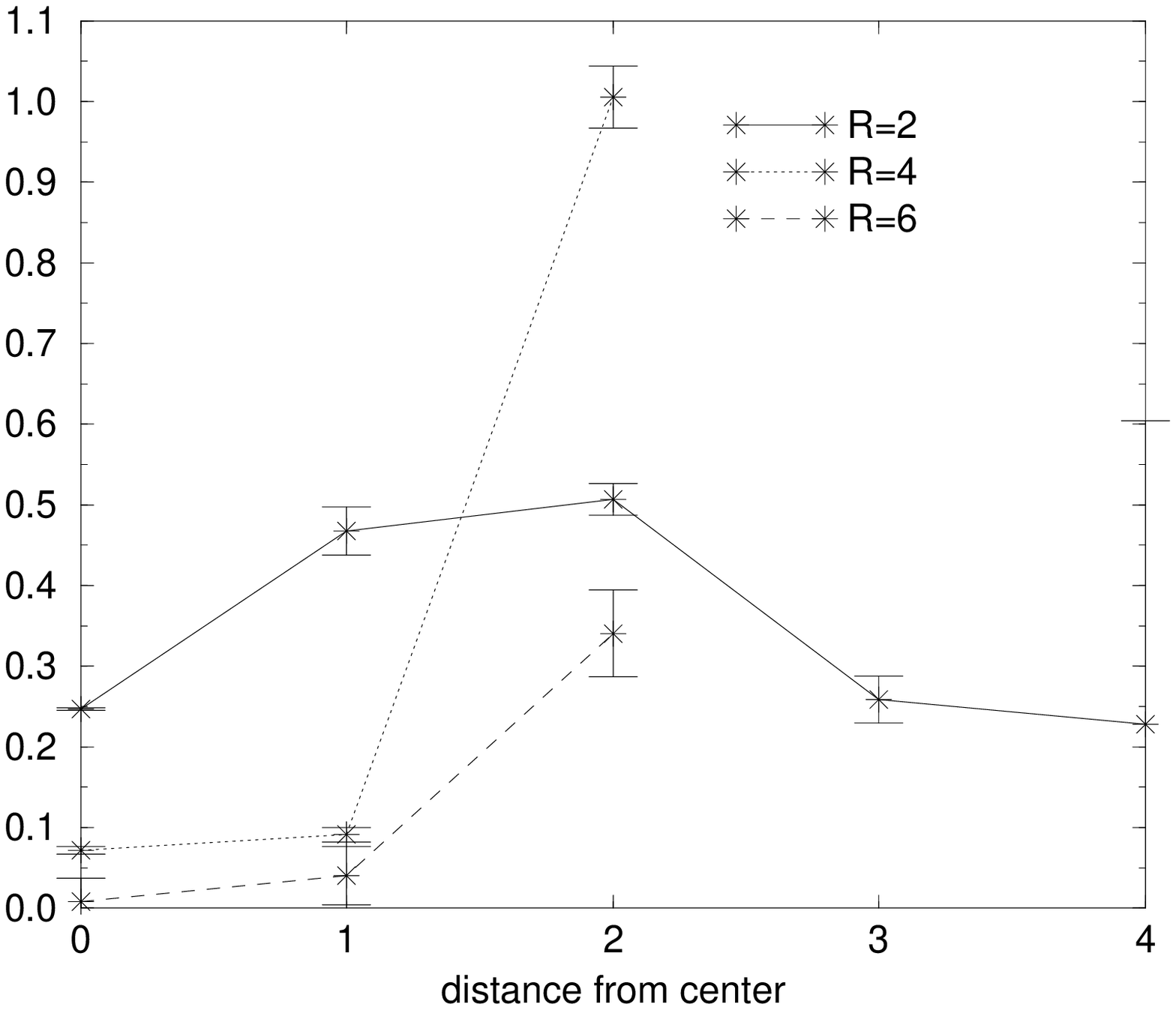} \epsfxsize=220pt\epsfbox{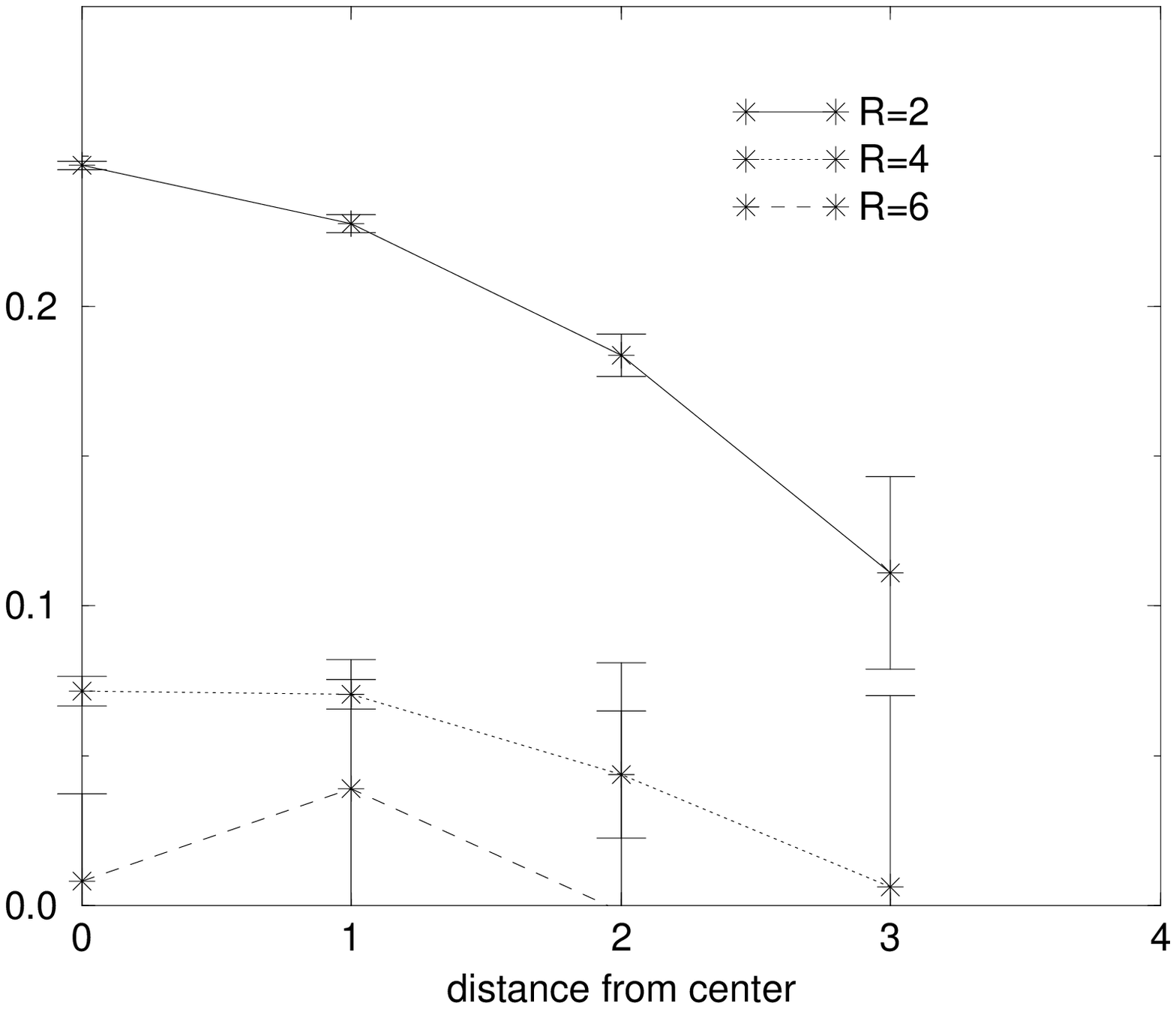} 
\caption{The ratio $2 FB(4)/[F(AB)-F(C)]$ on the same axis as in 
Fig.~\protect\ref{fa}.}
 \label{fa2}
\end{figure}

Our original hope when embarking on this aspect of the study was that a 
comparison could be made between the {\em integrands} in Eq.~\ref{efsum}, 
in order
to say more about the form of $f$. However, this has had only limited success.
The outcome is summarized in Figs.~\ref{fa} and \ref{fa2}. Figure~\ref{fa} 
shows 
the microscopic distribution of $FB(4)$ and 
$F(AB)-F(C)$, measured at the center point in between
the four quarks and moving away a) along the quark plane through the flux
tube in between two quarks or b) up in the direction normal to the
plane of the quarks. Figure~\ref{fa2} shows the ratio $2 FB(4)/[F(AB)-F(C)]$
on these same axes; the ratio 
$2 FB(4)/[F(AB)-F(C)]/(1-FB(4)/[F(AB)-F(C)])$ which is analogous to 
Eq.~\ref{eff}, but 
involves the integrands instead of the integrals in Eq.~\ref{efsum}, has
similar profiles (not shown).

In Figs.~\ref{fa} a,b) it is seen 
that the $0.5 [F(AB)-F(C)]$ profile drops away more rapidly than that for 
$FB(4)$. Therefore, if $-f/(1+0.5f)$ is interpreted as a form factor acting on 
the basic two-body profiles, it should not have a spatial extent larger
than the profile it is modifying. This interpretation should become
clearer for the larger values of $R$, where lattice artefacts play less of a 
role. Looking at $R=4,6$ it is seen that $0.5[F(AB)-F(C)]$, in fact, drops by
almost an order of magnitude on reaching the side of the square. Therefore 
the "extent" of $0.5[F(AB)-F(C)]$ is less than $R\times R$ -- 
being more like $(R-1)\times (R-1)$. This suggests that  the "extent" of 
$-f/(1+0.5f)$ and, likewise, $f$ should have the same limit.
Consequently, when $f$ is parametrized as in Eq.~\ref{ef}, it could be more
realistic to use, in place of the area $A=R^2$ contained
by the four quarks, an effective area that is somewhat less. This
interpretation fits in with
the value of $k_A < 1$ in Eq.~\ref{ef} obtained in Ref.~\cite{pen:96b}. In the 
continuum limit, however, $k_A$ was there found to approach one.

An alternative view that is more in line with the interpretation that $f$ is 
a form factor is to rewrite Eq.~\ref{ef} as
$f=\exp(-A/A_{\rm effective})$, where  $A_{\rm effective}=(k_A b_S)^{-1}$. Here
the perimeter term has been forgotten.
For the case of squares a sensible definition of "range" is then
$R_{\rm effective}=\sqrt{A_{\rm effective}}$. As stated after Eq.~\ref{ef}, in 
Ref.~\cite{gre:98} 
for the two basis state model the value of $k_A$ is 0.57(1)  giving
$R_{\rm effective}=5.0$.

\begin{table}[htb]
 \begin{center}
\begin{tabular}{l|l|ccc}
$R$ & type   & $T=2$ & $T=3$ & $T=4$ \\ \hline
2   & energy &  0.7393(26) & 0.7586(33) & 0.7635(39) \\
    & sum 1  &  0.69(12)   & 0.63(16)   & 0.56(21)   \\
    & sum 2  &  0.70(13)   & 0.59(17)   & 0.55(25) \\
4   & energy &  0.4760(23) & 0.4688(36) & 0.4655(56) \\
    & sum 1  & 0.9(2.7) &  \\
    & sum 2  & 0.9(3.9) & 0.39(33) &  \\
6   & energy &  0.1504(61) & 0.1931(81) & 0.190(24) \\
    & sum 1  &  0.42(81) &  &  \\
8   & energy &  0.0373(22) & 0.0541(91) & 0.048(48)\\ 
\end{tabular}
 \caption{The value of $f$ from energies and energy sums \label{tf}}
\end{center}
\end{table}

However, the main weakness in the above comparison of integrands is that, 
although 
the two basis state model ($A,B$ in Fig.~\ref{fpair}) is able to give a 
good fit to 
much of the four-quark data in Ref.~\cite{glpm:96}, it is unable to explain 
other 
data -- in particular that of four quarks at the corners of a regular
tetrahedron. In Refs.~\cite{gre:97,gre:98} it is shown that a more successful 
model
utilizes six basis states -- $A,B,C$ in Fig. 1 and $A^*,B^*,C^*$ where each
quark pair is now in an excited gluonic state. For this model  
the $A^*,B^*,C^*$ contribution begins to dominate as the interquark distances
increase. For example, with $\beta=2.4$ the $A,B,C$ component contributes
only 40\% (10\%) to the binding energy of four quarks at the corners of a 
square with sides $R=4(6)$. Another feature of this extended
version is that $k_A$ in Eq.~\ref{ef} becomes 1.51(8), giving 
$R_{\rm effective}=3.1$.
This implies that the longer range in the two basis state model is merely 
simulating the effect of $A^*,B^*,C^*$ and that, when these three states 
are treated explicitly, the basic interaction containing $f$ is of 
shorter range. But it is beyond the scope of the present study
to pursue this further, since it requires ingredients that are not
available from the present calculation -- in particular for two quarks the 
profiles of fields where the glue is in an excited state with $E_u$ symmetry.

\section{Conclusions \label{sconc}}

We have measured the full flux distributions of two quarks and four quarks 
at the corners of a square
in quenched SU(2) lattice gauge theory with $\beta=2.4$ on a $20^3\times 32$ 
lattice.
Multihit variance reduction was used to improve the signal on temporal links 
and switched off at the quark lines for proper measurement of self-energies.
The effect from the multihit was helpful, but not as dramatic as expected.
Using values of generalized $\beta$-functions from Ref.~\cite{pen:97b} we 
were able to use lattice sum rules, giving either the observed energy or 
zero, to see
where the measurements were expected to be most accurate. This strategy was 
particularly useful
after self-energies were removed by subtracting two distributions, and 
enabled us to choose the best data to analyse. 

The four-quark distributions in Figs.~\ref{fqb}--\ref{fqsld9} show how
the interaction pulls the distribution to the middle of the quarks. This 
effect decreases when the quarks move further apart. 

The distributions corresponding to the binding energies of the quarks, obtained
by subtracting the distributions of the lowest-lying two-quark pairings 
from the four quark one, are shown in Figs.~\ref{fqba9}--\ref{fqslda9}.
They can be seen to form a ``cushion'' of approximately constant density and 
height in between the quarks with tubes of larger density in between nearest 
neighbor quarks. 

For the first excited state of the four quarks we observe -- after 
subtraction -- an energy field structure that is 
much more complicated (Fig.~\ref{fea9}) than that for the ground state 
(Fig.~\ref{fqba9}).
This presumably arises because the states $A$ and $B$ are basically
combined as $A+B$ in the ground state and $A-B$ in the excited state, the
latter leading naturally to cancellations. As a general statement it 
is seen that the energy profiles in Figs.~\ref{fqba9}--\ref{feslda9} show an 
increasing amount
of fine detail -- all of which is 'real' in the sense of being larger
than the statistical errors. This data is a real challenge for any model that
claims to simulate the original gauge field theory. Unfortunately, at
present, such  theories are in their infancy. For example, the dual
model of Ref.~\cite{bak:96} has had some success in describing -- for two 
quarks --
the energy profile for the gluon field in its ground state. However, so far
it has been unable to say anything about four quarks or excited gluon 
fields in the two-quark case.

Our original hope was that these residual fields would give some 
guidance when constructing models that are explicitly dependent only on the 
quark positions. In the case of the model presented in 
Sect.~\ref{sf} the main
conclusion was that the model was seen to be qualitatively consistent
with the data. The data was unable to say anything about the actual form
of the multi-quark interaction term $f$ in Eq.~\ref{ef} besides the fact that
it should be contained inside the area of the four quarks -- suggesting
an effective interaction area somewhat smaller than the full $R^2$ area
of the square. Such a smaller area is consistent with our earlier fit results.
However, only when the six basis state model  
has replaced the two basis state analysis of Section~\ref{sf} can more definite
statements be made.

Apart from this paper, we are not aware of any theoretical attempts to 
understand the four-quark flux distribution. Hopefully the data presented
here will be useful for such attempts. 

\section{Acknowledgement}

The authors wish to thank P. Kurvinen for developing much of the analysis code.
Funding from the Finnish Academy and M. Ehrnrooth foundation (P.P) is 
gratefully acknowledged. Our simulations were performed on the Cray C94 at 
the Center for Scientific Computing (CSC) in Espoo. We thank M. Gröhn from 
CSC for help in visualizing the color fields.

\newcommand{\href}[2]{#2}\begingroup\raggedright\endgroup

\end{document}